\documentclass{aa}  
\bibpunct{(}{)}{;}{a}{}{,} % to follow the A&A style

\usepackage[usenames, dvipsnames]{color}
\usepackage{graphicx}
\usepackage{float}
\usepackage{placeins}
\usepackage{multicol}
\usepackage{subfigure} %Can I use?
\usepackage{multirow} % For table 8
\usepackage{booktabs} % For table 8
\usepackage{longtable} % For longtable
\usepackage{txfonts}
\usepackage{amsmath}
\usepackage{hyperref}

\begin{document}

   \title{Modelling simple stellar populations in the near-ultraviolet to near-infrared with the X-shooter Spectral Library (XSL)}
    \titlerunning{The X-shooter Spectral Library (XSL) simple stellar population models}
   \subtitle{}
     
      \author{Kristiina Verro\inst{\ref{inst1}}
             \and S.~C. Trager\inst{\ref{inst1}}
            \and R.~F. Peletier\inst{\ref{inst1}}
            \and A. Lan\c{c}on\inst{\ref{inst4}}
            \and \\A. Arentsen\inst{\ref{inst4}}
            \and Y.-P. Chen\inst{\ref{inst6}}
            \and P.~R.~T. Coelho\inst{\ref{inst11}}
            \and M. Dries\inst{\ref{inst1}}
            \and J. Falc{\'o}n-Barroso\inst{\ref{inst8},\ref{inst9}}
            \and \\A. Gonneau\inst{\ref{inst2},\ref{inst4}}
            \and M. Lyubenova\inst{\ref{inst3},\ref{inst1}}
            \and L. Martins\inst{\ref{inst12}}
            \and P. Prugniel\inst{\ref{inst5}}
            \and P. S{\'a}nchez-Bl{\'a}zquez\inst{\ref{inst10}}
            \and A. Vazdekis\inst{\ref{inst8},\ref{inst9}}                }

\institute{Kapteyn Astronomical Institute, University of Groningen, Landleven 12, 9747 AD Groningen, the Netherlands \\
\email{verro@astro.rug.nl} \label{inst1}
    \and
    Observatoire Astronomique de Strasbourg, Universit\'e de Strasbourg, CNRS, UMR 7550, 11 rue de l'Universit\'e, \\F-67000 Strasbourg, France\label{inst4}
    \and
	Institute of Astronomy, University of Cambridge, Madingley Road, Cambridge CB3 0HA, United Kingdom \label{inst2}
	\and
    ESO, Karl-Schwarzschild-Str. 2, D-85748 Garching bei  M\"unchen, Germany \label{inst3}
    \and
    CRAL-Observatoire de Lyon, Universit\'e de Lyon, Lyon I,  CNRS, UMR5574, France \label{inst5}
    \and
    New York University Abu Dhabi, Abu Dhabi, P.O. Box 129188, Abu Dhabi, United Arab Emirates\label{inst6} 
    \and 
    Instituto de Astrof\'isica de Canarias, V\'ia L\'actea s/n, La Laguna, Tenerife, Spain\label{inst8}
    \and
    Departamento de Astrof\'isica, Universidad de La Laguna, E-38205 La Laguna, Tenerife, Spain\label{inst9}
    \and
    Departamento de F\'isica de la Tierra y Astrof\'isica, UCM, 28040 Madrid, Spain\label{inst10}
    \and
    Universidade de S\~{a}o Paulo, Instituto de Astronomia, Geof\'isica e Ci\^{e}ncias Atmosf\'ericas, Rua do Mat\~{a}o 1226, 05508-090, S\~{a}o Paulo, Brazil\label{inst11}
    \and
    NAT - Universidade Cidade de São Paulo, Rua Galvão Bueno, 868, São Paulo, Brazil
    \label{inst12}
    %\email{verro@astro.rug.nl}
    }

  \date{Received 7 October 2021; Accepted ?? ? ????}
    
    \abstract{We present simple stellar population models based on the empirical X-shooter Spectral Library (XSL) from near-ultraviolet (NUV) to near-infrared (NIR) wavelengths. The unmatched characteristics of relatively high resolution and extended wavelength coverage (350--2480\,nm, $R\sim10\,000$) of the XSL population models bring us closer to bridging optical and NIR studies of intermediate and old stellar populations. It is now common to find good agreement between observed and predicted NUV and optical properties of stellar clusters due to our good understanding of the main-sequence and early giant phases of stars. However, NIR spectra of intermediate-age and old stellar populations are sensitive to cool K and M giants. The asymptotic giant branch, especially the thermally pulsing asymptotic giant branch (\mbox{TP-AGB}), shapes the NIR spectra of 0.5–2 Gyr old stellar populations; the tip of the red giant branch defines the NIR spectra of populations with ages larger than that. We therefore construct sequences of the average spectra of static giants, variable \mbox{O-rich} giants,  and \mbox{C-rich} giants to include in the models separately. The models span the metallicity range $-2.2<\mathrm{[Fe/H]}<+0.2$ and ages above 50 Myr, a broader range in the NIR than in other models based on empirical spectral libraries. We focus on the behaviour of colours and absorption line indices as a function of age and metallicity. Our models can reproduce the integrated optical colours of the Coma cluster galaxies at the same level as  other semi-empirical models found in the literature. In the NIR, there are notable differences between the colours of the models and Coma cluster galaxies. Furthermore, the XSL models expand the range of predicted values of NIR indices compared to other models based on empirical libraries. Our models make it possible to perform in-depth studies of colours and spectral features consistently throughout the optical and the NIR range to clarify the role of evolved cool stars in stellar populations.}

    % 5 {} token are mandatory


   \maketitle
%
%________________________________________________________________
\section{Introduction}

Stellar population models are fundamental in determining the basic properties of unresolved stellar systems. Those properties include the initial mass function (IMF), star formation rate, star formation history, total mass in stars, and stellar metallicity and abundance patterns \citep[see review by][]{Conroy2013}. 
With the next generation wide-field spectroscopic facilities, such as the upcoming WEAVE for the William Herschel Telescope \citep[Jin et al., in prep.]{WEAVE1}, MOONS for the Very Large Telescope \citep{MOONS}, and 4MOST for the Visible and Infrared Survey Telescope for Astronomy \citep{4MOST}, spectroscopic information of different types of galaxies in various environments will increase in quantity and in quality. Furthermore, with recent advances in near-infrared instrumentation on large telescopes, such as X-shooter \citep{Vernet2011} and KMOS \citep{KMOS1, KMOS2} on ESO's VLT, or the forthcoming HARMONI on the ELT \citep{HARMONI}, the domain of evolved cool stars in stellar populations will be increasingly more accessible. Stellar spectral libraries and associated stellar population models need to keep up with these developments.

An increasing effort has been put into developing better stellar population models that are based on empirical stellar libraries. The goals are to build models with higher resolution and longer wavelength ranges, based on stellar spectral libraries covering the Hertzsprung–Russell diagram (HR diagram henceforth) more extensively than ever before. For example, the widely used UV--IR \citet{BC2003} models are still based on theoretical spectra across large wavelength regions, while MILES stellar population models \citep{MILESI, MILESIII}, which are based on the fully empirical MILES library \citep{MILESlib2006,MILESII}, have been extended towards the NIR and UV over the years, resulting in the E-MILES models with a wavelength coverage of 1680--50\,000\,\AA\ \citep{MIUSCAT, EMILES, UVMILES}. The recent SDSS MaStar stellar population models \citep{Maraston20}, with a wavelength coverage of 3600--10\,300\,\AA\, are based on nearly 9000 stars, a 10-fold increase on the previous generation of models, although covering only the optical wavelength range. With these modern stellar population models, it is now common to find good agreement between the observed and the predicted NUV and optical properties of stellar clusters \citep[e.g][]{Peacock_2011,Ricciardelli2012,Conroy2018}. This consensus shows our good understanding of the main-sequence and early giant phases that constitute the near-ultraviolet and optical light of stellar populations. 

However, the existing optical-to-NIR stellar population models have problems. The NIR traces populations of a range of ages and suffers lower dust extinction than the optical. But we are far from a complete understanding of some stellar evolutionary phases which strongly affect the spectral energy distributions of stellar populations in the NIR \citep[e.g.][]{Mouhcine2002,UVMILES, Baldwin2018, Riffel2019}. The asymptotic giant branch (AGB), especially the thermally pulsating AGB (\mbox{TP-AGB}), shapes the NIR spectra of 0.5--2\,Gyr old stellar populations; the tip of red giant branch (RGB) defines the NIR spectra of older populations. Current stellar population models in the NIR are based on available empirical libraries such as \citet{Pickles1998, LW2000}, (E-)IRTF (see below), or theoretical stellar spectra such as MARCS \citep{MARCS}, PHOENIX \citep{Husser2013} or BaSEL \citep{Lejeune1997,Lejeune1998,Westera2002} models. The IRTF Spectral Library of \citet{IRTF} and the extended-IRTF of \citet{EIRTF} are empirical libraries of 0.8--5.0\,$\mu$m and 0.7--2.5\,$\mu$m (respectively) stellar spectra observed at a resolving power of $R = 2000$ with the SpeX spectrograph at the NASA Infrared Telescope Facility on Mauna Kea. The original IRTF library covers mainly solar-metallicity late-type stars (but also some oxygen-rich and carbon-rich AGB stars); the E-IRTF expands the metallicity coverage. The E-MILES, \citet{Conroy2018} and \citet{MenesesGoytia2015b} stellar population models take advantage of either the IRTF or E-IRTF library. The empirical library of \citet{LW2000} has a spectral resolution $R\sim 1000$, and is limited to cool giant and supergiant stars only. \citet{Mouhcine2002} and \citet{Maraston2005} have included these spectra in their stellar population models.

None of these empirical libraries have extensive coverage of the important stellar evolutionary stages needed for stellar population modelling in the NIR. Furthermore, (O- and C-rich \mbox{TP-)AGB} and RGB stars are rarely segregated in stellar population modelling. This leads to a large variety of optical-to-NIR stellar populations, which in turn lead to discrepancies between the star formation histories (SFH henceforth) derived from optical and NIR spectral ranges, or from different models. An example is the \citet{Maraston2005} set of SSP models, which have enhanced flux and strong molecular carbon and oxygen absorption features through-out the near-infrared spectra of intermediate age populations compared to, for example, \citet{BC2003}, E-MILES and \citet{Conroy2018} models. Such strong molecular bands predicted by the \citet{Maraston2005} models have been detected in some studies \citep[e.g][]{Lyubenova2012}, but not in others \citep[e.g][]{Zibetti2013}. Recent works in stellar evolution theory \citep{Girardi2013, Pastorelli2020} explain this observational discrepancy with the `AGB boosting' effect, which is linked to the physics of stellar interiors -- stellar populations in a narrow 1.57 and 1.66 Gyr age range at Magellanic Cloud metallicities have a factor of $\sim$2 increase of the \mbox{TP-AGB} contribution to the integrated luminosity of the stellar population. Some of the \citet{Lyubenova2012} globular clusters of the Magellanic Clouds are in this very narrow age and metallicity range and show clear spectral features of \mbox{TP-AGB} stars, while the post-starburst galaxies of \citet{Zibetti2013} are probably not within this range. Besides the inclusion of the \mbox{TP-AGB} phase into the stellar population models, the overall quality and coverage of the stellar spectral library is important. \citet{Baldwin2018} found that the largest differences in derived SFHs are caused by the choice of stellar spectral library and suggested the inclusion of high-quality NIR stellar spectral libraries into stellar population models should be a top priority for modellers. 

Furthermore, theoretical stellar spectra cannot be used at present to make accurate predictions for NIR spectra of stellar populations, as they have considerable problems in reproducing spectral energy distributions and molecular bands of observed cool stars \citep[e.g.][]{MartinsCoelho2007, Kurucz2011, Coelho2014, Knowles2019,Martins2019, Coelho2020,Lancon2021}. These stars are very difficult to model due to processes like hot-bottom burning, stellar winds, long-period pulsations, presence of circumstellar dust and the third dredge-up \citep{Pastorelli2019, Pastorelli2020}. 

Another common limitation of existing stellar population models based on empirical stellar libraries is their low spectral resolution. Stellar population models which are based on empirical stellar libraries, typically have resolutions of $R\sim2000$. For example, the commonly used spectral-line index system (LIS henceforth) of \citet{MILESI} suggests using LIS-5.0\AA{} ($R\sim1000$ at the Mg triplet at 5170\,\AA, which corresponds to a velocity dispersion of $127 ~\mathrm{km~s}^{-1}$) to study low-velocity-dispersion systems such as globular clusters or dwarf galaxies. Higher spectral resolution is required for more detailed modelling of emission and absorption lines; for example, higher-resolution spectroscopy can provide more accurate measurements of numerous absorption lines for many different chemical elements. A notable high-resolution empirical stellar population model is the Pegase.HR stellar population models \citep{LeBorgne2004,LeBorgne2011}, with $R=10\,000$ over 4000--6800\,\AA{} wavelength range, based on the ELODIE stellar spectral library \citep{ELODIE2001,ELODIE2004,ELODIE2007}. For example, \c{S}en et al. (in prep.) defined a set of line indices with a resolution of $\sigma =25 ~\mathrm{km~s}^{-1}$ using the Pegase.HR models and determined abundance ratios of 11 elements in small, unresolved galaxies outside the Local Group. They found that the majority of their dwarf galaxies have abundance ratios slightly less than solar.

Here we present simple stellar population models\footnote{The models are available in electronic form on the XSL web-page \url{http://xsl.astro.unistra.fr}} based on 639 stellar spectra from the X-shooter Spectral Library data release 3 (XSL DR3: Verro2021a). This new library is designed for stellar population purposes, with unprecedented simultaneous wavelength coverage of 3500\,\AA--2.48\,$\mu$m with a resolution of $\sigma \sim 13~\mathrm{km\,s^{-1}}$  (corresponding to $R\sim 10\,000$). XSL aims to cover the entire HR diagram as extensively as possible, with an emphasis on the advanced stellar evolutionary stages. We incorporate spectra of 44 oxygen rich, cooler than 4000\,K (quasi-)static stars, 39 oxygen rich \mbox{TP-AGB} stars and 26 spectra of carbon rich \mbox{TP-AGB} stars into our new stellar population models. With this development, the XSL simple stellar population models will help us to bridge the optical and the NIR studies of intermediate and old stellar populations and clarify the role of evolved cool stars in stellar population synthesis. The moderately high resolution of XSL population models creates new possibilities in the optical to NIR for absorption-line index studies. 

This paper is structured as follows: we review the main ingredients for stellar population models in Sect. \ref{sect:mainingredients} and describe the model calculation in Sect. \ref{sect:modelcalc}. We describe and analyse the general behaviour of the new XSL simple stellar population models in Sect. \ref{sect:generalbehaviour}. We compare colours and absorption-line indices with observed galaxies in Sect. \ref{sect:comacolours} and \ref{sect:Riffelindices} respectively. Furthermore,
we provide the stellar mass-to-light ratios in Sect. \ref{sect:MLratios} and further comment the effects of cool giant stars on the population models in \ref{sect:static/variable}. Finally, in Sect. \ref{sect:safety} we define the regions in age and metallicity where the XSL stellar population models are safe to use.

Throughout this paper, the $UBVRI$ magnitudes are in the Johnson-Cousins system, and $JHK$ magnitudes are in the homogenized Bessell system \citep{BessellBrett1988} (Vega system).

%__________________________________________________________________

%%%%%%%%%%%%%%%%%%%%%%%%%%%%
\section{Main ingredients for stellar population models}
\label{sect:mainingredients}
%%%%%%%%%%%%%%%%%%%%%%%%%%%%

The construction of simple stellar population (SSP, henceforth) models is rather straightforward, as it consists of only three ingredients -- stellar evolution theory (isochrones), an IMF, and a stellar spectral library. These ingredients are typically combined by Equation \ref{Eq:SSP}: \\
\begin{eqnarray} \label{Eq:SSP}
    f_\mathrm{SSP}(t,[\mathrm{Fe/H}]) &=& \int_{m_\mathrm{low}}^{m_\mathrm{high}(t)} f_* \left[T_\mathrm{eff} (M), \log g (M) |t,[\mathrm{Fe/H}] \right] \nonumber \\ 
    && \qquad\quad\times\,\Phi(M)\,dM,
\end{eqnarray}
where $M$ is the initial stellar mass, $\Phi(M)$ is the IMF, $f_*$ is the spectrum of a star of mass $M$ of effective temperature $T_\mathrm{eff}$ and surface gravity $\log g$ at metallicity $[\mathrm{Fe/H}]$, and $f_\mathrm{SSP}(t,[\mathrm{Fe/H}])$ is the resulting spectrum of a stellar population of a certain age ($t$) and metallicity $[\mathrm{Fe/H}]$\footnote{$[\mathrm{Fe/H}] =  [\mathrm{M}/\mathrm{H}] = \log(Z/X)-\log(Z/X)_\odot$, with $(Z/X)_\odot=0.0207$ and $Y=0.2485+1.78Z$ for the PARSEC/COLIBRI isochrones and $Z_\odot=0.019$ and $Y=0.23+2.25Z$ for the Padova00 isochrones (see Sect.~\ref{sect:isochrones}).}, and the integration runs from the lowest stellar mass in the IMF, $m_\mathrm{low}$, to the highest stellar mass still living at time $t$, $m_\mathrm{high}(t)$. The nuances of the ingredients themselves are what make population modelling difficult in practice. We recommend the review by \citet{Conroy2013} for an overview of this broad topic. We discuss the specific choices for the simple stellar population models in this work below. 

\subsection{Isochrones}\label{sect:isochrones}

Due to the extension of these SSP models to the NIR, the advanced evolutionary stages of stars become extremely important. XSL contains a large number of evolved cool giants, which makes synthesizing realistic stellar populations in the NIR possible, as long as we know how to integrate them into the models. 

An isochrone determines the location of stars with the same age and metallicity on the HR diagram and is constructed from stellar evolution calculations. On one hand, our selection of isochrones is motivated by the thorough treatment of the advanced evolved stages. With that in mind, we use the PARSEC/COLIBRI isochrones. The PARSEC version 1.2S models \citep{PARSEC2012,Tang2014,Chen2014MNRAS,Chen2015} describe the evolution of stars from pre-main sequence stars to the first thermal pulse in the helium shell, after forming an electron-degenerate carbon-oxygen core. Then the COLIBRI models \citep{Marigo2013,Rosenfield2016,Pastorelli2019,Pastorelli2020} add the \mbox{TP-AGB} evolution, from the first thermal pulse to the total loss of envelope. These isochrones have the most advanced handling of \mbox{TP-AGB} stars to date, based on high-quality observations of resolved stars in the Small Magellanic Cloud with detailed stellar population synthesis simulations computed with the TRILEGAL code \citep{TRILEGAL2005}. On the other hand, we aim to calculate stellar population models from simpler and more widely used stellar evolution tracks as well. The \citet[][Padova00 henceforth]{Girardi2000} isochrones allow us to compare directly our models with the E-MILES stellar population models. These tracks include a simple but synthetic treatment until the tip of the AGB, but they do not include a third dredge-up. Therefore there is no transition from \mbox{O-rich} to \mbox{C-rich} \mbox{TP-AGB} stars in the tracks, and so these isochrones are missing these stars.
\subsection{IMF}

An IMF describes the initial distribution of masses for a population of stars formed at the same time. XSL stellar population models are calculated using a \citet{Salpeter1955} or a \citet{Kroupa2001} IMF. The Salpeter IMF is a single power law with an $\alpha= 2.35$ slope, and is valid for $0.4 < m/M_\odot < 10$. The Kroupa IMF is a double power law, with $\alpha= 1.35$ slope for $m/M_\odot < 0.5$ stars, and $\alpha= 2.35$ for higher mass stars. In both cases, we use the relation in the mass range $0.09 < m/M_\odot < 120$, using extrapolation to lower masses. More XSL SSP models, calculated with various IMFs, will be presented and discussed in Verro et al. (in prep.). 

%%%%%%%%%%%%%%%%%%%%%%%%
\subsection{The X-shooter Spectral Library (XSL)}
%%%%%%%%%%%%%%%%%%%%%%%%
\label{XSL}
XSL is a moderate-resolution ($R\sim10\,000$) NUV--NIR stellar spectral library intended for stellar population modelling. We are using the XSL DR3 data to construct stellar population models. In Verro et al. (2021a), we provided 830 spectra of 683 stars, which are corrected for galactic extinction and merged to the full wavelength range of X-shooter, 350--2480 nm. The data were homogeneously reduced and calibrated in \citet{DR2}. XSL spectra are given in rest-frame, at a resolution $\sigma = 13,\,11,\,16\,\mathrm{km\,s}^{-1}$ in UVB, VIS and NIR arms respectively \citep{DR2}. \citet{Arentsen2019} provided a uniform set of stellar atmospheric parameters -- effective temperatures, surface gravities, and metallicities -- for 754 spectra of 616 XSL stars. We adopt these stellar parameters for the DR3 spectra. Our sample has many stars with multiple observations. We regard these observations as separate stars with slightly different stellar parameters, as determined in \citet{Arentsen2019}.

Not all stars in DR3 are useful for stellar population modelling. With the exception of the red giants, we select only non-variable non-peculiar stars with complete X-shooter spectra from the DR3 data set. We exclude spectroscopic binary stars. We only include XSL spectra that have not been corrected for slit flux losses and galactic extinction when constructing the `static', \mbox{O-rich} \mbox{TP-AGB} and \mbox{C-rich} \mbox{TP-AGB} star sequences in Sect. \ref{Sect:coolevolvedstars}, otherwise we use spectra that are corrected. We only include DR3 spectra for which stellar parameters have been estimated in \citet{Arentsen2019} or in Verro et al. (2021a). Furthermore, as described in Sect.~\ref{Sect:coolevolvedstars}, we remove supergiants from the library, because we aim to model stellar populations older than 50 Myr, in which supergiants do not occur. These selections result in 639 spectra of 534 stars (from the 830 stellar spectra of 683 stars of DR3) which are used to create the XSL stellar population models.

%%%%%%%%%%%%%%%%%%%%%%%%%%%%
\section{Model calculation}
%%%%%%%%%%%%%%%%%%%%%%%%%%%%
\label{sect:modelcalc}
\subsection{Spectral interpolator}
Each point on the isochrone needs a representative stellar spectrum, when generating an SSP model. The limited coverage of the HR diagram by empirical libraries requires a method to assign the stars in the library to the isochrones. Commonly, an interpolator is used to do this. An interpolator creates a synthetic spectrum at a given set of parameters (e.g.\ $T_\mathrm{eff}$, $\log g$ and $[\mathrm{Fe/H}]$) from a library of empirical or theoretical spectra. A local interpolator (e.g.\  \citealt{Vazdekis2003,Sharma2016A&A,Dries2018}) interpolates spectra using its local neighbourhood: library stars in the vicinity of the point for which we
want to create a spectrum are weighted and combined to create a representative spectrum for that point. A global interpolator (e.g.\ \citealt{ELODIE2001, Koleva2009, Prugniel2011,Wu2011}) fits polynomials of $T_\mathrm{eff}$, $\log g$ and $[\mathrm{Fe/H}]$ at each wavelength point to the whole or a large subset of the spectra in the library. Here we use a combination of the two in different areas of the HR diagram. Moreover, static giants, \mbox{O-rich} \mbox{TP-AGB} stars, and \mbox{C-rich} \mbox{TP-AGB} stars are treated separately.

%%%%%
\subsubsection{Global interpolation}

\begin{figure}
    \centering
    \includegraphics[width = 0.5\textwidth]{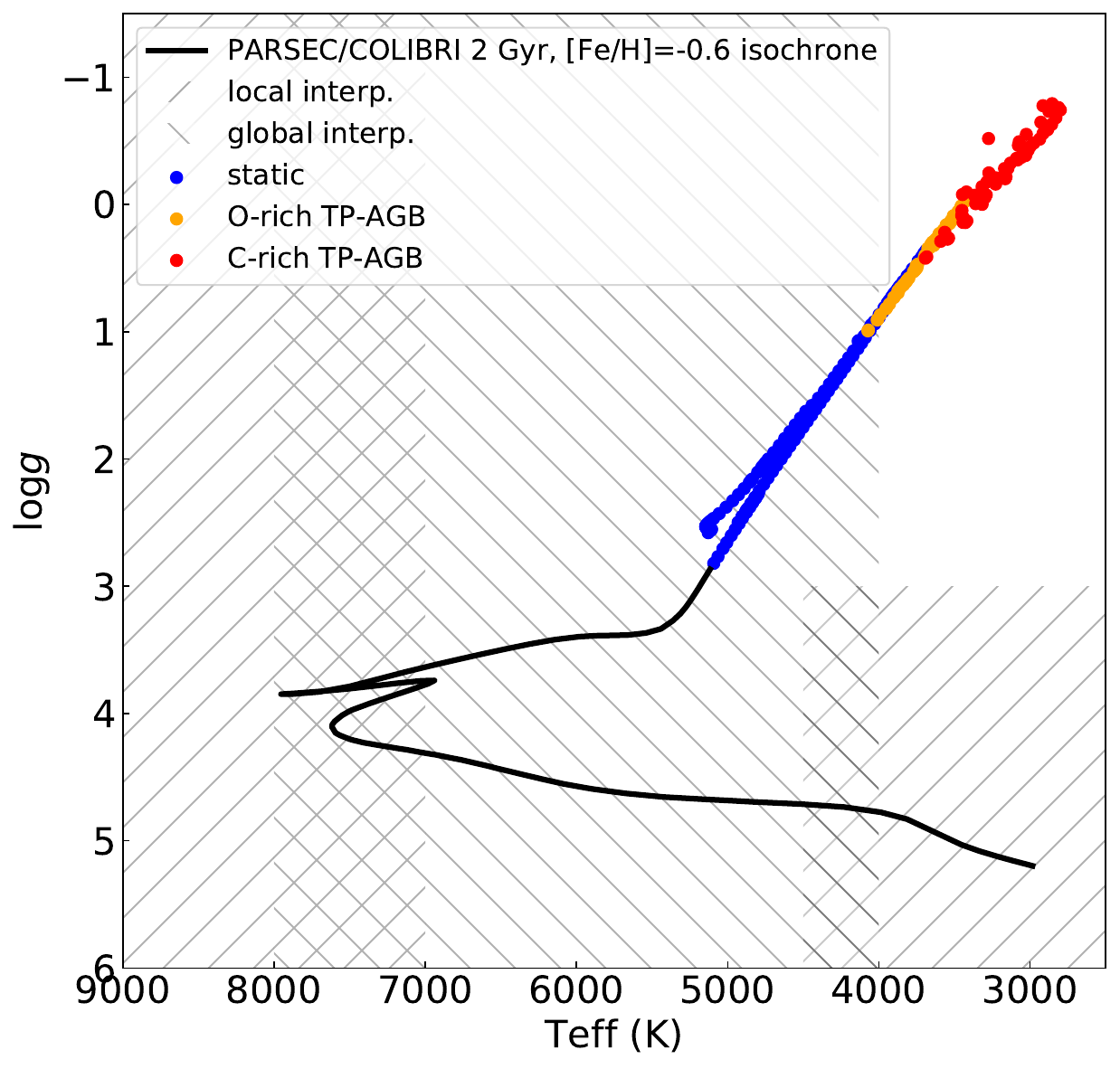}
    \caption{A 2~Gyr, $\mathrm{[Fe/H]}=-0.6$ PARSEC/COLIBRI isochrone and the locations on the HR-diagram where spectra are generated by local interpolation, global interpolation, or potentially taken from the `static' giant, \mbox{O-rich} \mbox{TP-AGB}, or \mbox{C-rich} \mbox{TP-AGB} sequences. We only switch to the respective sequences when we have reached the bluest average spectrum on that sequence (according to the colour--temperature relation). For example, only the coolest ($T_\mathrm{eff}\lessapprox4000\,\mathrm{K}$) RGB stars are represented by a spectrum originating from the `static' sequence, and the spectra of warmer RGB stars are created by the global interpolator.}
    \label{fig:interp_methods}
\end{figure}

The global interpolator consists of polynomial expansions for each wavelength pixel in powers of the three stellar parameters. This type of interpolator was first introduced in \cite{ELODIE2001} and used with the ELODIE stellar library in \citet{ELODIE2001} and  \citet{Wu2011} and the MILES stellar library in \citet{Prugniel2011} and \citet{Sharma2016A&A}. The polynomial coefficients $\beta_i$, $i=0,1,\ldots,25$, fitted for each wavelength pixel $\lambda$ over the subset of spectra are as follows:

\begin{multline}
Y(x,y,z, \lambda ) =  \beta_0(\lambda) + \beta_1(\lambda)\cdot x
+ \beta_2(\lambda)\cdot z
+ \beta_3(\lambda)\cdot y \\
\qquad + \beta_4(\lambda)\cdot x^2
+ \beta_5(\lambda)\cdot x^3 
+ \beta_6(\lambda)\cdot x^4
+ \beta_7(\lambda)\cdot x \cdot z \\
\qquad + \beta_8(\lambda)\cdot x \cdot y
+ \beta_9(\lambda)\cdot  x^2 \cdot y 
+ \beta_{10}(\lambda)\cdot  x^2 \cdot  z
+ \beta_{11}(\lambda)\cdot  y^2 \\
\qquad + \beta_{12}(\lambda)\cdot   z^2
+ \beta_{13}(\lambda)\cdot  x^5 
+ \beta_{14}(\lambda)\cdot  x \cdot y^2
+ \beta_{15}(\lambda)\cdot  y^3 \\
\qquad + \beta_{16}(\lambda)\cdot   z^3 
+ \beta_{17}(\lambda)\cdot  x \cdot  z^3
+ \beta_{18}(\lambda)\cdot  y \cdot  z 
+\beta_{19}(\lambda)\cdot  y^2 \cdot  z \\
\qquad +\beta_{20}(\lambda)\cdot  y \cdot  z^2 
+\beta_{21}(\lambda)\cdot  x^3\cdot y
+\beta_{22}(\lambda)\cdot  x^4 \cdot y \\
\qquad+\beta_{23}(\lambda)\cdot  x^3 \cdot  z
+\beta_{24}(\lambda)\cdot  x^2 \cdot y^2
+\beta_{25}(\lambda)\cdot  x^2 \cdot y^3,\\
\label{Eq:poly}
\end{multline}

\noindent where $x = \log T_\mathrm{eff}$, $y =\log g$ and $z = [\mathrm{Fe/H}]$. A weighted linear least squares solution is found for each wavelength pixel. The weight of each XSL spectrum in the sum of squared differences is made up of two factors, one being the signal-to-noise ratio ($S/N$ hereafter) of the spectrum, and the other reflecting how isolated this XSL spectrum is in parameter space. The latter is described by the inverse density of XSL spectra in a box of size ($T_\mathrm{eff}\pm 2500\,\mathrm{K}$, $\log g \pm 1.5\,\mathrm{dex}$, $\mathrm{[Fe/H]}\pm 1.0\,\mathrm{dex}$) surrounding the desired parameter. We assign a uniform weight of $S/N = 10$ to spectra which have been corrected for slit flux losses in DR3 with a polynomial\footnote{That is, not corrected for flux loss at the slit using a separate wide-slit observation, which most but not all XSL spectra had available; see \mbox{\citet{Chen2014}} and \citet{DR2} for more information. Slit flux loss correction with a polynomial relies on estimated stellar parameters and the interpolation scheme; see Verro2021a.}. This interpolator assumes a smooth transition of spectra in the stellar atmosphere parameter space. Using information from a large subset of stars, an interpolated spectrum is little affected by issues of individual stars (e.g.\ poor flux calibration, dichroic contamination, inaccurate stellar parameter estimation, extinction correction issues, or XSL arm-merging inaccuracies). We use this type of interpolation in the `warm' star regime ($4000\,\mathrm{K} < T_\mathrm{eff} < 8000\,\mathrm{K}$); see Sect. \ref{sect:combining} and Fig. \ref{fig:interp_methods} for more details.

%%%%%
\subsubsection{Local interpolation}
\label{local}

Unfortunately, global interpolation fails in poorly-covered parameter-space regions of the library, including at the edges of the parameter space, due to the use of polynomials. In these regions, the local interpolation comes to assist. The local interpolator averages stellar spectra in a box of parameters around the desired point, so it works better in lower density regions of the HR diagram. We use this type of interpolation in the cool dwarf ($T_\mathrm{eff} < 4500\,\mathrm{K}$, $\log g > 4.0 \,\mathrm{dex}$) and hot star ($T_\mathrm{eff} > 7000\,\mathrm{K}$) regime; see Sect. \ref{sect:combining} and Fig. \ref{fig:interp_methods} for more details.

The local interpolation combines weighted spectra in eight cubes in the stellar parameter space, all with one corner at $\theta_0$ ($\equiv5040/T_{\mathrm{eff},0}$), $\log g_0$,  $[\mathrm{Fe/H}]_0$. The initial size of each three-dimensional cube of $\Delta \theta_0$, $\Delta \log g_0$ and  $\Delta[\mathrm{Fe/H}]_0$ is $3\sigma_{\theta_m} \times 3\sigma_{\log g_m} \times 3\sigma_{\mathrm{[Fe/H]}_m}$, where $\sigma_{\mathrm{param}_m}$ corresponds to the minimum uncertainty in the determination of the respective stellar parameters. Following stellar parameter estimations from \citet{Arentsen2019}, we adopted the following values as uncertainties [$\sigma_{\theta_m}$, $\sigma_{\log g_m}$, $\sigma_{\mathrm{[Fe/H]}_m}$]: hot stars -- [0.018, 0.20, 0.1]; cool dwarfs -- [0.012, 0.14, 0.1]. If no stars are found, the box is enlarged along each of its axes in steps of $1 \sigma$, until at least one star is found. This interpolation scheme is described in detail in \citet[][Appendix B]{Vazdekis2003} and \citet{Dries2018}.

\subsubsection{Interpolation quality}

\begin{table}
    \centering
    \caption{X-shooter dichroic contamination regions and main telluric bands}
    \begin{tabular}{lr}
        \hline\hline
        \multicolumn{1}{c}{Dichroic region} & \multicolumn{1}{c}{wavelengths (nm)} \\
        \hline
         UVB--VIS & 545--590 \\
         VIS--NIR & 994--1150 \\
         \hline
         \multicolumn{1}{c}{Telluric region} & \multicolumn{1}{c}{} \\
         \hline
         VIS  & 930--960\\
         NIR a & 1110--1160\\
         NIR b & 1350--1410\\
         NIR c & 1810--1930\\
        \hline
    \end{tabular}
    \label{tab:dichroic_bands}
\end{table}

\begin{figure*}
        {\includegraphics[width=\hsize]{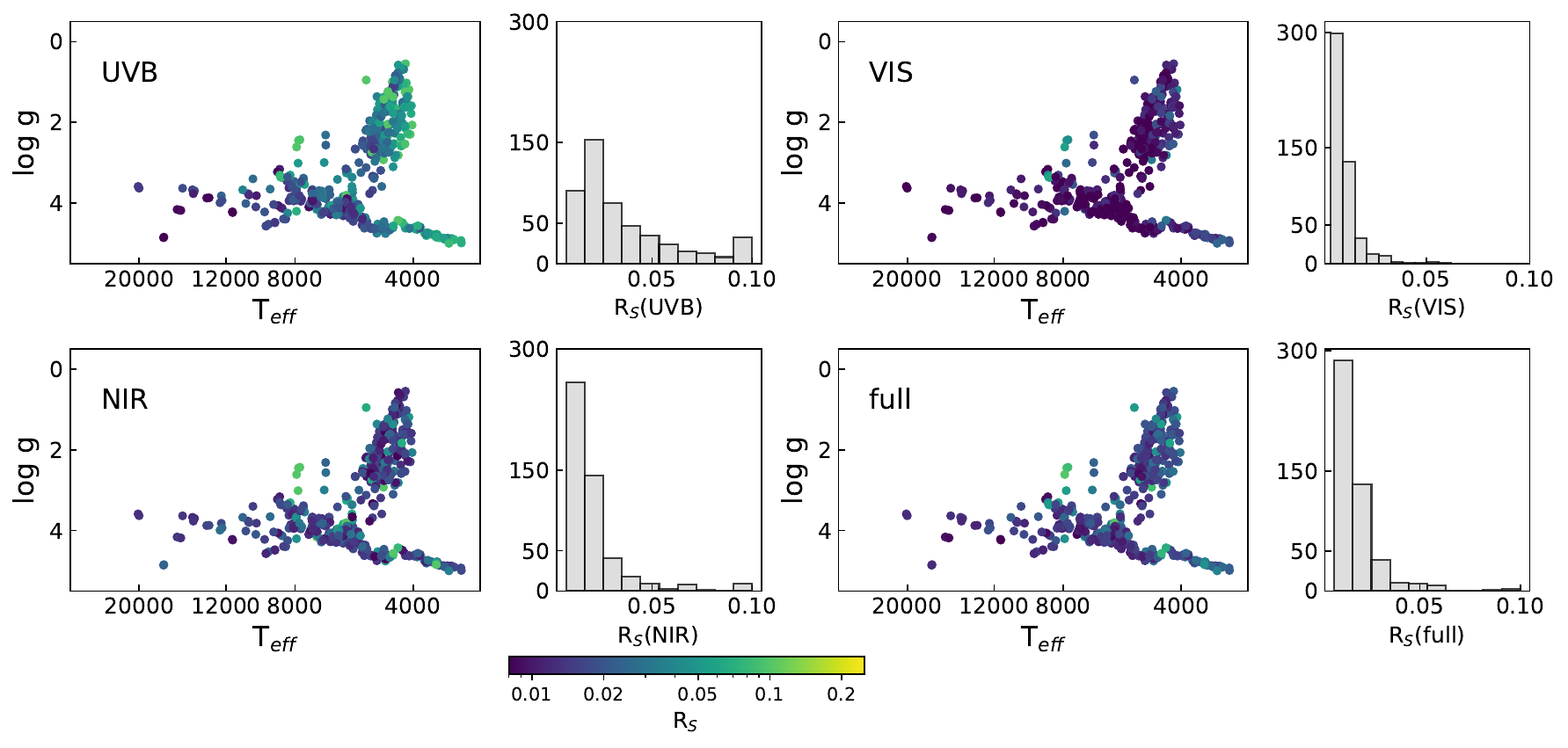}    }
    \caption{The weighted median residual $R_\mathrm{S}$ between the original spectrum and the interpolated spectrum for the full wavelength range, and the X-shooter UVB, VIS, and NIR arms separately, as a function of position in the HR diagram. The colour-bar is logarithmic. Histograms show the distributions of $R_\mathrm{S}$ calculated within these spectral ranges at the full XSL resolution. For ease of visualisation, spectra with $R_\mathrm{S}>0.1$ are placed into the $R_\mathrm{S}=0.1$ bin in the histograms.} 
    \label{fig:Rs_interpol} 
\end{figure*}

To test the local and global interpolator, we have created an interpolated spectrum for each star in XSL DR3 data set (excluding cool giants, which are incorporated into the models separately), in such a way that the original XSL star is excluded from the data set that we use to build the interpolator. We calculate the median residual $R_\mathrm{S}$ between the original $S_\mathrm{or}$ and the interpolated spectrum $S_\mathrm{int}$:
\begin{equation}
R_S  = \mathrm{median} \left( \frac{\mathrm{abs}(S_\mathrm{or}-S_\mathrm{int})}{S_\mathrm{or}} \right)
\end{equation}
The main telluric bands, as well as the XSL dichroic areas (shown in Table~\ref{tab:dichroic_bands}) are masked out when calculating $R_\mathrm{S}$. The median residuals $R_\mathrm{S}$ for the XSL stars used in stellar population modelling though the usage of the global and local interpolators are shown in Fig.~\ref{fig:Rs_interpol} as a function of positions in the HR diagram for the full wavelength range, and for the X-shooter UVB, VIS, and NIR arms separately.
Figure \ref{fig:fourlines_HRD} shows a similar plot for the median residuals around four spectral line indices (CaHK, H$\beta$, CaT and CO1.6).
We note however that we only discuss the incorporation of the very cool giant stars to the models in Sect. \ref{Sect:coolevolvedstars}; therefore the very cool giants are missing from Fig.~\ref{fig:Rs_interpol}.

The median residuals should be taken as a rough quality measure, considering the variety of spectral types and long wavelength range of XSL -- cool stars have relatively less signal (smaller $S/N$) in the UVB than hot stars and hot stars have relatively less signal in the NIR than cool stars. A large mismatch between an interpolated spectrum and the corresponding original spectrum may be the result of very low $S/N$. Large median residual values can also indicate poorer reproduction of the star by the interpolation due to uncertain stellar parameters, extinction correction, DR3 merging factors, peculiarity of the spectrum, residual telluric lines, or due to the interpolation scheme itself. We expect poorer matches at the edges of the parameters space and areas with low density, due to the nature of the interpolators. As seen from Fig.~\ref{fig:Rs_interpol}, the residuals in all wavelength ranges are mostly less than 5\%. For the VIS and NIR spectra, the median residuals are of the order of a few percent. The regions around H$\beta$, CaT, CO1.6 lines in the VIS and NIR arms, the median residuals are also of the order of a few percent, while region around the CaHK line in the UVB arm has higher residuals. The UVB-arm spectrum is the most difficult to interpolate for most spectral types, due to the multitude of spectral features compared to the VIS and NIR arms and the continuum shape changes rapidly with stellar parameters and cool stars have near-zero fluxes in the UVB arm. The CaHK line (3800-4000 \AA{}) is in the region of the UVB spectrum where these difficulties are most evident.
We gave some examples of XSL DR3 and its interpolated counterpart in Verro (2021a), where we used the same interpolation scheme.

\subsection{Incorporating cool evolved stars}
\label{Sect:coolevolvedstars}
In the XSL stellar population models, we give extra attention to the cool ($T_\mathrm{eff}<4000\,\mathrm{K}$) evolved giants. We divide these stars into (\mbox{O-rich}) static giants, \mbox{O-rich} \mbox{TP-AGB} stars and \mbox{C-rich} \mbox{TP-AGB} (`carbon') stars. Individual observed spectra of such stars vary strongly in their (NIR) spectral energy distributions (SEDs) as a function of time, and from star to star, so they cannot be used directly in the synthesis of galaxy spectra. An ideal stellar library should include spectra of individual variable stars observed over their pulsation cycles. In reality, the light curves and phases are generally not accurately known. Furthermore, the stellar parameters are not accurately known. The stellar parameter estimation should be done based on spectral type and temperature-sensitive spectral features. Full-spectrum fitting with theoretical \citep{Lancon2019,Lancon2021} or interpolated empirical spectra \citep{Arentsen2019} for these stars is unreliable. Re-evaluating the stellar parameters for these stars and conducting additional observations of stars in different pulsation stages are well beyond the scope of the current paper. Instead, we have used the approach described in \citet{Lancon2002} -- using average spectra of static giants, \mbox{O-rich} \mbox{TP-AGB} stars, and \mbox{C-rich} \mbox{TP-AGB} stars, binned by broad-band colour, and relying on empirical relations to dictate where an average spectrum of a star of a certain colour should occur -- to incorporate these stars into our stellar population models. This method allows us to also use XSL giant stars with $T_\mathrm{eff}<4000\,\mathrm{K}$, for which the parameter estimation by \citet{Arentsen2019} is inadequate.

\subsubsection{Differentiating between \mbox{O-rich} static giants, supergiants and variable stars}
Differentiating between \mbox{O-rich} static/quasi-static giants, supergiants and variable stars is difficult. The spectra of long period variables with small visual amplitudes are very similar to those of static stars. Supergiants and giants can have similar optical features, although supergiants have redder SEDs \citep{LW2000, Alvarez2000}. All of these type of stars can have the same broad band colours, so binning by a certain temperature sensitive broad-band colour without separating by spectral type first could result in a very red old stellar population model with strong supergiant or \mbox{TP-AGB} features. We separate the (quasi-)static from the high amplitude variables using the $(I-K)$ colour and the $H$-band H$^-$/H$_2$O feature. 
\begin{figure}
    \centering
    \includegraphics[width = 0.45\textwidth]{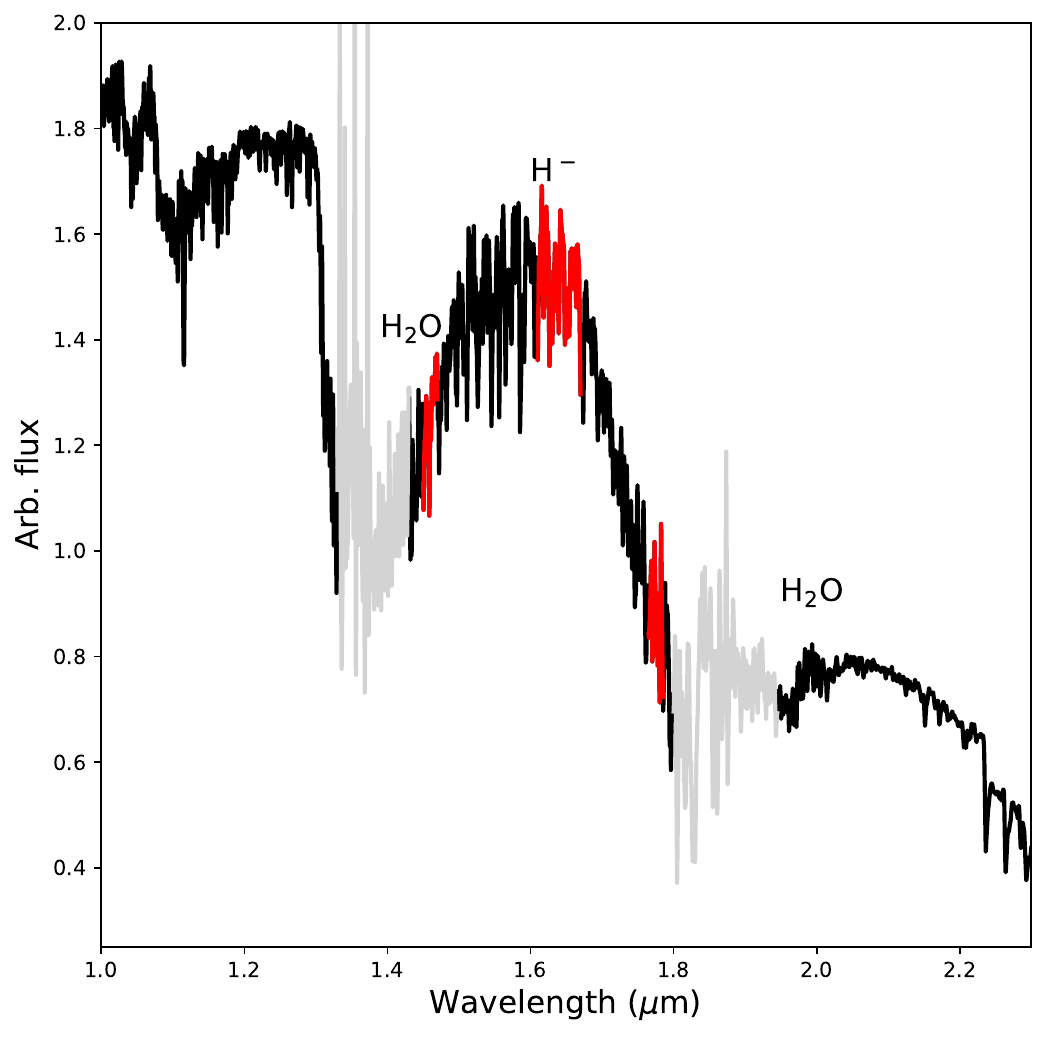}
    \caption{The spectral H$^-$/H$_2$O feature is a combination of the 1.6\,$\mu$m `bump' of the minimum opacity of H$^-$ion and H$_2$O vapour absorption bands around 1.4\,$\mu$m and 1.9\,$\mu$m. The index bands are marked in red. The telluric absorption is marked in gray. This spectrum is an average of \mbox{O-rich} \mbox{TP-AGB} stars with $(I-K) = 3.19$. }
    \label{fig:Hbump}
\end{figure}
\begin{table}
    \centering
    \caption{Definition of the H$^-$/H$_2$O index at the native XSL resolution; wavelengths in $\mu$m.}
    \begin{tabular}{cc|cc|cc}
        \hline\hline
         \multicolumn{2}{c|}{blue}& \multicolumn{2}{c}{central}&\multicolumn{2}{|c}{red}\\
         %\hline
         left & right & left & right & left & right\\
         \hline
         1.450 & 1.470 & 1.610 & 1.670 & 1.765 & 1.785\\
         \hline
%         \multicolumn{6}{|c|}{$\mu$m}\\
%         \hline
    \end{tabular}
    \label{tab:Hindex}
\end{table}

The H$^-$/H$_2$O feature is a combination of the 1.6\,$\mu$m `bump' of the minimum opacity of H$^-$ ion and H$_2$O vapour absorption bands around 1.4\,$\mu$m and 1.9\,$\mu$m. These H$_2$O bands create the curved shape of the spectrum illustrated in Fig.~\ref{fig:Hbump}, which is a characteristic feature of long-period variable M stars \citep{Bessell1989,Matsuura1999, Alvarez2000}. Although these water bands are contaminated by telluric lines, the overall feature is still distinctive. We create an H$^-$/H$_2$O index to describe the feature, defined in Table~\ref{tab:Hindex}. We measure the index at the native XSL resolution, in magnitudes, following the index equation of \citet{Worthey1994}:
\begin{equation}
I = -2.5 \log \left[ \left(\frac{1}{\lambda_1 - \lambda_2} \right)  \int_{\lambda_1}^{\lambda_2} \frac{F_\lambda}{F_\mathrm{cont}} \,d\lambda \right],
\end{equation}
where $F_c$ is the pseudo-continuum flux defined by drawing a straight line from the midpoint of the blue continuum level to the midpoint of the red continuum level, and $F_\lambda$ is the flux of the index. 

\begin{figure}
    \centering
    \includegraphics[width = 0.5\textwidth]{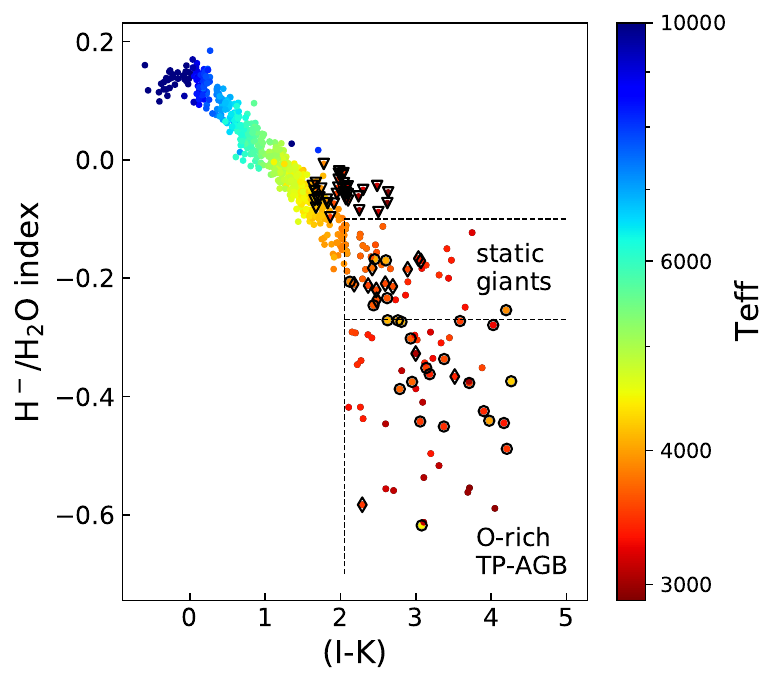}
\caption{H$^-$/H$_2$O index strengths of XSL stars as a function of their $(I-K)$ colours. Points are colour-coded by their effective temperatures from \citet{Arentsen2019}, saturated at 10\,000\,K. Carbon stars are marked with circles. Supergiants are marked with diamonds. Both \mbox{C-rich} \mbox{TP-AGB} stars and supergiants are excluded from the division into static and variable stars. M dwarfs have weak H$^-$/H$_2$O index strengths but can have red $(I-K)$ colours, and they have been marked with triangles.}
    \label{fig:IK_Hbump}
\end{figure}

Figure \ref{fig:IK_Hbump} shows the XSL stars on this colour--index plane, colour-coded by $\log T_\mathrm{eff}$ from \citet{Arentsen2019}. The stronger the $H$-band feature, the more negative the index. While the rest of the stars in the XSL follow a linear relation in this plane, stars redder than $(I-K) = 2$, corresponding to stars cooler than 4000\,K, show a wide variety of H$^-$/H$_2$O index strengths. \mbox{C-rich} \mbox{TP-AGB} stars are plotted in this figure but have been removed when dividing red giants into static and variable stars. \mbox{C-rich} stars are very recognizable due to their distinctive SEDs, with bands of carbon compounds and an absence of oxide bands. XSL \mbox{C-rich} \mbox{TP-AGB} stars have been studied in detail by \citet{Gonneau2016, Gonneau2017}. The NIR bands of oxygen-rich H$_2$O and carbon rich CN and C$_2$ overlap in wavelength. Carbon stars have strong H$^-$/H$_2$O index strengths due to CN and C$_2$ absorption, not because of H$_2$O. 

To remove supergiants from this data set, we use the CN1.10 index defined by \citet{Rock2015}. Supergiants display prominent NIR CN absorption (in particular at 1.10\,$\mu$m), while other \mbox{O-rich} giants do not. However, the 1.15\,$\mu$m H$_2$O band, which is also heavily blended with TiO and VO bands in the coolest long-period variables, can be confused with the CN band \citep{LW2000}. Those long-period variables should have strong H$^-$/H$_2$O features, while supergiants should not, and the two can be separated. Furthermore, supergiants \emph{may not} have a strong CN1.10 feature. We also remove three spectra of stars which are in the \citet{Massey2002} catalog of supergiants.

%%%%%%%%%%%%%%
\subsubsection{Average spectra of `static' red giants}
\begin{figure}
    \centering
    \includegraphics[width = 0.5\textwidth]{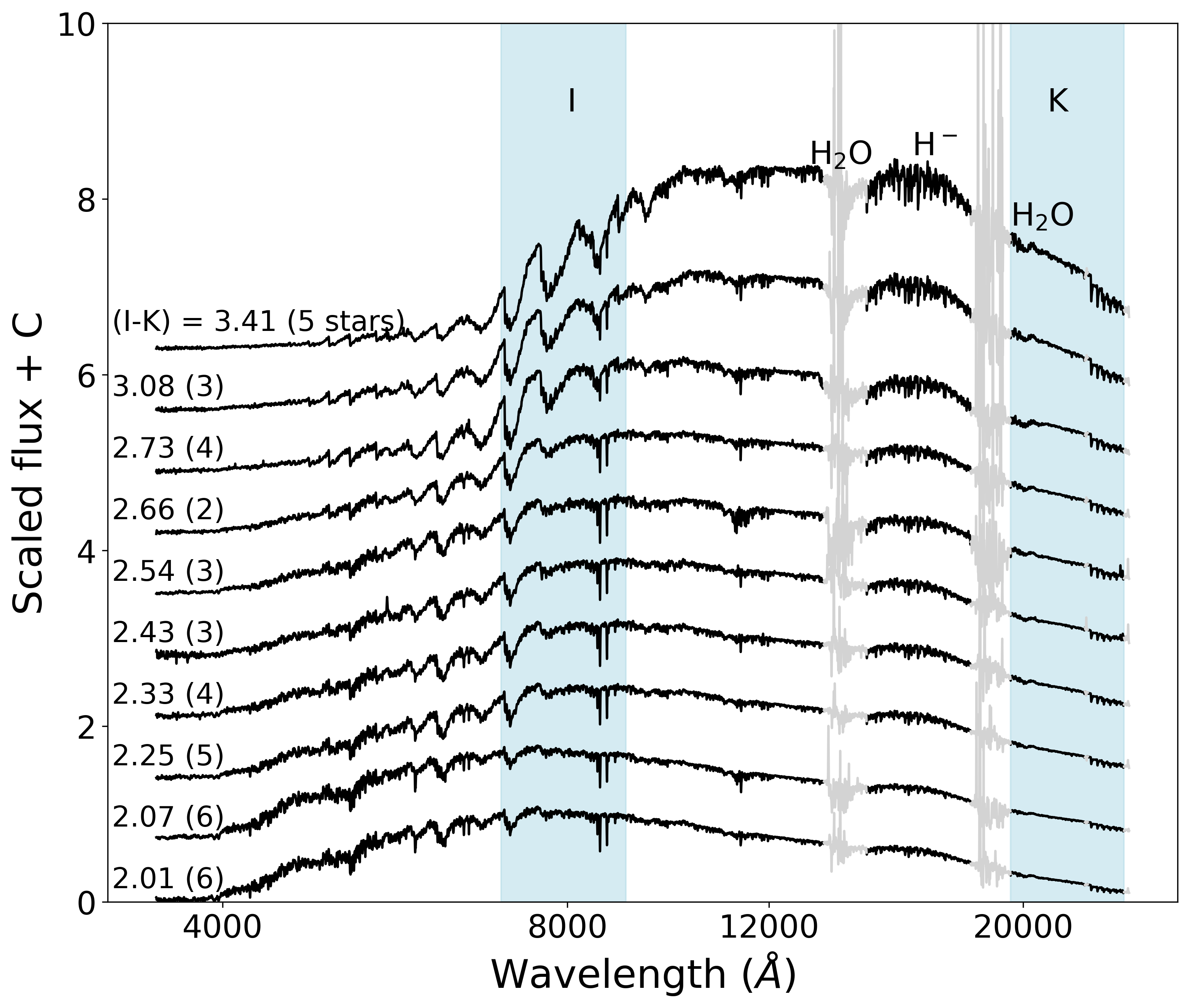}
    \caption{The sequence of `static' giant spectra, sorted according to $(I-K)$ (on the left of each spectrum). The number of spectra the average is comprised of is given in the brackets. The spectra have been smoothed to lower resolution for clarity.}
    \label{fig:IKseqRGB}
\end{figure} 

We select 44 oxygen-rich stars from Fig.~\ref{fig:IK_Hbump}, which are \mbox{(quasi-)static}. We create a sequence of average spectra with each average spectrum consisting of stars with similar $(I-K)$ colour. $(I-K)$ colour is known to correlate with the effective temperature in M stars \citep[e.g][]{Bessell1989,Bessell1998,Lancon2002,Lancon2019}. We combined stars in one bin using weighted averaging (with $S/N$ in the $I$-band as the weight). This `static sequence' is shown in Fig.~\ref{fig:IKseqRGB}. The selected 44 stars are listed in Table \ref{tab:static_giants} and shown individually in Fig.~\ref{fig:staticstars}. 

\subsubsection{Average spectra of O-rich TP-AGB stars}
\label{Sec:TPAGBO}
The $(I-K)$ colour is known to correlate with the effective temperature also for \mbox{TP-AGB} stars; $(V-K)$ and $(R-K)$ could be used for this purpose as well \citep{Ridgway1980,Lancon2002}. However, \mbox{TP-AGB} stars do not follow a simple colour--temperature relation. Stars with different pulsation properties can be found at the same \mbox{TP-AGB} temperature. When their temperatures decrease, \mbox{TP-AGB} luminosities rise, their radii increase,  their masses decrease due to mass loss, and their stellar pulsation properties change \citep[e.g.\ the DARWIN models for M-type AGB stars][]{Bladh2019}. Therefore using a simple colour--temperature relation means we assume that the spectrum of an individual variable star, averaged over its cycle, is similar to an average spectrum of many stellar spectra of various masses, amplitudes and phases, but with a common colour-inferred temperature. 

\begin{figure}
    \centering
    \includegraphics[width = 0.5\textwidth]{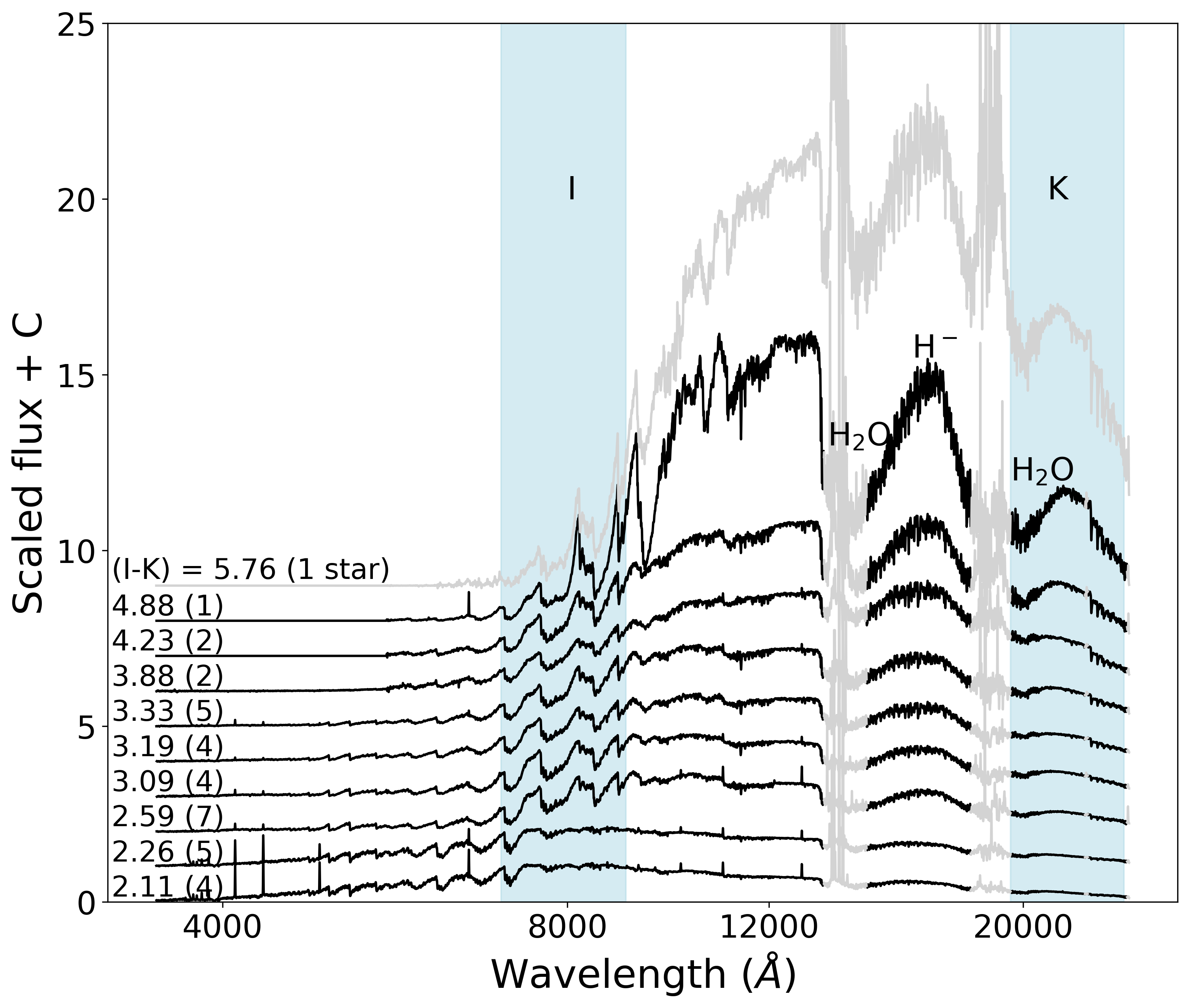}
    \caption{The sequence of \mbox{O-rich} \mbox{TP-AGB} spectra, sorted according to $(I-K)$ (on the left of each spectrum). The number of spectra the average is comprised of is given in the brackets. The spectra have been smoothed to lower resolution for clarity. The reddest average spectrum on this sequence, displayed in grey, is not used in the stellar population models due to its extreme SED.}
    \label{fig:IKseq}
\end{figure}

We select 39 XSL spectra as \mbox{O-rich} \mbox{TP-AGB} stars. These stars are listed in Table~\ref{tab:variable_giants} and shown individually in Fig.~\ref{fig:AGBstars}. This selection produces a sequence of \mbox{O-rich} spectra with continually evolving properties, as seen in Fig.~\ref{fig:IKseq}. We note the deepening of the NIR $H$-band H$^-$/H$_2$O feature with increasing colour. We call this the oxygen-rich `variable sequence'. We note that the reddest average spectrum on this sequence, marked with grey in Fig.~\ref{fig:IKseq}, consists of only one star, X0145 (OGLEII DIA BUL-SC41 3443), and has a very extreme colour of $(I-K) = 5.76$. We do not use this star due to it being the only star with such an extreme colour. Moreover, according to the colour--temperature relation of \citet{Worthey2011}, we do not need it (see Sect.~\ref{sect:combining} and \ref{sect:ctrelations}). The average spectrum $(I-K) = 4.88$ is also a single star, X0020 (ISO-MCMS J005714.4-730121). There are no other stars with such red colours. Because we need a spectrum with such extreme colours, we \emph{do} use X0020 in the stellar population models.

\subsubsection{Average spectra of C-rich TP-AGB stars}
\label{TPAGBC}
Some \mbox{O-rich} \mbox{TP-AGB} stars will become \mbox{C-rich} through convective dredge-up of newly synthesized carbon from their cores. This third dredge-up is induced by thermal pulses \citep[e.g.][]{1983ARA&A..21..271I} and depends on the initial mass and metallicity of these stars. Their spectra differ radically from those of other of cool giants. The spectrum of a \mbox{C-rich} \mbox{TP-AGB} star is characterized by bands of carbon compounds, such as CN and C$_2$ bands, and by the absence of oxide bands such as TiO and H$_2$O. As with \mbox{O-rich} AGB stars, \mbox{C-rich} \mbox{TP-AGB} stars are variable in nature and so difficult to include in a stellar population model. However, they are essential contributors to the NIR light of 1--3\,Gyr old stellar populations, especially at subsolar metallicities \citep[e.g.][]{Ferraro1995, Lancon1999,Mouhcine2003,Pastorelli2019,Pastorelli2020}. 

\begin{figure}
    \centering
    \includegraphics[width = 0.5\textwidth]{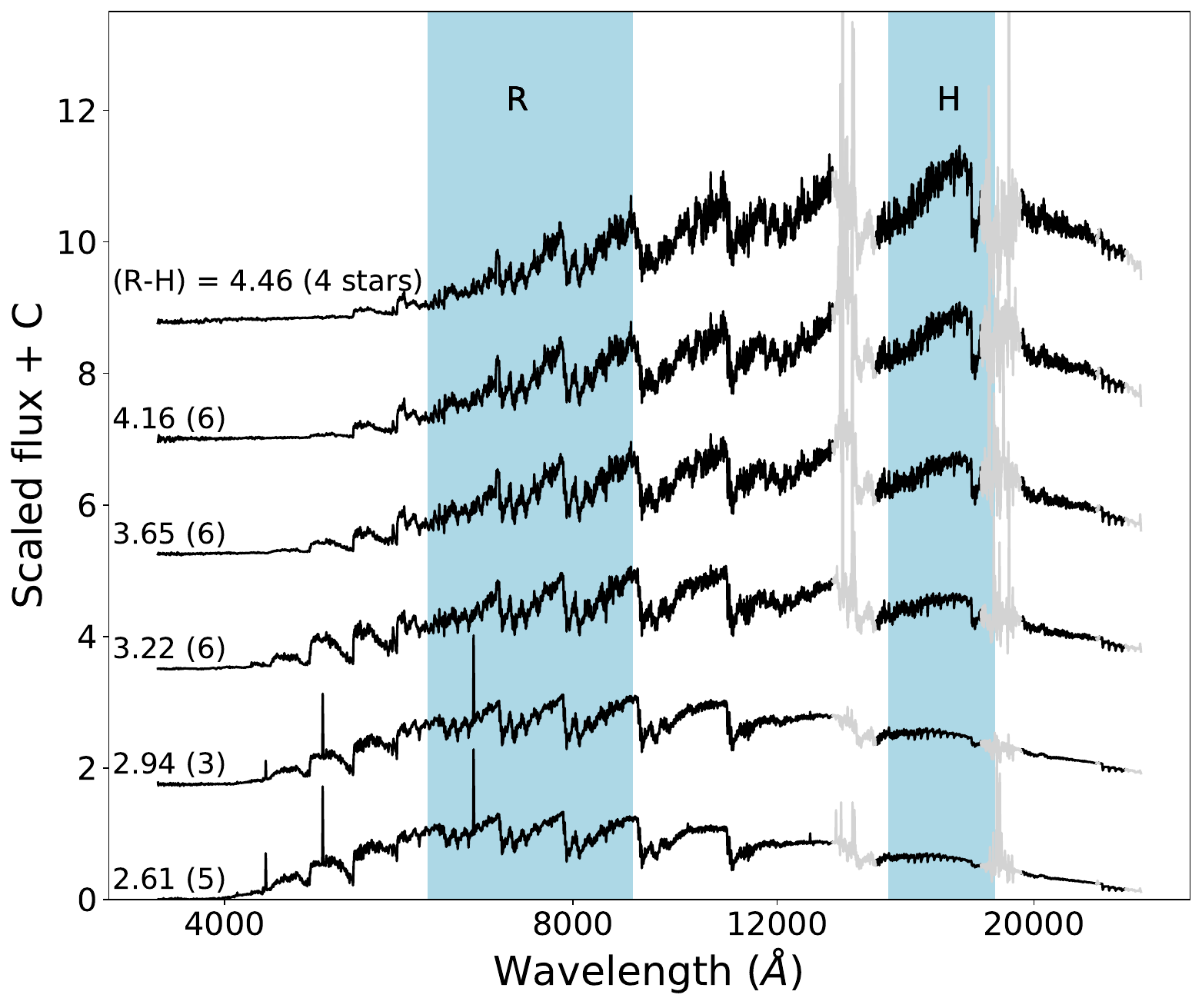}
    \caption{The sequence of \mbox{C-rich} \mbox{TP-AGB} spectra, sorted by their $(R-H)$ colours (on the left of each spectrum). The number of spectra the average is comprised of is given in the brackets. The spectra have been smoothed to lower resolution for clarity.}
    \label{fig:Cstar_averages}
\end{figure}

Similar to \mbox{O-rich} \mbox{TP-AGB} stars, \citet{Lancon2002} suggested using a NIR colour as a classification parameter for the \mbox{C-rich} \mbox{TP-AGB} star spectra in stellar population models but warned that this disregards information such as metallicity, carbon-to-oxygen ratio or pulsation properties. Although there are 51 spectra of \mbox{C-rich} \mbox{TP-AGB} stars in XSL, we have selected 26 of them. The chosen spectra have the full spectrum available and are corrected for flux losses in \citet{DR2}. These stars are listed in Table \ref{tab:carbon_stars}.
\citet{Loidl2001} showed that $(R-J)$ and $(R-H)$ are among the best effective temperature indicators for these stars, but also $(V-K)$, $(J-K)$, and $(H-K)$ have been shown to correlate well with temperature \citep{Bergeat2001}. We have constructed six average spectra of XSL \mbox{C-rich} \mbox{TP-AGB} stars based on these 26 XSL carbon rich stars, sequenced and averaged based on their $(R-H)$ colours. We prefer $(R-H)$ broadband colour, as this results in the cleanest sequence of XSL \mbox{C-rich} \mbox{TP-AGB} stars. 

We show the sequence of the average \mbox{C-rich} \mbox{TP-AGB} star spectra in Fig.~\ref{fig:Cstar_averages} and the spectra inside individual bins in Fig.~\ref{fig:Cstar_bins}. The number of stars in each bin varies, as the 26 spectra do not cover the $(R-H)$ colour sequence uniformly and we aim to combine together the closest spectra in this broadband colour.

\subsection{Combining the interpolation methods and the average spectra of evolved giants}
\label{sect:combining}

Figure \ref{fig:interp_methods} shows an example of how the global, local, and the three sequences of evolved giant star spectra are used to generate the representative spectra in different regions on a HR diagram. The cool dwarf stars are generated by the local interpolator below 4000\,K. Between 4000\,K and 4500\,K, the resulting spectrum is the linear combination of the spectra produced by local interpolation and global interpolation, weighted by
\begin{equation}
\label{eq:q}
    q = \frac{\log(T_\mathrm{eff}) - \log(T_\mathrm{lower})}{\log(T_\mathrm{higher}) - \log(T_\mathrm{lower})},
\end{equation}
where $T_\mathrm{lower} = 4000\,\mathrm{K}$ and $T_\mathrm{higher} = 4500\,\mathrm{K}$. Spectra of stars with effective temperatures between 4500\,K and 7000\,K are generated by the global interpolator, and  star hotter than 8000\,K by the local interpolator. The transition from the warm/global to the hot/local regime is from 7000\,K to 8000\,K using the weights in Eq.~\ref{eq:q} of $T_\mathrm{lower} = 7000\,\mathrm{K}$ and $T_\mathrm{higher} = 8000\,\mathrm{K}$. 

We use isochrone keywords to determine where the isochrone track enters into the relevant evolutionary stage where the `static', \mbox{O-rich} \mbox{TP-AGB} or \mbox{C-rich} \mbox{TP-AGB} spectra are used. On Padova00 isochrones, we model the bottom of the RGB (`RGBb') until the first thermal pulse (`1TP') with the static sequence; and from the first thermal pulse and beyond with the \mbox{O-rich} \mbox{TP-AGB} sequence. On the PARSEC/COLIBRI isochrones, we model stages 3 (RGB) to 7 (E-AGB, including) using the `static' sequence, and stage 8 (\mbox{TP-AGB}) with the \mbox{O-rich} \mbox{TP-AGB} sequence until the given carbon over oxygen ratio becomes one. Stars with $\mathrm{C/O} \geq 1$ are modelled using the \mbox{C-rich} \mbox{TP-AGB} sequence.

 However, we only switch to the sequences when we have reached the bluest average spectrum on the sequence. Hence, only the coolest ($T_\mathrm{eff}\lessapprox4000\,\mathrm{K}$) giants are represented by a spectrum originating from either the `static',  \mbox{O-rich} \mbox{TP-AGB} or \mbox{C-rich} \mbox{TP-AGB} star sequence. Warmer stars are created with a global interpolator. There is no transition region when switching from global interpolation to the static sequence, or from the static to the \mbox{O-rich} \mbox{TP-AGB} variable star sequence, or from the \mbox{O-rich} \mbox{TP-AGB} sequence to the \mbox{C-rich} \mbox{TP-AGB} sequence. We linearly interpolate between the spectra on each sequence to infer a representative spectrum for a point on an isochrone with a given colour. 

The choice of the colour--temperature relation is important in NIR stellar population modelling, and can change the NIR colours of SSPs considerably (see Sect. \ref{sect:static/variable} for a discussion). For the \mbox{O-rich} \mbox{TP-AGB} sequence, we use the empirical surface-gravity-dependant $(I-K)$ colour--temperature relation of \citet{Worthey2011}. We use the colour--temperature relation of \citet{Bergeat2001} to assign a $(J-K)$ colour for the \mbox{C-rich} \mbox{TP-AGB} sequence stellar parameters. We note that the broadband colour we use to construct the \mbox{C-rich} \mbox{TP-AGB} sequence differs from the broadband colour we use here, because \citet{Bergeat2001} does not provide a $(R-H)$--temperature relation. We prefer their relation, because it is based on the measurements of angular diameters of 52 stars available from lunar occultations and interferometry, the largest set to date.

Old solar-metallicity and metal-rich populations need a template spectrum at the tip of the RGB which is redder than the reddest spectrum on the static sequence. However, as the tip of the RGB dominates the NIR light of these populations, we cannot switch to the more red \mbox{TP-AGB} spectra, as this would introduce strong \mbox{TP-AGB} features into the population models. This issue is discussed further in Sect.~\ref{sect:static/variable}.
 
\subsection{Bolometric corrections}
We employ the $V$-band bolometric corrections (BC$_\mathrm{V}$ henceforth) given in \citet{Worthey2011} for all stars except the \mbox{C-rich} \mbox{TP-AGB} stars. \citet{Worthey2011} reviews literature bolometric corrections and uses a combination of sources: \citet{VandenBerg2003} for the middle of the temperature range, supplemented by the \citet{Vacca1996} formula for $4.40<\log (T_\mathrm{eff}\,\mathrm{K^{-1}}) <4.75$ for the hottest dwarfs and supergiants; \citet{Bessell1998} for giants; and \citet{Leggett2001} for cool dwarfs. For giants with $T_\mathrm{eff}<4000\,\mathrm{K}$, we switch from $V$- to $I$-band using the $(V-I)$ colour provided by \citet{Worthey2011}, because these stars can have little to no flux in the $V$-band. We use the \citet{Kerschbaum2010} $K$-band BC for carbon-rich giants. Within the bolometric corrections, we adopt BC$_{\mathrm{V},\odot}=-0.09$, BC$_{\mathrm{I},\odot}=0.61$, BC$_{\mathrm{K},\odot}=1.42$ and a bolometric magnitude of 4.72 for the Sun \citep{Torres2010}.

\section{General behaviour of the models}
\label{sect:generalbehaviour}

\begin{figure*}
    \centering
    \begin{subfigure}[]
        \centering
        \includegraphics[width = \textwidth]{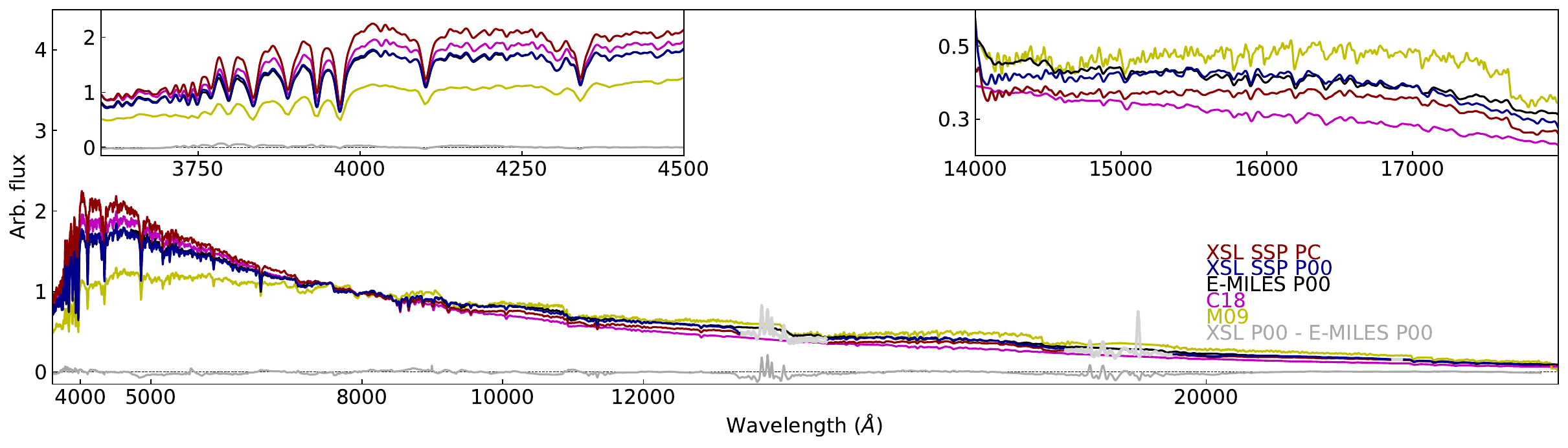}
    \end{subfigure}
    ~ 
     \begin{subfigure}[]
        \centering
        \includegraphics[width = \textwidth]{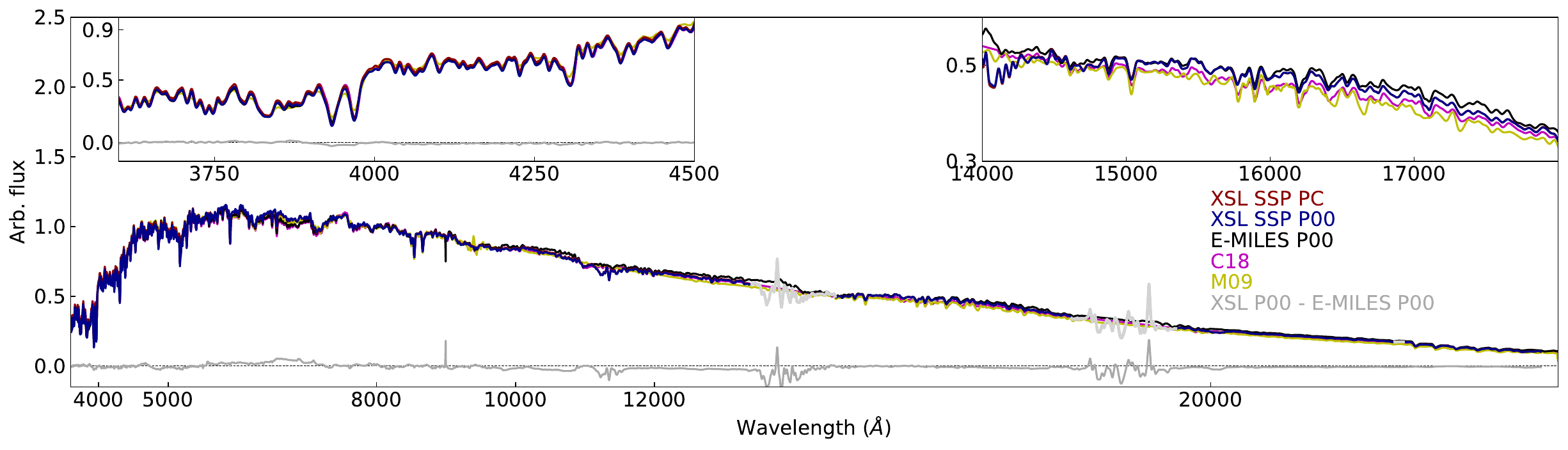}
    \end{subfigure}
    \caption{XSL (`PC': PARSEC/COLIBRI; `P00': Padova00), E-MILES P00, \citet[][M09]{Maraston2009} and \citet[][C18]{Conroy2018} SSP model spectra of (a) 1\,Gyr and (b) 10\,Gyr old solar-metallicity stellar populations. The spectra are smoothed to $R=500$.} The M09 spectra are displayed at their original ($R\approx500$) resolution. All spectra are normalised to a common $I$-band flux. The residual of the XSL P00 model from the E-MILES P00 model is shown in grey in each panel.
     \label{fig:fullSSPs}
\end{figure*}

In this section, we focus on predictions of colours and absorption-line indices of our SSP models. We compare them with the E-MILES models \citep{MILESI,MILESIII,EMILES}, the \citet[][M09 hereafter]{Maraston2009} models  and the \citet[][C18 hereafter]{Conroy2018} models. Example spectra of these models are shown in Fig.~\ref{fig:fullSSPs}. Further examples of XSL SSP models are shown in Appendix \ref{fig:SSP_examples}. Here, we use the XSL SSP models calculated using the Salpeter IMF.

The comparison with the E-MILES is relevant because these models are widely used in the study of intermediate and old stellar populations \citep[][to name a few recent works]{Neumann2021, RodriguezBeltran2021,Barbosa2021,Lonoce2021}.  Here, we use the E-MILES models calculated using the Salpeter IMF and both the Padova00 and BaSTI \citep{Pietrinferni2004,Pietrinferni2006,Cordier2007,Percival2009} isochrones. The E-MILES models cover an extensive 1680--50\,000\,\AA\ wavelength range. Unlike the XSL models, they do not consist of stars observed simultaneously at all wavelengths. Instead, E-MILES models are a combination of separately generated UV, optical and NIR population models, merged at overlapping wavelengths. They make use of the Indo-US \citep{IndoUS}, MILES \citep{MILESlib2006, MILESII}, CaT \citep{CaT,CaT2,CaT3}, NGSL \citep{NGSL2006,NGSL_reso2012} and IRTF \citep{IRTF} stellar libraries at different wavelengths. Because of this mixture, the resolution of E-MILES models varies from a constant $\mathrm{FWHM}=2.5$\,\AA\ in the NUV and optical to a constant $\sigma = 60\,\mathrm{km\,s}^{-1}$ in the NIR wavelengths.

The importance of \mbox{TP-AGB} stars in SSP models was emphasised by \citet{Maraston2005,Maraston2006,Maraston2009}. This is why we include M09 models in some comparisons here. These solar metallicity models extend from the UV to NIR (1150-–25000\,\AA) and have low resolution ($R\approx500$). The M09 models make use of the \citet{Pickles1998} library of empirical stellar spectra. The M09 models were calculated for the isochrone sets of \citet{Cassisi1997,Cassisi1997II, Cassisi2000}. M09 used the ‘fuel consumption theorem’ with the average spectra of \mbox{TP-AGB} stars of \citet{Lancon2002} to include the \mbox{TP-AGB} stars into the SSP models. They calibrated the flux contribution of this phase against optical and NIR photometry of globular clusters in the Magellanic Clouds. Due to this particular treatment of \mbox{TP-AGB} stars, the NIR flux of stellar populations of ages between 0.5 and 1.5\,Gyr is enhanced. This can be seen in Fig.~\ref{fig:fullSSPs}a -- M09 models have clearly stronger carbon star features than other SSP models. 

The C18 models are similar to the E-MILES models, as they also use the MILES and the Extended IRTF spectral libraries to synthesise the optical and NIR part of the SSP. However, the C18 models are based on isochrones of the MIST project \citep{MIST02016, MIST12016} and use different interpolation methods than the E-MILES models. C18 models have a constant $\sigma=100\,\mathrm{km\,s}^{-1}$ resolution.

The hottest turn-off star determines the shape of the optical population model. The stars on the tip of the RGB dominate the NIR light of the 10~Gyr population, but the TP-AGB stars dominate the NIR light of the 1~Gyr population. This leads to larger differences between different SSP models for 1~Gyr populations than for the 10~Gyr populations that are seen in Fig. \ref{fig:fullSSPs} -- TP-AGB stars are more difficult to incorporate into SSP models than RGB stars. This will be further discussed in Sect. \ref{sect:static/variable}.

\subsection{Colours measured from our models}

\begin{figure*}
    \centering
    \includegraphics[width = 0.85\textwidth]{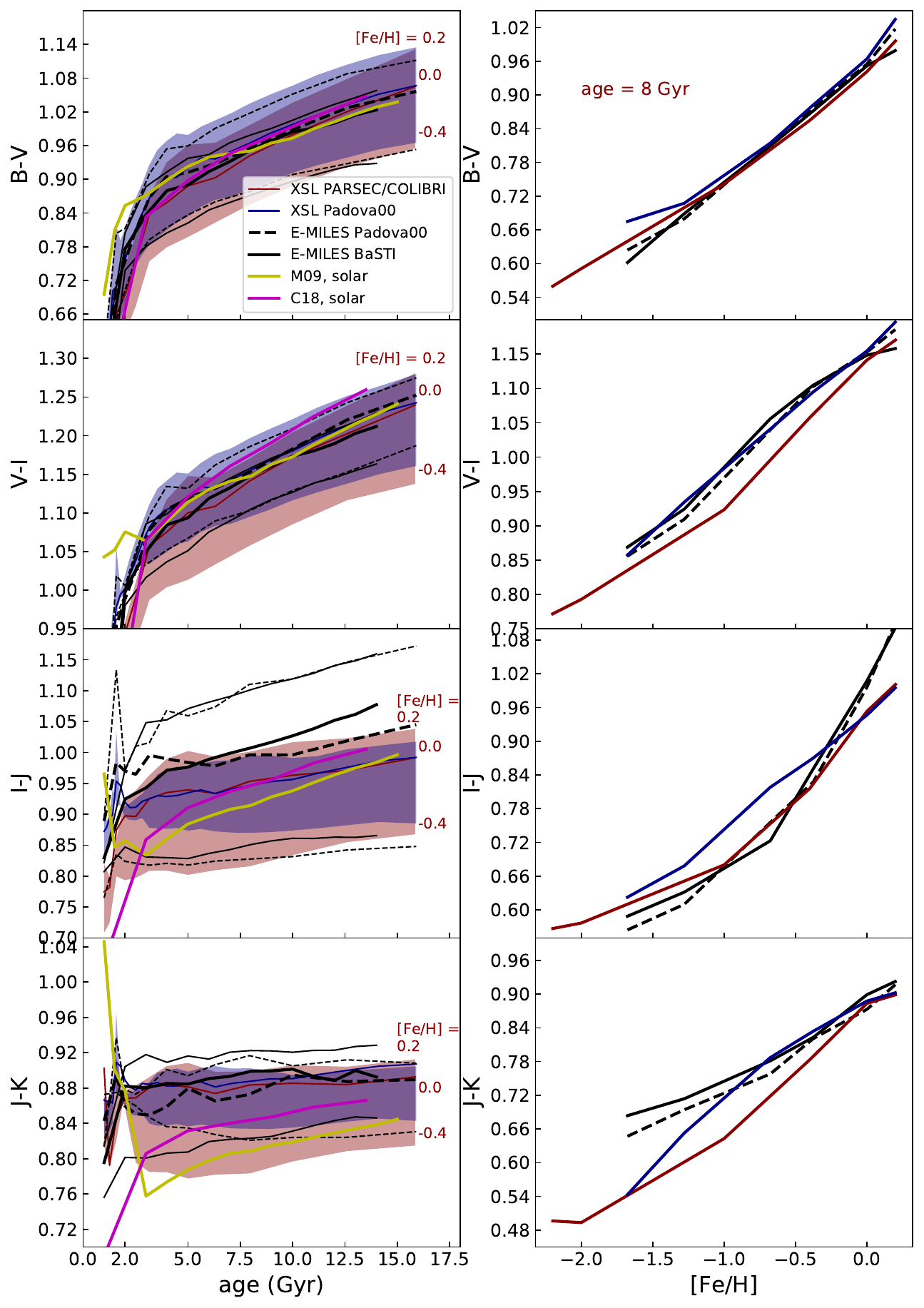}
    \caption{Comparison of the behaviour of colour as a function of age and metallicity for the E-MILES, M09,  C18 and XSL SSP models (see legend). Left panels: colour as a function of age; right panels: colour as a function of metallicity. Top row: $(B-V)$; middle top row: $(V-I)$; middle bottom row: $(I-J)$; bottom row: $(J-K)$. Ages span from 1\,Gyr to 16\,Gyr, and metallicities span from $\mathrm{[Fe/H]} = -0.4$\,dex to $\mathrm{[Fe/H]} = +0.2$\,dex (left panels); $\mathrm{[Fe/H]} = -2.2$\,dex (XSL PARSEC/COLIBRI) or $\mathrm{[Fe/H]} = -1.7$\,dex (other) to $\mathrm{[Fe/H]} = +0.2$\,dex (right panels). We note that E-MILES models are not safe to use in the NIR below $\mathrm{[Fe/H]} = -0.4$\,dex, but are included for illustrative purposes. Solar-metallicity E-MILES models are shown in heavier line-strengths than sub- and super-solar models of the left panels. XSL SSP sub- and super-solar models are represented by shaded areas, centered in the solar metallicity. We note the different colour-scale values between the same colour panels.}
     \label{fig:SSP_colourage}
\end{figure*}

Figure \ref{fig:SSP_colourage} shows the behaviour of the optical/NIR colours measured from our Padova00- and PARSEC/COLIBRI-based SSP models and from the other models discussed above. We show the colour behaviour as a function of age (left panels) and metallicity (right panels). Ages span from 1\,Gyr to 16\,Gyr, and metallicities span from $\mathrm{[Fe/H]} = -2.2$\,dex (XSL PARSEC/COLIBRI) or $\mathrm{[Fe/H]} = -1.7$\,dex (other) to $\mathrm{[Fe/H]} = +0.2$\,dex. We note that E-MILES models are not safe to use in the NIR below $\mathrm{[Fe/H]} = -0.4$\,dex, but are included for illustrative purposes.

\textit{Age--colour relations: }All models follow the same trends as our SSP models and become redder in $(B-V)$, $(V-I)$ and $(I-J)$ with increasing age. The NIR $(J-K)$--age relation is flat for old ages. There are some notable differences between models: at supersolar metallicities, E-MILES BaSTI models have $(B-V)$ and $(V-I)$ colours similar to XSL, C18 and M09 solar models. NIR colours of E-MILES supersolar models are redder ($\Delta(I-J)\approx 0.1$) than XSL models. Even the $(I-J)$ colours of E-MILES solar metallicity models are $\sim0.05$ redder than other models. Furthermore, the M09 and C18 solar-metallicity models are somewhat similar to the $\mathrm{[Fe/H]} = -0.4$ models of XSL and E-MILES in $(J-K)$. $(I-J)$ and $(J-K)$ have model-dependent behaviour in the \mbox{TP-AGB} regime (ages $< 3$\, Gyr).

\textit{Metallicity--colour relations: }All models follow the same trends as our SSP models and become redder in all colours with increasing metallicity.
While the $(B-V)$--metallicity relation is almost identical for XSL and E-MILES models, differences arise towards the NIR. XSL Padova00 models have bluer $(V-I)$, $(I-J)$ and $(J-K)$ colours than XSL PARSEC/COLIBRI models. Considering the range of metallicities where E-MILES models are safe to use ($[Fe/H] \in [-0.4,0.0,-0.2]$), $(I-J)$ colour stands out having a steeper metallicity--colour relation than XSL models.

It is hard to pinpoint a single reason for these model discrepancies, specially in the NIR. Differences in used empirical libraries is one of them, but E-MILES and C18 models do not agree as well. Issues arising from E-MILES or C18 SSP model merging or XSL DR3 merging of stellar spectra are another possible source of disagreements between models. Moreover, we include cool giants into SSP models differently than other groups, with the use of the `static' and `variable' sequences. The NIR colour differences between XSL Padova00 and PARSEC/COLIBRI reflect the usage of isochrones with different levels of sophistication for the description of the TP-AGB phase. Sub-solar XSL PARSEC/COLIBRI SSP models, which have more thorough description of the TP-AGB phase than the XSL Padova00 models, show bluer NIR colours ($\Delta(I-J)\approx 0.06$ and $\Delta(I-J)\approx 0.07$ at $\mathrm{[Fe/H]} = -1.0$); differences are small for solar metallicity models, but noticeable in Fig. \ref{fig:fullSSPs}. 

We concentrate on the comparison with E-MILES, as those models are widely applied and their behaviour studied. \citet{EMILES} has presented a thorough analysis of E-MILES optical and NIR colours.

\subsection{Optical absorption line indices measured from our models}
\label{sect:LIS}
\begin{figure*}
    \centering
    \begin{subfigure}[]
        \centering
        \includegraphics[width = 0.3\linewidth]{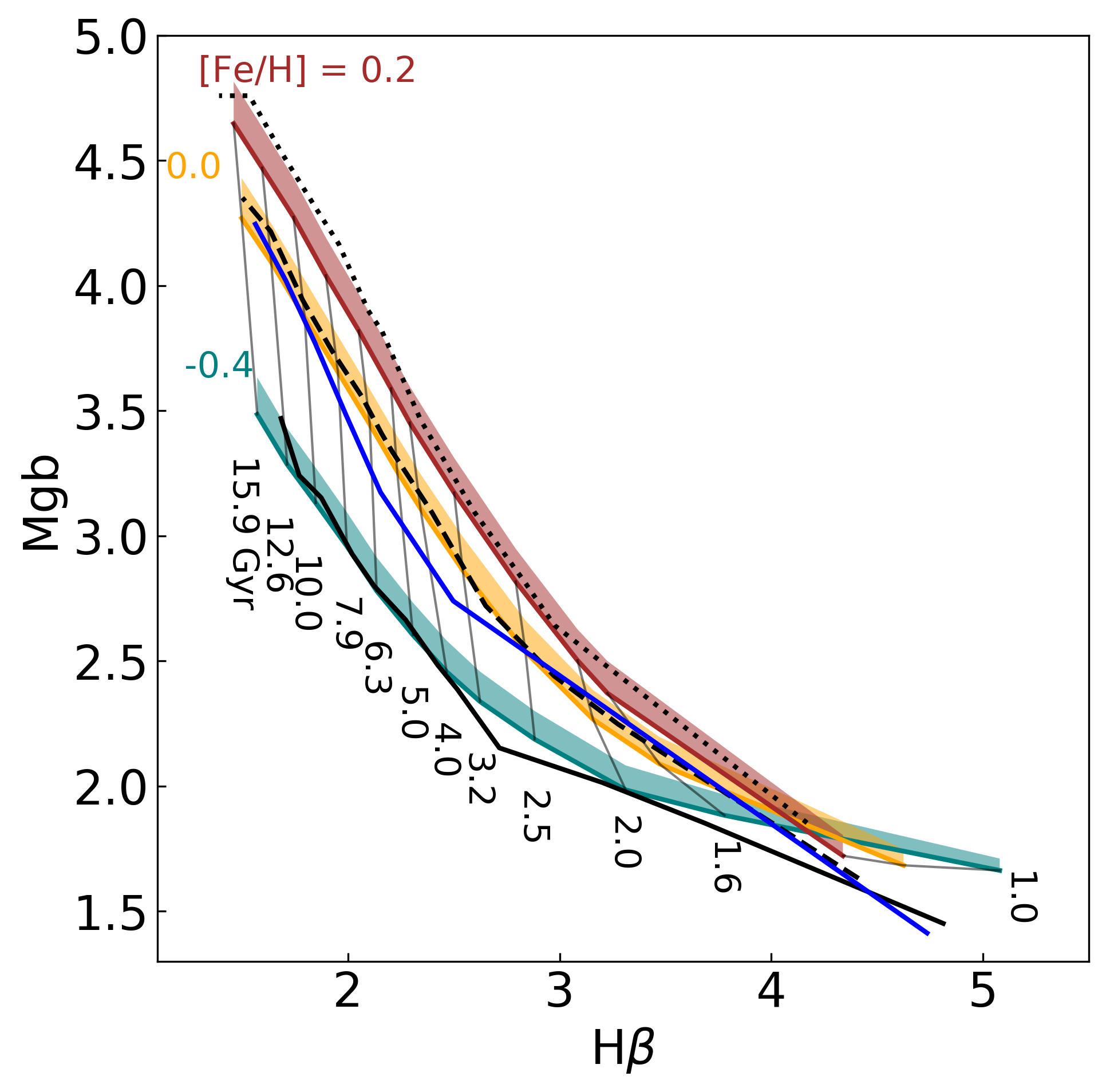}
    \end{subfigure}%
    ~ 
    \begin{subfigure}[]
        \centering
        \includegraphics[width = 0.3\linewidth]{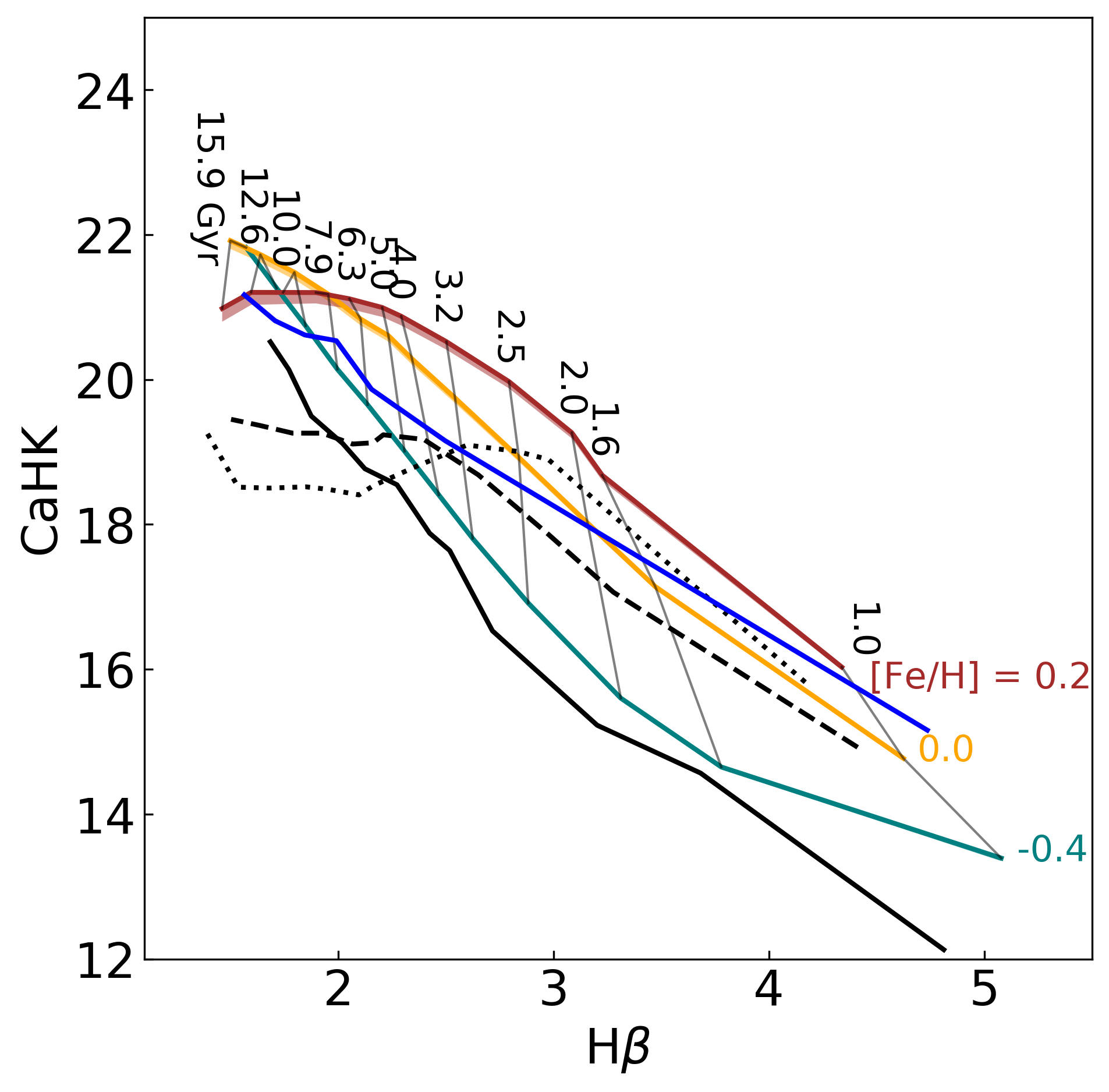}
    \end{subfigure}%
    ~ 
    \begin{subfigure}[]
        \centering
        \includegraphics[width = 0.3\linewidth]{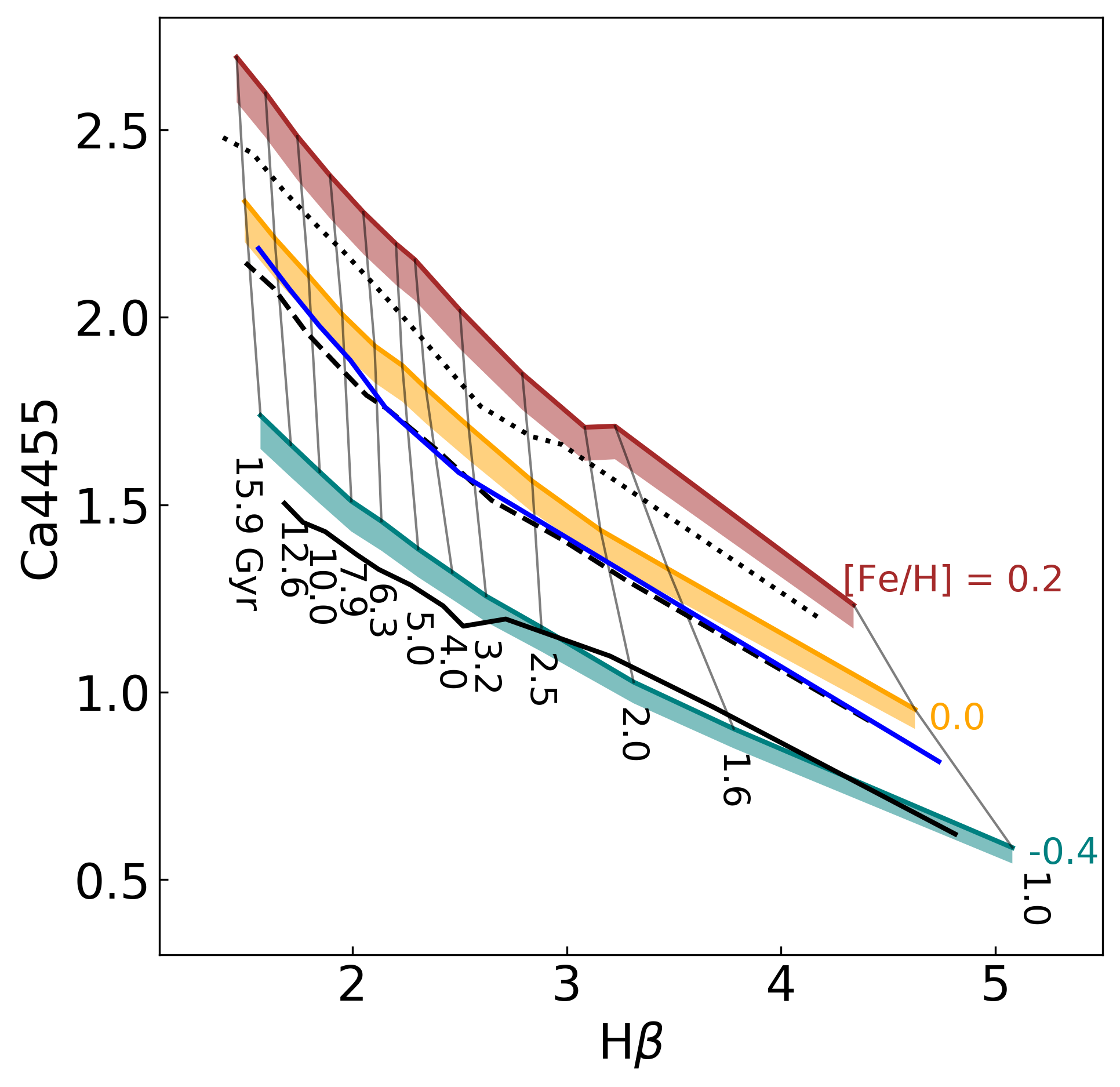}
    \end{subfigure}
    ~ 
    \begin{subfigure}[]
        \centering
        \includegraphics[width = 0.3\linewidth]{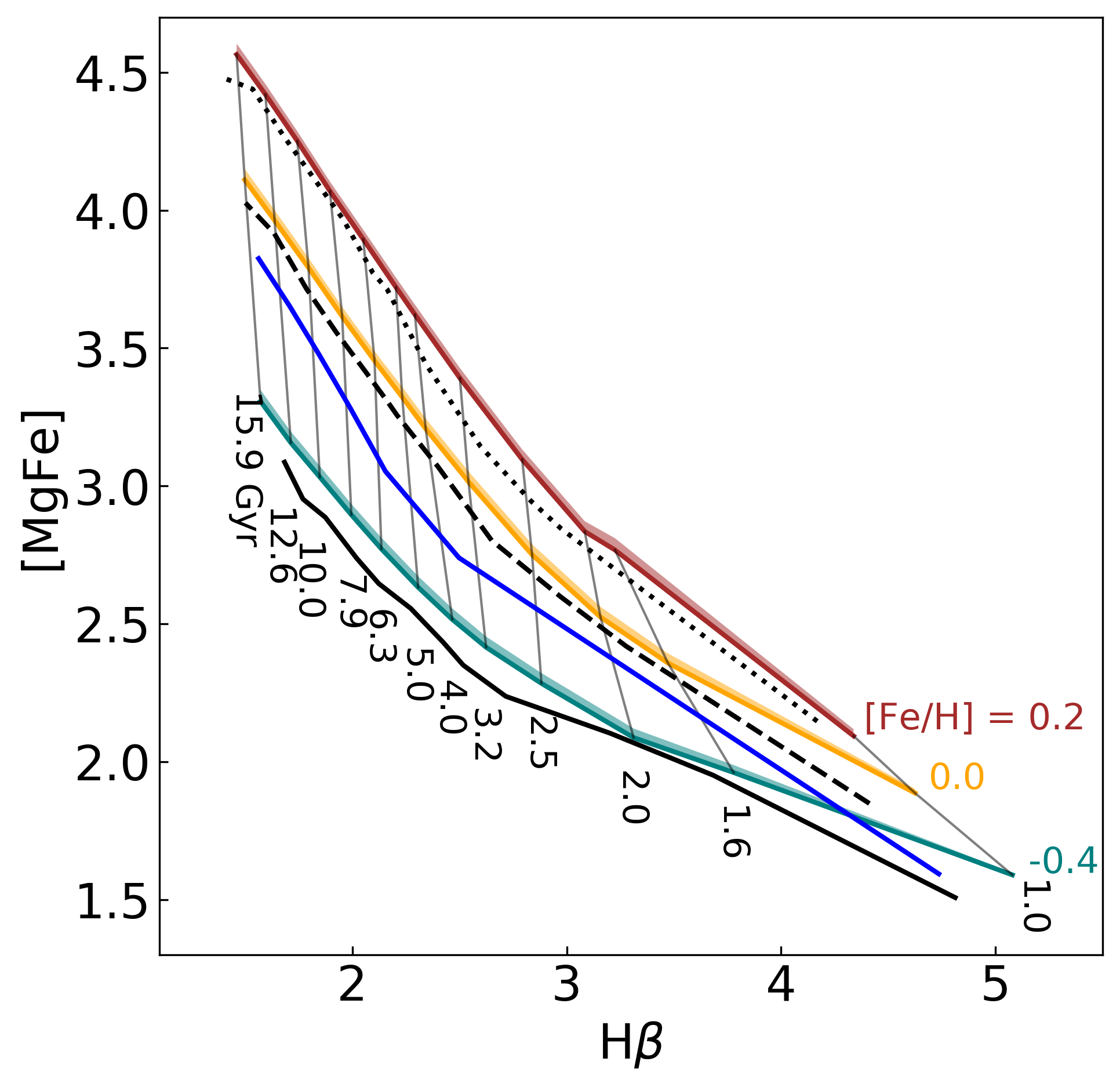}
    \end{subfigure}
    ~ 
    \begin{subfigure}[]
        \centering
        \includegraphics[width = 0.3\linewidth]{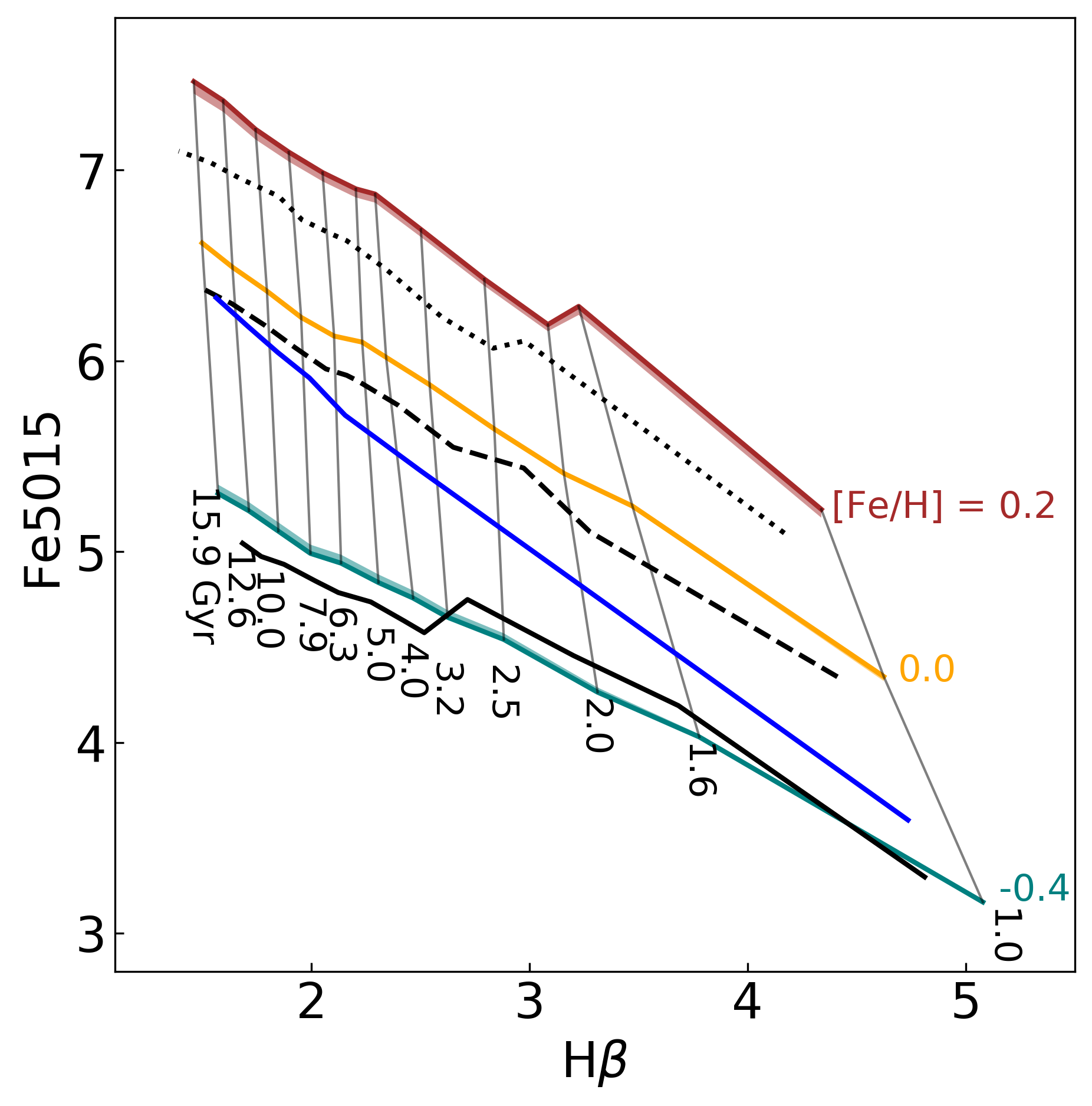}
    \end{subfigure}
    ~ 
     \begin{subfigure}[]
        \centering
        \includegraphics[width = 0.3\linewidth]{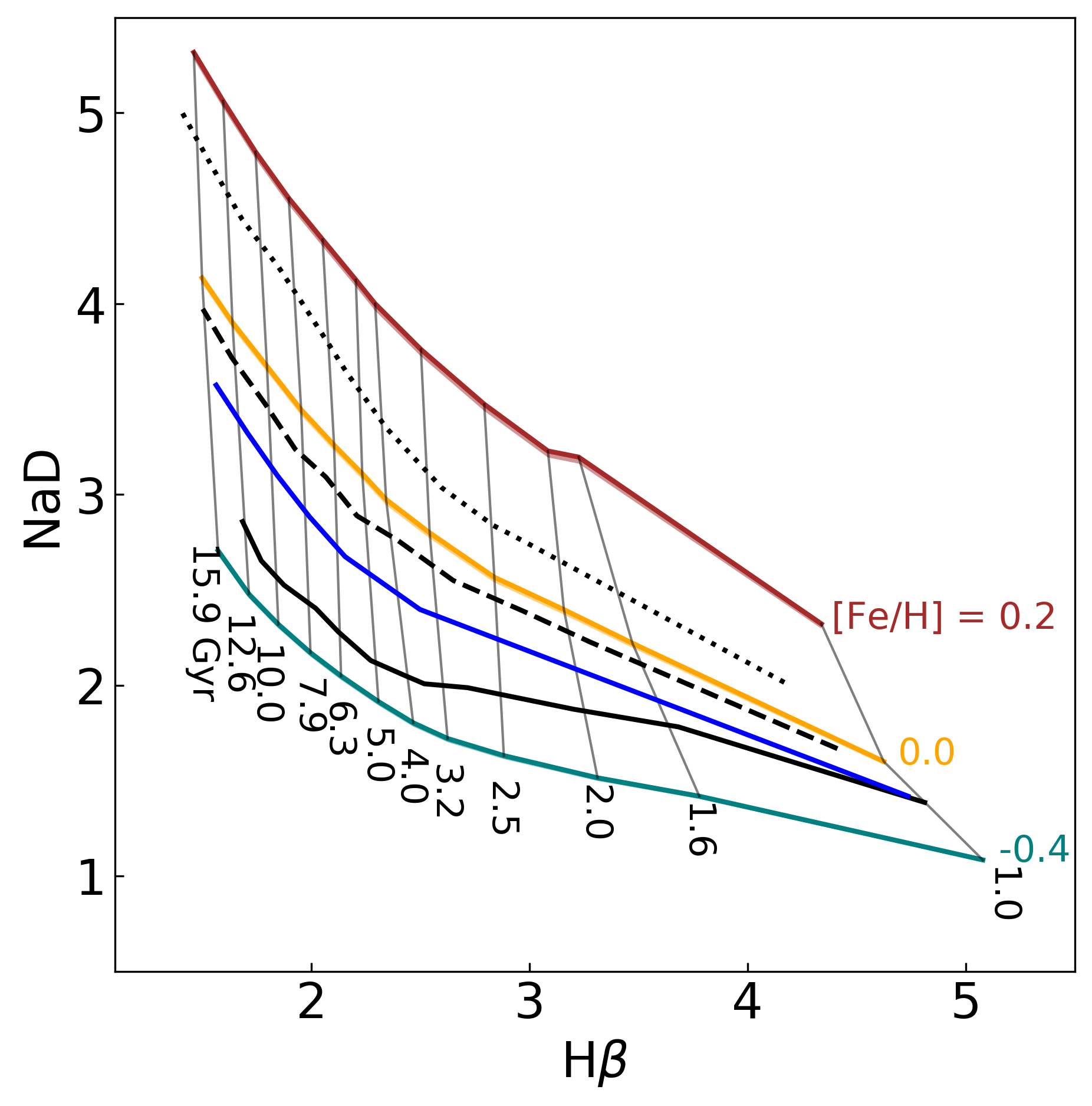}
    \end{subfigure}
    ~ 

    \caption{Comparison of the behaviour of model Mgb, CaHK, Ca4455, [MgFe], Fe5015, and NaD absorption line index strengths as a function of the model H$\beta$ index strength. The shaded area represents XSL Padova00 models with varying spectral resolution ($\sigma$) from $\sigma=13\,\mathrm{km\,s}^{-1}$ (the native XSL resolution) to $\sigma=60\,\mathrm{km\,s}^{-1}$ (the minimum E-MILES resolution).  Black lines represent E-MILES Padova00 model predictions with dotted, dashed and solid lines representing $\mathrm{[Fe/H]}=+0.2$, $0.0$ and $-0.4$\,dex, respectively, measured at the original E-MILES resolution. The blue solid line represents predictions from the C18 solar metallicity models.}
    \label{fig:indices_comparison}
     
\end{figure*}

We compare the widely used optical absorption line indices measured from the XSL and E-MILES SSPs, using diagnostic plots such as H$\beta$ \textit{vs} Mgb, Ca4455, Fe5015, NaD \citep{Trager1998}, CaHK \citep{Serven2005}, and [MgFe] indices in Fig.~\ref{fig:indices_comparison}. [MgFe] is defined by \citet{Thomas2003} as
\begin{equation}
    \mathrm{[MgFe]} \equiv  \sqrt{\mathrm{Mg}b \times (0.72 \times \mathrm{Fe5270} + 0.28 \times \mathrm{Fe5335})}
\end{equation}

\noindent In Fig.~\ref{fig:indices_comparison}, we show measurements from the XSL Padova00 models and E-MILES P00 models older than 1\,Gyr and with metallicities $\mathrm{[Fe/H]} \in [-0.4,0.0,+0.2]$\,dex. Similar grids for XSL PARSEC/COLIBRI and E-MILES BaSTI models are shown in Fig.~\ref{fig:indices_comparison_PC}, but with an extension towards the lowest metallicities of XSL SSP models. Furthermore, we have added absorption line indices measured from solar C18 models with ages between 1 and 13\,Gyr, measured at its original $\sigma = 100\,\mathrm{km\,s}^{-1}$ resolution.

The optical absorption line index trends of different models are similar. The comparison of XSL and E-MILES Padova00 models shows that some differences arise from the different stellar spectra and interpolation methods. However, the differences between C18 and E-MILES models illustrate how well models using the same stellar library (MILES/IRTF) but different stellar population modelling techniques compare. 

There are a few notable differences between the grids. On one hand, the \ion{Ca}{ii} absorption-line index CaHK shows different behaviours at older ages. There is a saturation seen for the oldest population models, but this saturation happens at different metallicities for the varying models. XSL spectra have stronger index values than E-MILES models, while the C18 models are closer to XSL in this index. This is a prominent spectral feature in SSPs, coming from F, G, and K stars. On the other hand, there is roughly less than a 0.15\,\AA\ disagreement between the models for \ion{Ca}{i} line index Ca4455. There is also a crossing of \ion{Mg}{} (Mgb and [Mg/Fe]) index values of young SSP models with different metallicities. Furthermore, XSL models show a larger spread in NaD index values at a given H$\beta$ strength than the other models; however, this line lies in the dichroic contamination region in the XSL spectra, and should be used with (extreme) caution.

\section{Colours of Coma cluster galaxies}
\label{sect:comacolours}

\begin{figure*}
    \centering
    \begin{subfigure}[]
        \centering
        \includegraphics[width = 1\linewidth]{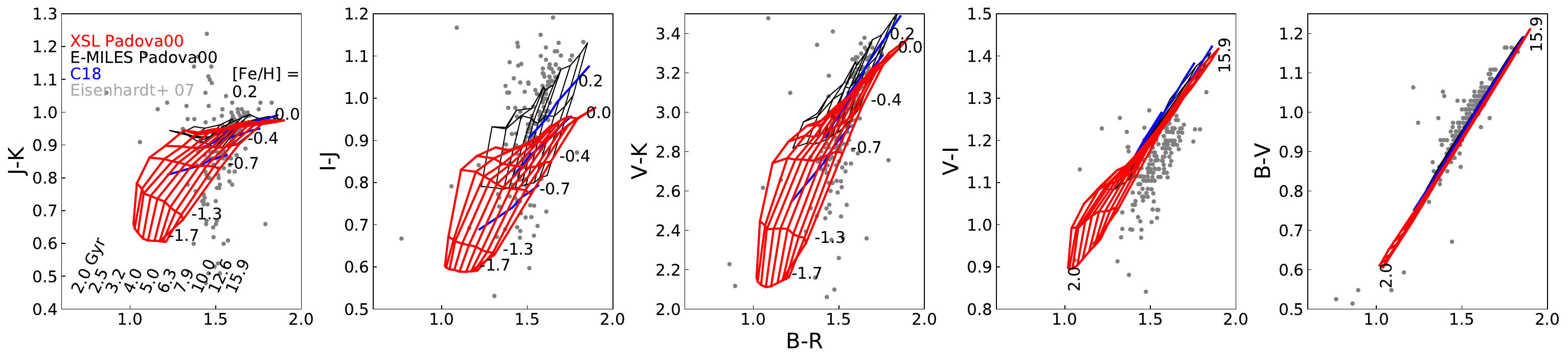}
    \end{subfigure}%
    ~ 
    \begin{subfigure}[]
        \centering
        \includegraphics[width = \linewidth]{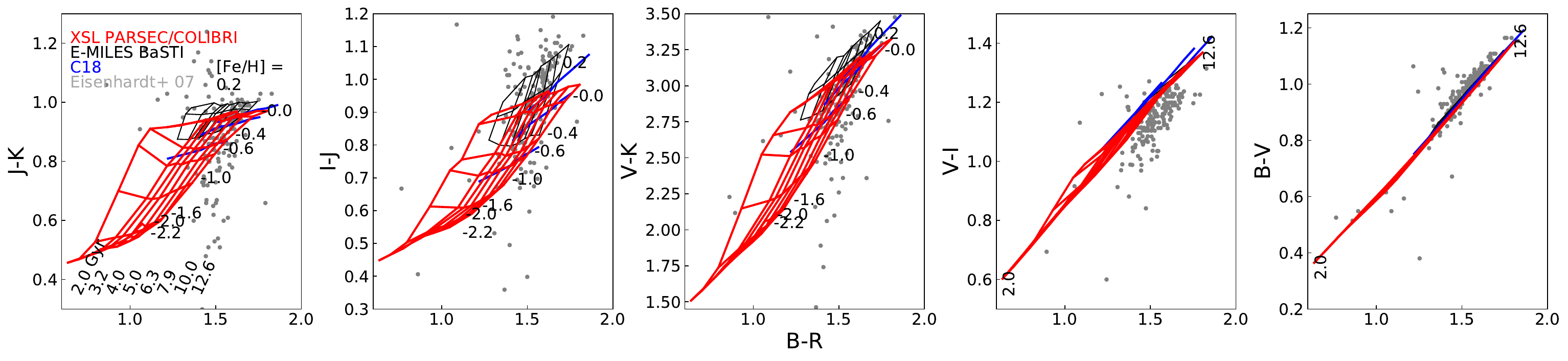}
    \end{subfigure}%
    ~ 
 
    \caption{Colour--colour diagrams of galaxies in the Coma cluster with model predictions overlaid.  Gray points show the \citet{Eisenhardt2007} data for galaxies in the Coma cluster. (a) XSL and E-MILES Padova00 models; (b) XSL PARSEC/COLIBRI and E-MILES BaSTI models. Only $\mathrm{[Fe/H]}=+0.2$, $0.0$ and $-0.4$\,dex E-MILES models are shown. C18 models with $\mathrm{[Fe/H]}=+0.2$, $0.0$ and $-0.5$\,dex models are denoted by blue solid lines. XSL models extend to even lower metallicities.}
    \label{fig:Comacolours}
\end{figure*}

To show the potential of our new XSL models, we compare model colour predictions with photometry of galaxies in the Coma cluster on the colour--colour planes in Fig.~\ref{fig:Comacolours}. We use these galaxies since many galaxies in a rich cluster show colors consistent with old SSP models \citep[e.g.][]{Bower1992}. The photometry is taken from \citet[Table 9]{Eisenhardt2007}. We only show the photometry of galaxies with redshift within 3$\sigma$ of the average redshift of the Coma cluster galaxies from \citet{Upadhyay2021}, $z = 0.0224 \pm 0.0033$. This results in 180 galaxies, the majority of which are ETGs. We have redshifted the models to $z = 0.0224$ and have used the response functions provided by \citet{Eisenhardt2007} for the spectrophotometry.

The XSL SSP models reproduce optical colours of ETGs well in general. However, there are some cases where models do not match the data. For example, all models shown in Fig.~\ref{fig:Comacolours} have redder $(V-I)$ ($\sim0.1$\,mag) colours than the galaxies at fixed $(B-R)$. In the solar and metal-rich regime, the XSL SSP models are most cases bluer than E-MILES and C18 in the NIR. This is very apparent in the $(I-J)$ or $(V-K)$ colours at fixed $(B-R)$, where XSL SSP models are roughly $\sim0.1$\,mag bluer. The colours containing the $I$- and $J$-filter are particularly interesting, since they encompass the joining region the MIUSCAT and IRTF-based SSP models into the E-MILES models and the merging of the individual spectral arms, respectively. The colour offsets, especially in $(I-J)$, can be due to merging of the XSL DR3 VIS--NIR spectra. It can also be due to inclusion of cool giant stars using separate giant sequences and the colour--temperature relation. The NIR colours are constrained by the reddest `static' and `variable' giant templates.

Model offsets in optical colours have been discussed in detail by \citet{Ricciardelli2012}, who tested the MIUSCAT models, the optical part of the E-MILES models, on nearby ETGs. None of the MIUSCAT SSP models are able to match some of the observed optical colour distribution (namely $(u-g)$ or $(r-i)$ colours at fixed $(g-r)$) of nearby ETGs, while the colours of Milky Way globular clusters are reproduced remarkably well. They suggest that the ETGs of their sample are not necessarily simple old stellar populations, and need small contributions from either young or/and metal-poor stellar populations. Furthermore, the impact of $\alpha$-enhancement and the choice of IMF on galaxy colours cannot be neglected.

\section{Optical/NIR absorption line indices}
\label{sect:Riffelindices}

To date, stellar population studies of unresolved galaxies have used mainly the optical absorption line indices, but the NIR spectral features can provide insights into the stellar populations dominated by cool stars \citep{Lancon1999,Mouhcine2002NGC7252,Riffel2007, Riffel2008,Lancon2008,MQ2009,Kotilainen2012,Lyubenova2012,Riffel2015,Riffel2019}. \citet{Riffel2019} presented 47 correlations among the different absorption features in the optical and NIR for 16 star-forming galaxies (SFGs henceforth) and for 19 ETGs. They found that the models consistently agree with the observations for the optical absorption features, but not so much for the NIR indices. 

\begin{figure*}
    \centering
    \begin{subfigure}[]
        \centering
        \includegraphics[width = 0.3\linewidth]{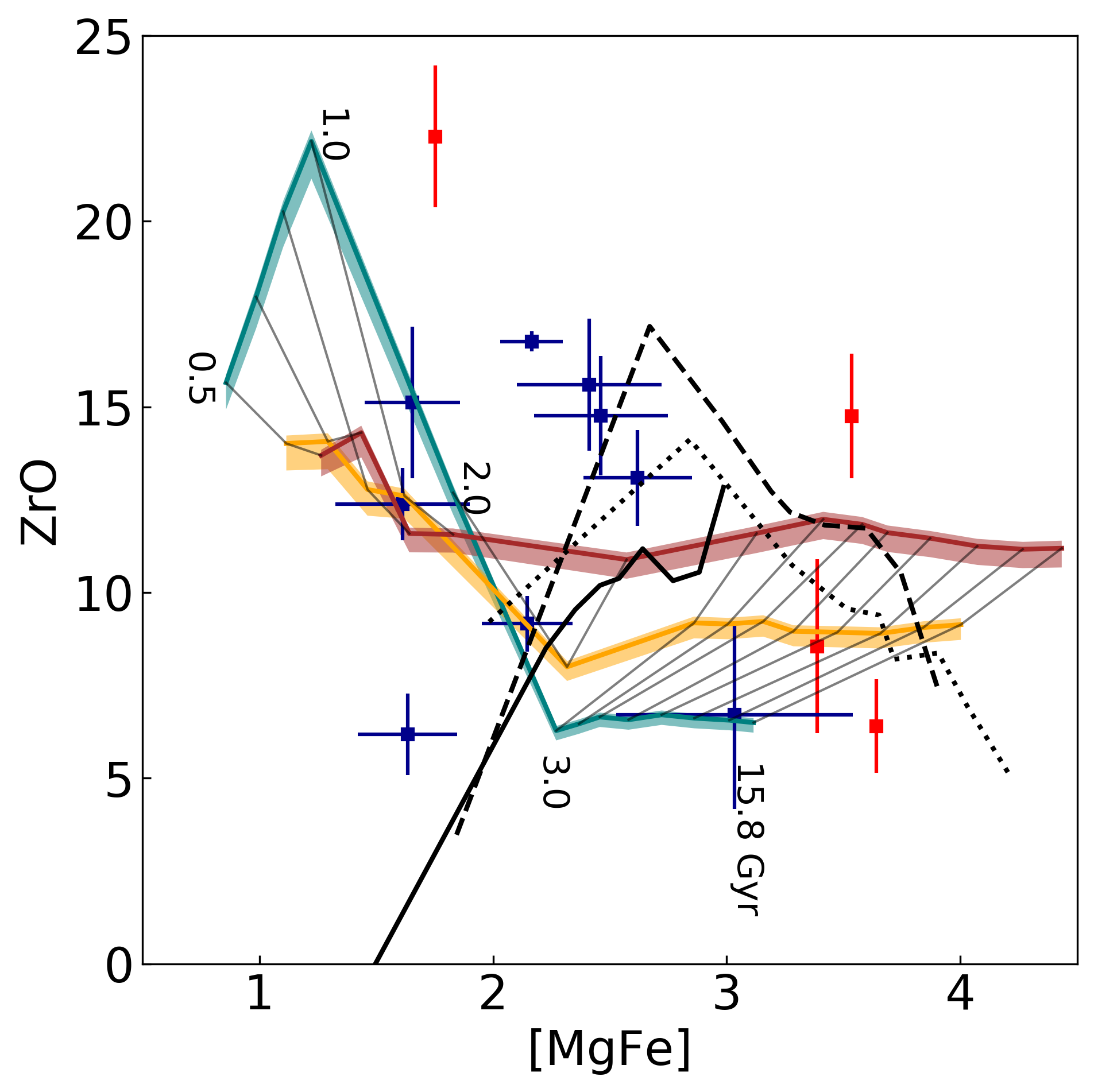}
    \end{subfigure}%
    ~ 
    \begin{subfigure}[]
        \centering
        \includegraphics[width = 0.3\linewidth]{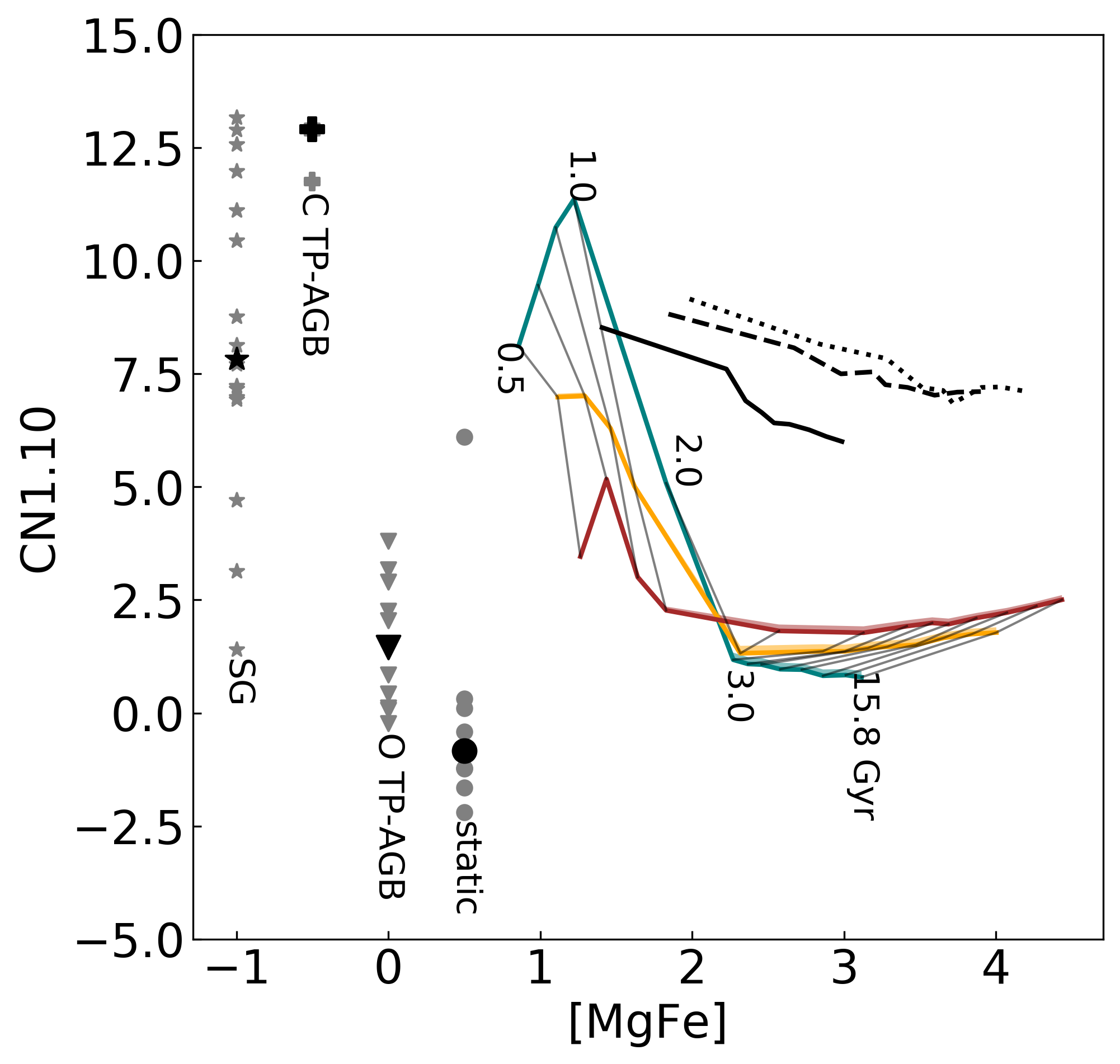}
    \end{subfigure}%
    ~ 
     \begin{subfigure}[]
        \centering
        \includegraphics[width = 0.3\linewidth]{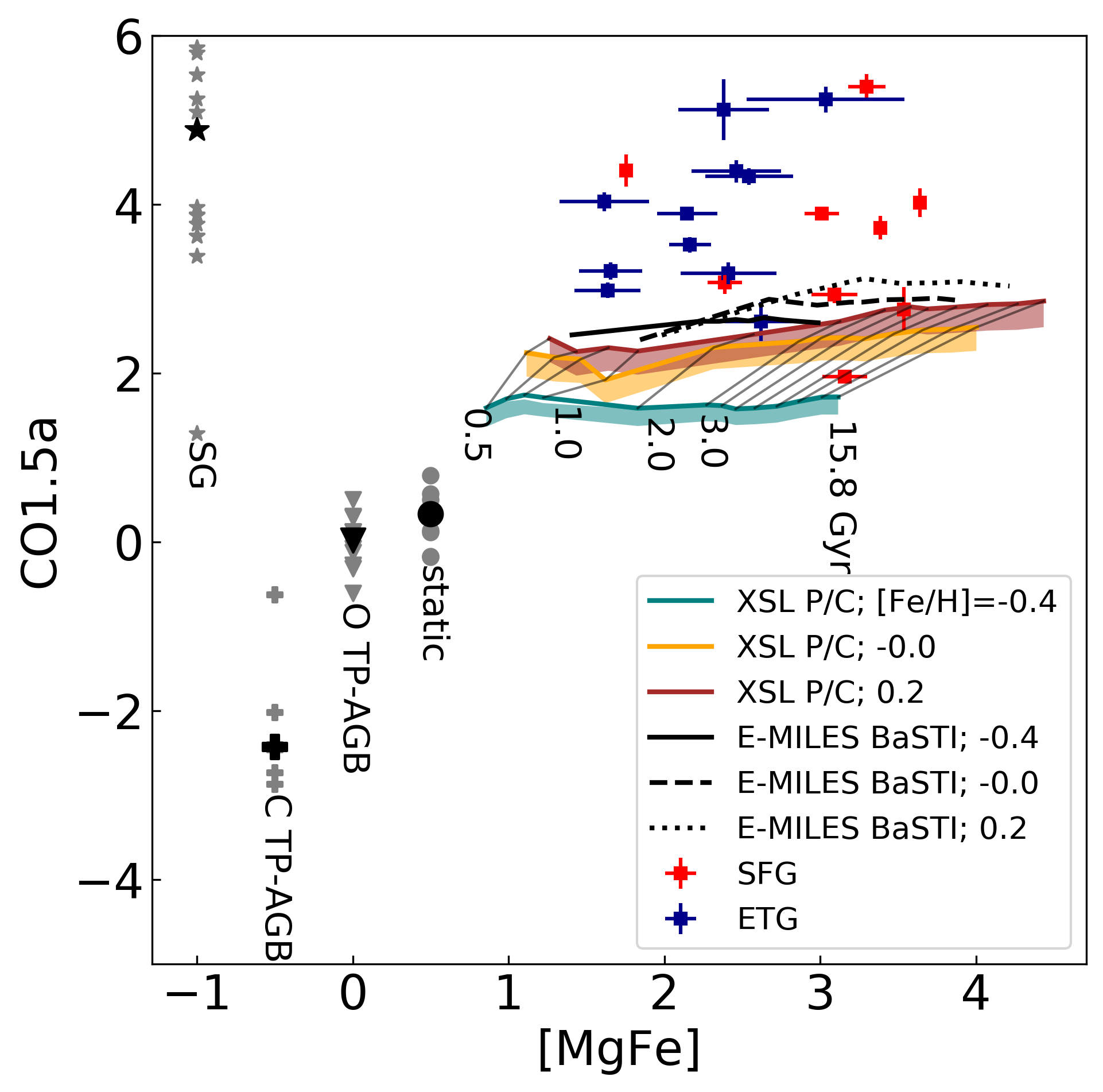}
    \end{subfigure}
    ~ 
    \begin{subfigure}[]
        \centering
        \includegraphics[width = 0.3\linewidth]{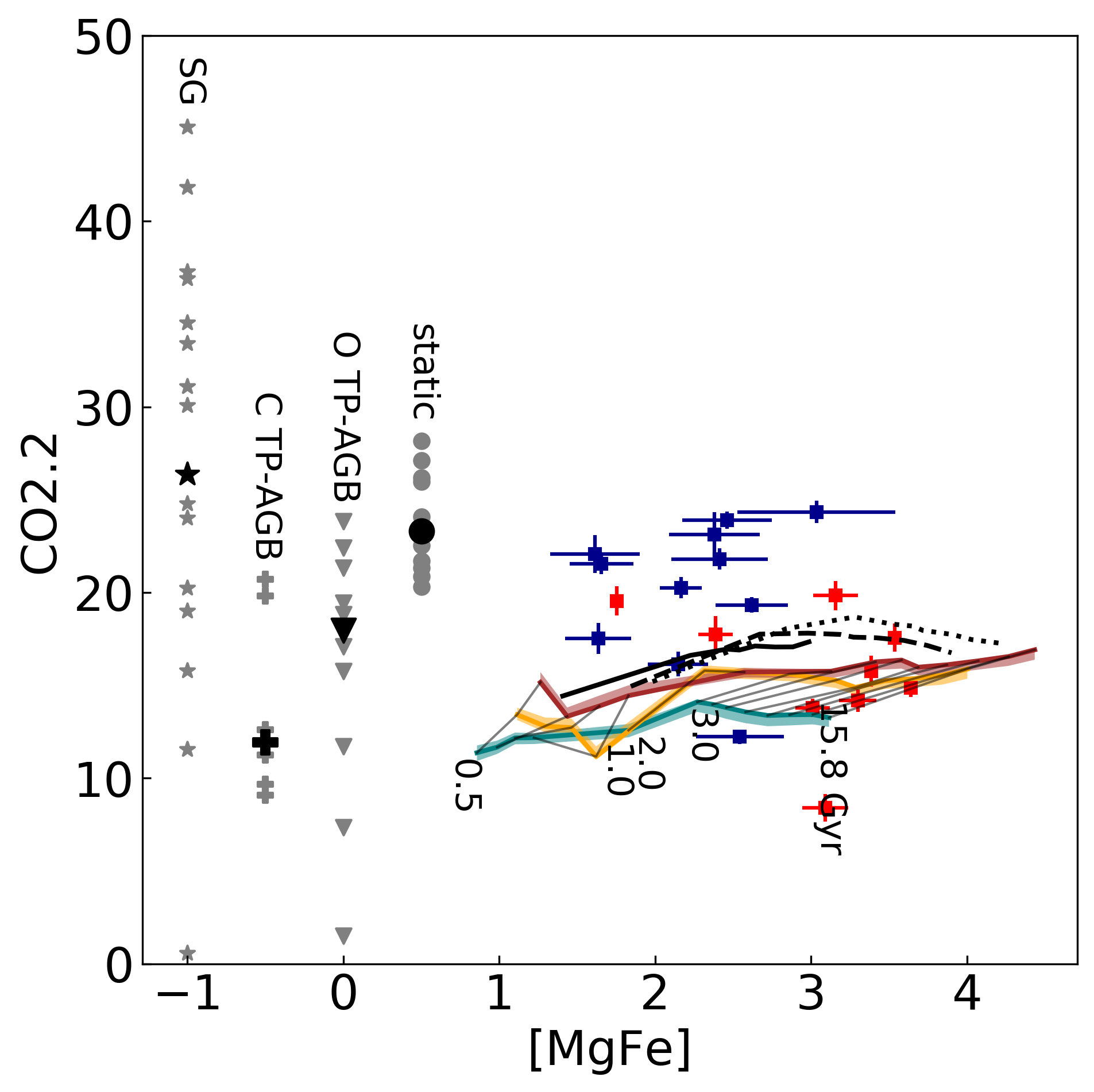}
    \end{subfigure}
    ~ 
    \begin{subfigure}[]
        \centering
        \includegraphics[width = 0.3\linewidth]{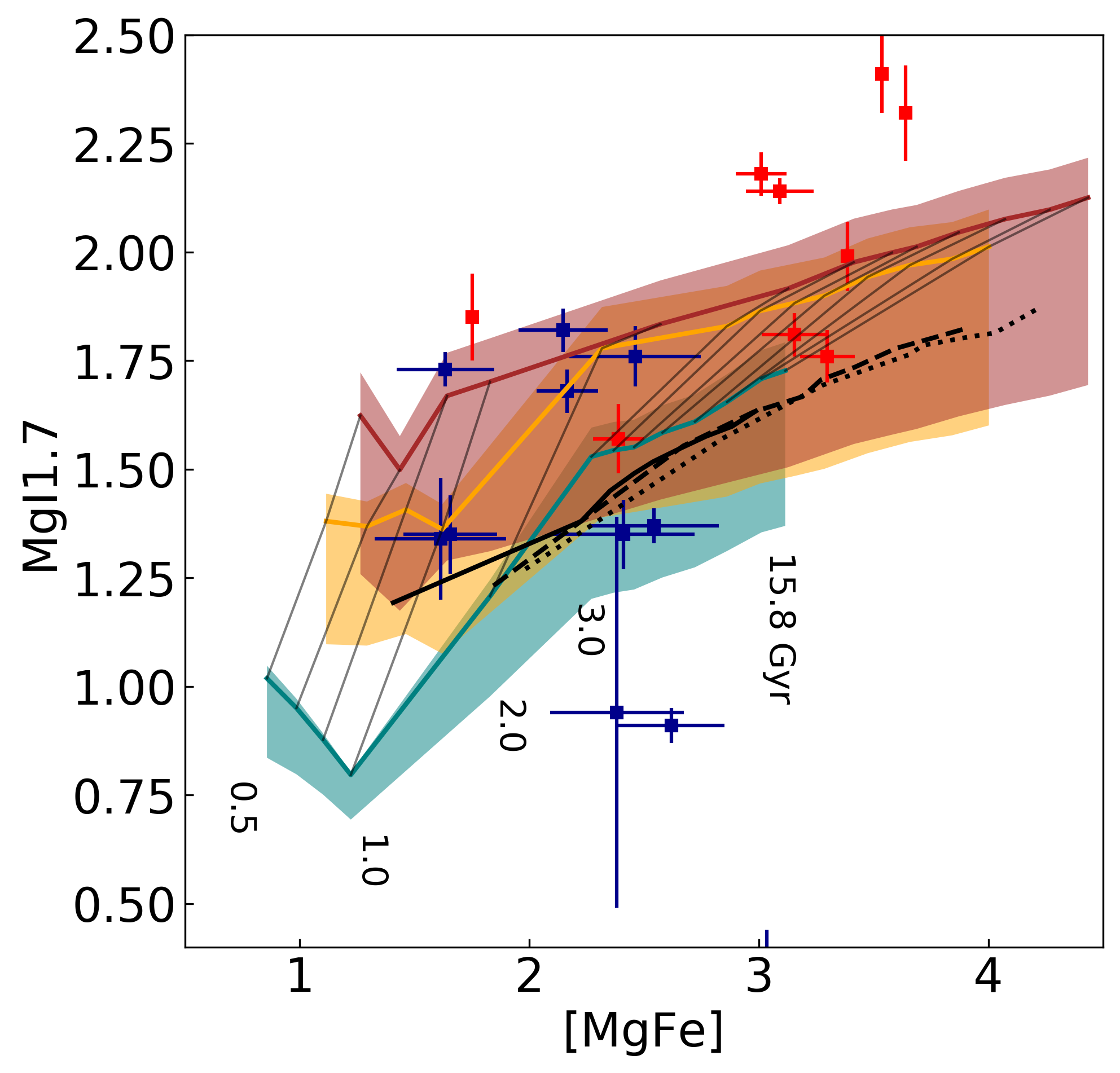}
    \end{subfigure}
     ~ 
      \begin{subfigure}[]
        \centering
        \includegraphics[width = 0.3\linewidth]{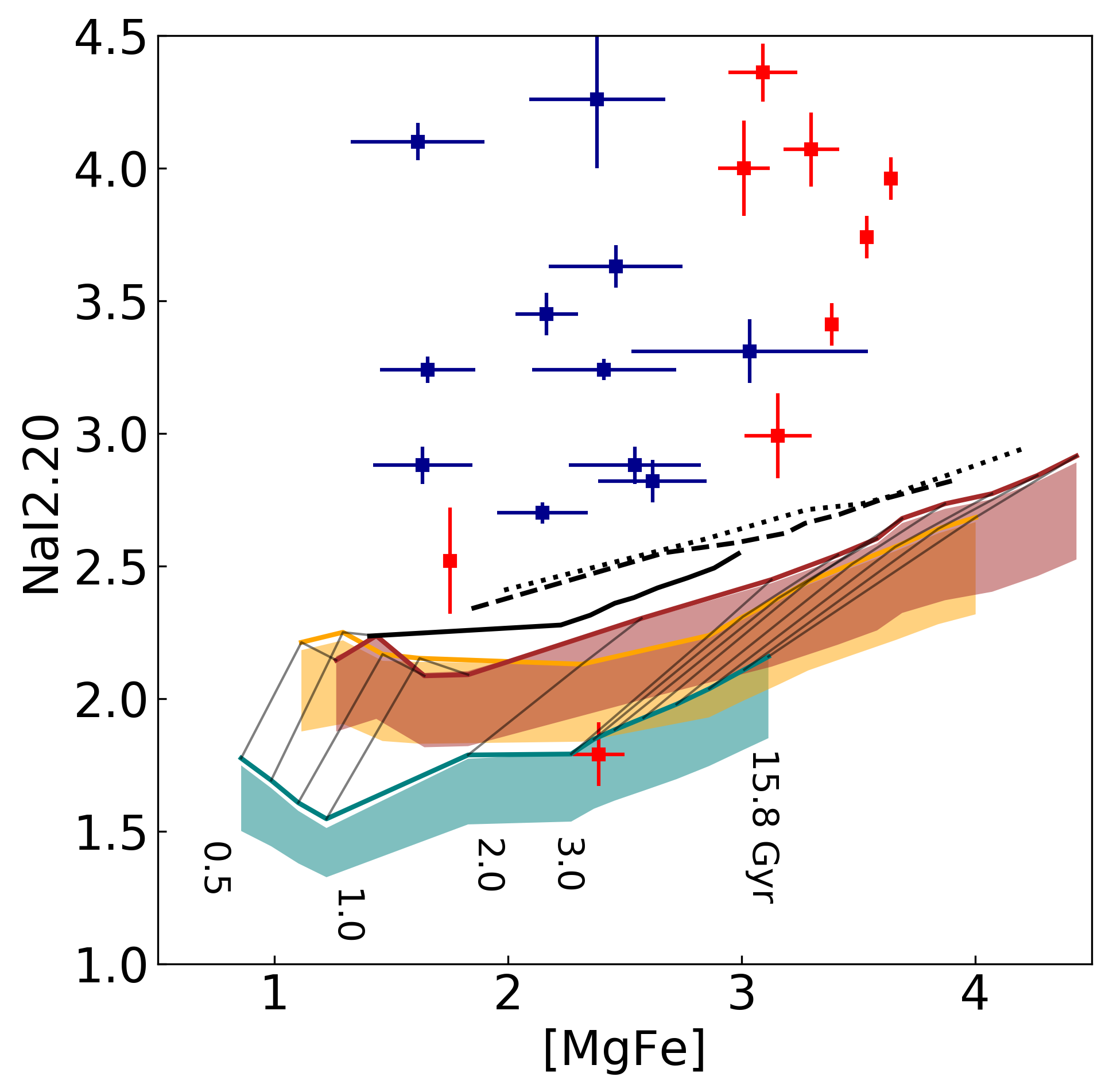}
    \end{subfigure}
    
    \caption{Selected index--index comparisons from \citet{Riffel2019}. Shaded areas represent XSL PARSEC/COLIBRI SSP model predictions with red, yellow and teal indicating $\mathrm{[Fe/H]}=+0.2$, $0.0$ and $-0.40$\,dex, respectively. Shaded areas represent models with a spectral resolution of $\sigma=16\,\mathrm{km\,s}^{-1}$ (the native XSL resolution) to $\sigma=228\,\mathrm{km\,s}^{-1}$, the resolution of the \citet{Riffel2019} spectra, centred on $\sigma=60\,\mathrm{km\,s}^{-1}$ (the E-MILES NIR resolution). Black lines represent E-MILES BaSTI model predictions with dotted, dashed and solid lines representing $\mathrm{[Fe/H]}=+0.26$, $0.06$ and $-0.35$\,dex, respectively (roughly the same metallicities as the XSL models). We note that the age range differs for the E-MILES models, as the NIR spectra of E-MILES are only reliable above 1\,Gyr.
    In panel (b), CN1.10$^{*}$ \citep{Rock2015} index definition is used instead of the \citet[][]{Riffel2019} definition, which is affected by residuals from telluric absorption correction. We have omitted the SFG and ETG measurements of CN1.1 due to differences in index definitions. In panels (b)-(d) XSL static, \mbox{O-rich} \mbox{TP-AGB} and \mbox{C-rich} \mbox{TP-AGB} sequences, and XSL supergiants (which \emph{are not} included in the XSL SSP models) are shown in grey at arbitrary optical index values (as these stars lack optical features) and their median values with larger black symbols. Indices of star-forming galaxies are marked in blue and of ETGs in red. These values are taken from Tables 6--7 and B1--B3 of \citet[][]{Riffel2019} respectively.}
    \label{fig:Riffelindexes}
\end{figure*}

Motivated by this discrepancy, we look at some of the suggested indices from \citet{Riffel2019} and compare them with the XSL PARSEC/COLIBRI SSP model predictions and those from the E-MILES BaSTI models. Although \citet{Riffel2019} found correlations among the different absorption features in the optical and NIR, which seemingly suggest an evolution from a SFG to a ETG, multiple stellar populations are likely to be an issue when attempting to compare optical and NIR indices of SFGs.

We have selected six NIR indices and plot them against [MgFe]. These index--index diagrams are shown in Fig.~\ref{fig:Riffelindexes}. We limit the comparisons to XSL PARSEC/COLIBRI Salpeter and E-MILES BaSTI Salpeter SSP models only, as they have a more up to date handling of cool giant evolutionary phases. We note that \citet{Riffel2019} used E-MILES Padova00 models in their comparisons. 

The ZrO--[MgFe] comparison in Fig.~\ref{fig:Riffelindexes}a shows that this line is affected by the CN features of C-rich TP-AGB stars in XSL SSP models. Metal poor stellar population models with ages less than 3\,Gyr show a steep increase in the strength of this index due to dominance of \mbox{TP-AGB} stars, especially the \mbox{C-rich} stars. Carbon stars have very high ZrO index values, around 70 \AA{}, but we have omitted the giants from Fig.~\ref{fig:Riffelindexes}a for clarity. The E-MILES models show a different behaviour of ZrO -- SSP models with ages less than 2\,Gyr showing a steep decrease in the strength of this index.

We have also included the CN1.10--[MgFe] comparison to illustrate the behaviour of this important NIR index. However, in Figure Fig.~\ref{fig:Riffelindexes}b, we use the CN1.10 index definition of \citet{Rock2015}. We have omitted the SFG and ETG measurements of CN1.10 due to differences in index definitions between \citet{Rock2015} and \citet{Riffel2019}. \citet{Riffel2019} defined the CN1.10 red continuum band at 11310--11345 \AA, coinciding with the region of severe telluric absorption. We see some residuals in the telluric absorption region of XSL spectra, which also affects the NaI1.14 index. The CN1.10 \citep{Rock2015} index definition, the same definition we used to remove supergiants, has the red continuum placed at 11100--11170 \AA, away from the telluric contamination. The CN1.10--[MgFe] comparison in Fig.~\ref{fig:Riffelindexes}b shows a systematic offset between the E-MILES and XSL models in CN1.10 index values. The smaller XSL predictions are a direct consequence of separating \mbox{C-rich} \mbox{TP-AGB} stars and removing supergiant stars, as described in Sect.~\ref{Sect:coolevolvedstars}. The supergiants \emph{are not} included in the XSL SSP models, but are shown in Fig.~\ref{fig:Riffelindexes}b-c for illustrative purposes. The XSL SSP models have mostly shallower CN1.10 features. However, stellar population models with ages less than 3\,Gyr show a steep increase in the strength of this index due to \mbox{C-rich} stars. 

As seen in Fig.~\ref{fig:Riffelindexes}c-d, SFGs show similar, if not stronger, CO1.5a and CO2.2 index features compared to ETGs. However, none of the SSP models reproduce these strong CO features. Both carbon and oxygen are abundant elements in cool giants, and these molecules are formed and can be observed in both M and C stars. But CO2.2 lines and CO1.5 lines originate from different regions within the extended atmospheres of cool stars \citep{Nowotny2005}. Furthermore, stellar population $H$-band CO lines are blends. The CO1.5a line is a blend of CO and \ion{Mg}{i}. The CO2.2 line is an almost pure CO feature \citep{Riffel2019}. The static sequence spectra have strong CO2.2 indices, influencing older models to have strong CO2.2 values. \mbox{TP-AGB} stars show a variety of CO2.2 line strengths -- the \mbox{TP-AGB} phase does not substantially influence the CO index. This was also concluded by \citet{Rock2015}, and the same can be seen for the CO1.5a index. CO is expected to be enhanced in younger ($<50$~Myr) stellar populations \citep[e.g.][]{Lancon2008,Riffel2007,Riffel2015} due to the presence of supergiant stars, which are not included in the XSL SSP models. 

If a SFG hosts even a small population of supergiants, the NIR CO and CN indices \emph{will} be affected, but the optical indices \emph{might not} be affected by this younger population component. This is clearly seen from Fig.~\ref{fig:Riffelindexes}b-d, where the CO and CN index strengths of the XSL supergiants are much stronger than of the other cool giants.

It is now possible to perform in-depth studies of spectral features in the NIR. With moderate-high resolution of and high amount of spectra of cool giant stars, the XSL SSP models are useful tools. On the one hand, XSL SSP models improve the model range of some lines, such as the MgI1.7 line in Fig.~\ref{fig:Riffelindexes}e. On the other hand, XSL models expand the range of predicted values of the NaI2.2 index (Fig.~\ref{fig:Riffelindexes}f), but towards lower index values, contrary to the strong index values of SFGs and ETGs.  Individual elemental abundance variations, velocity dispersion broadening, wavelength shifts, residuals from telluric absorption correction, signal-to-noise ratio, flux calibration, IMF, inclusion of cool giant stars, and the presence of multiple stellar populations can all influence NIR spectral line indices. Indeed, \citet{Rock2017} and \citet{LaBarbera2017} suggested that for ETGs the large values obtained for the NaI2.2 index are due to a combination of a bottom-heavy IMF and enhanced sodium abundances. Further research is needed for the majority of the NIR spectral features, using purposefully defined NIR indices, such as those of \citet{Eftekhari2021}. A full analysis of the colors and indices of the galaxies of \citet{Riffel2019} over the X-shooter range of wavelengths requires models with non-trivial star formation histories and configurations (for instance a prescription for the spatial distribution of dust relative to young and old stars), and lies outside the scope of this paper.

\section{Stellar mass-to-light ratios}
\label{sect:MLratios}
% Placement of a figure of age-Z-MLratio of PC XSL SSPs
\begin{figure*}
    \centering
    \begin{subfigure}[]
        \centering
        \includegraphics[width = 0.25\linewidth]{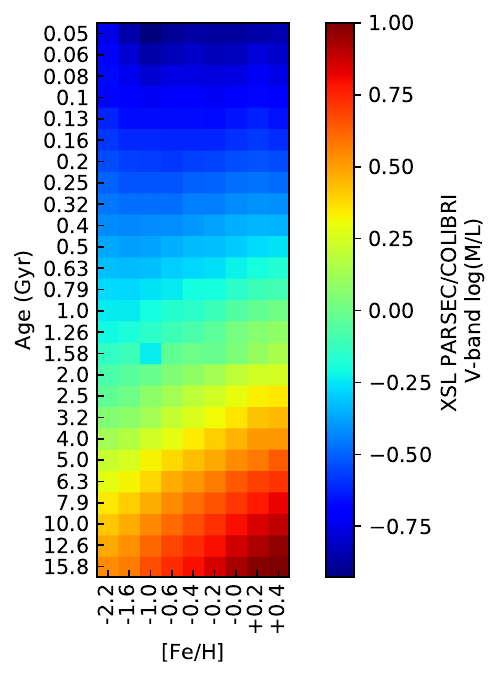}
    \end{subfigure}%
    ~ 
    \begin{subfigure}[]
        \centering
        \includegraphics[width = 0.25 \linewidth]{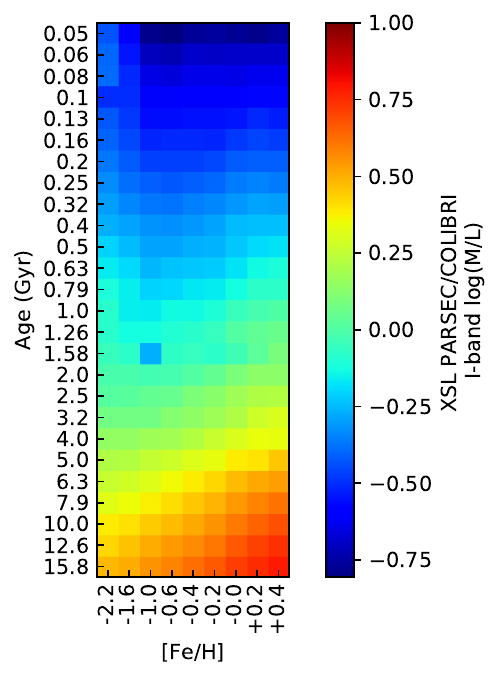}
    \end{subfigure}%
    ~ 
        \begin{subfigure}[]
        \centering
        \includegraphics[width = 0.25 \linewidth]{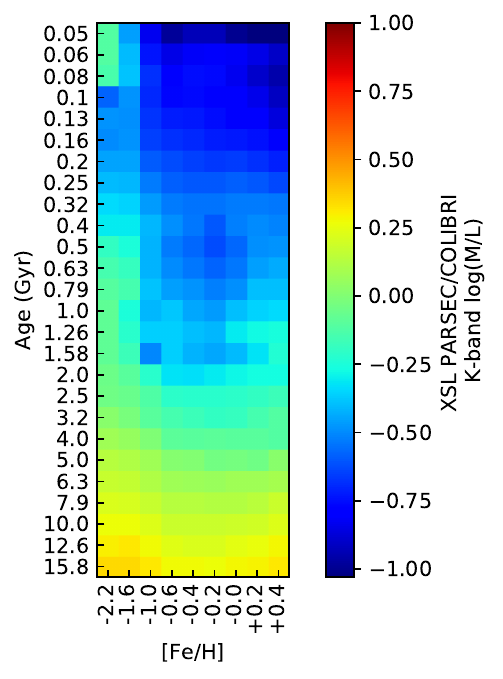}
    \end{subfigure}%
    ~ 
 
    \caption{ The evolution of synthetic (log) stellar mass-to-light ratios for the $V$-band (left), $I$-band (middle) and $K$-band (right), measured from the XSL PARSEC/COLIBRI Salpeter models.}
    \label{fig:MLratios}
\end{figure*}

The stellar mass-to-light ($M_*/L$ hereafter) ratio is an important characteristic of a stellar population. Many of the population properties (e.g. morphology or star formation history) are correlated with the stellar mass \citep[e.g.][to name some recent works]{Bernardi2020,Telford2020, Ge2021,deGraaff2021,DEugenio2021}. The stellar mass of a population is not a directly observable quantity but its luminosity is. One way of estimating population mass is through the synthetic $M_*/L$ ratio: in such a case, the population light is converted into a mass using a stellar $M_*/L$ ratio derived from stellar population models \citep[see reviews by][]{Conroy2013, Courteau2014}. 

Existing stars in a stellar population contribute to the mass and luminosity of that population. But stars progressively die and turn into stellar remnants (white dwarfs, neutron stars and black holes) as the stellar population ages. Those remnants contribute to the mass but not to the luminosity. The total mass of an SSP with a certain age and metallicity is the sum of stellar and remnant masses, weighted by the IMF. The weight of the IMF is determined by the initial mass of the star, but the mass of the star/remnant at that time is what contributes to the mass budget. 

In Fig.~\ref{fig:MLratios} we present the synthetic $M_*/L$ ratios derived from the XSL PARSEC/COLIBRI models with Salpeter IMFs in the $V$, $I$ and $K$ photometric bands. The luminosity is given in units of solar luminosity in the respective photometric band. The solar magnitudes used are: ($V$, $I$, $K$)=(4.81,4.11,3.30) mag, measured from the Solar spectrum of \citet{SUNspec1996}. We use the relation in the mass range $0.09<m/M_\odot<120$. The PARSEC/COLIBRI models describe the mass loss of stars and provide both initial and actual stellar masses for existing stars. We use the metallicity-dependent initial--remnant mass relation descriptions provided in \citet{Fryer2012} for massive stars ($9$--$120\,M_\odot$) for non-solar metallicities and the \citet{Sukhbold2016} relation for solar metallicity. For low- and intermediate-mass stars ($0.87< M_{*,init} < 8.2\,M_\odot$), we use the PARSEC-based white dwarf initial--final mass relation of \citet{Cummings2018}, extrapolating the relation to $8.2$--$9\,M_\odot$. We assume that the mass lost in the form of ejected gas is blown out of the stellar population and does not contribute to the mass budget.

The dominant driver of SSP luminosity is its age, as the most-massive stars have short lifetimes but are orders of magnitude more luminous than the less massive stars. The luminosity of an SSP changes rapidly with time. The mass of an SSP is dominated (for the Salpeter IMF considered here) by the least-massive stars. These stars live a long time, and thus the mass of an SSP changes little after the first few Gyrs. As seen from Fig.~\ref{fig:MLratios}, the $M_*/L$ ratio changes rapidly until about 2\,Gyr, with the most massive and luminous stars dying off. The effect of metallicity on the $M_*/L$ ratio is weaker. For stellar populations older than a few Gyrs, the higher the stellar population’s metallicity, the higher the $M_*/L$ ratios for optical passbands, but the (slightly) lower the NIR $M_*/L$ ratio. 

The differences between the $M_*/L$ ratios in the $V$, $I$ and $K$ photometric bands are expected, as the hottest turn-off star determines the $V$-band luminosity; the stars at the tip of the RGB determine the $K$-band luminosity of old stellar populations, and the TP-AGB stars determine the $K$-band luminosity of 50\,Myr to 2\,Gyr populations. Furthermore, the influence of \mbox{TP-AGB} stars on the $M_*/L$ ratio peaks for populations with ages between 0.4 to 1.58\,Gyr and metallicities between $\mathrm{[Fe/H]}=-0.6$ and $0$. These stars emit mostly in the NIR, increasing the NIR luminosity and lowering the NIR $M_*/L$ ratio. This can be clearly seen in Fig.~\ref{fig:MLratios}c. The $M_*/L$ ratio is dependent on the stellar evolutionary phases accounted for in the modelling. Without the \mbox{TP-AGB} stars, the $M_*/L$ would increase monotonically with age.

The $M_*/L$ ratio is strongly dependent on the IMF. We provide discussion of the $M_*/L$ ratios from XSL PARSEC/COLIBRI models calculated with other IMFs in an upcoming paper (Verro et al. in prep.). Furthermore, there are differences between $M_*/L$ ratios determined from different models. We discuss this briefly in Appendix \ref{MLratiomodelcomparison}.

\section{On the separation of static and variable giants}
\label{sect:static/variable}
Separating `static' cool giant (from RGB to E-AGB) stars from the variable \mbox{TP-AGB} stars with the use of the `static' and `variable' sequences in XSL SSP models is an important step towards understanding the source of NIR flux in stellar populations. These stars lie very close to each other on the HR diagram, but their spectral shapes can be very different, as discussed in Sect. \ref{Sect:coolevolvedstars}. There is an ongoing debate as to their impact on the integrated spectra of even simple stellar populations. Clear \mbox{C-rich} \mbox{TP-AGB} signatures have been detected in some of the $J$- and $H$-band spectra of globular clusters in the LMC \citep{Lyubenova2012}. These globular clusters are intermediate age (1--2\,Gyr) and have metallicities around $\mathrm{[Fe/H]}=-0.4$\,dex. However, \citet{Zibetti2013} explored a set of post-starburst galaxies, with luminosity-weighted ages between 0.8 and 1.6\,Gyr and metallicities between $\mathrm{[Fe/H]}=-0.68$ and $+0.3$\,dex and found no strong spectral signatures of these stars. This discrepancy has been explained by \citet{Girardi2013} by `AGB boosting' effect, which is linked to the physics of stellar interiors -- stellar populations in a narrow 1.57 and 1.66 Gyr age range at MC metallicities have \mbox{TP-AGB} contribution to the integrated luminosity of the stellar population increase by a factor of $\sim$2. This was recently confirmed by \citet{Pastorelli2020}; their modelling showed a 80\% peak in $K$-band flux coming from (mainly \mbox{C-rich}) \mbox{TP-AGB} stars.

\subsection{RGB and \mbox{TP-AGB} light fractions in the NIR}
\label{sect:RGBAGBlight}
\begin{figure*}
    \centering
    \begin{subfigure}[]
        \centering
        \includegraphics[width = 0.24\linewidth]{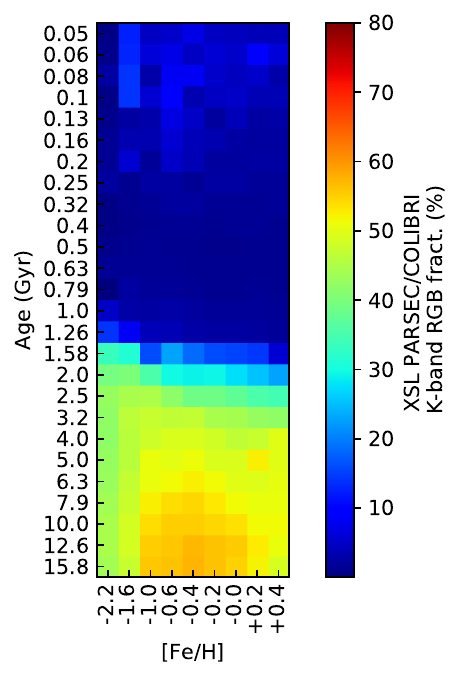}
    \end{subfigure}%
    ~ 
    \begin{subfigure}[]
        \centering
        \includegraphics[width =0.24 \linewidth]{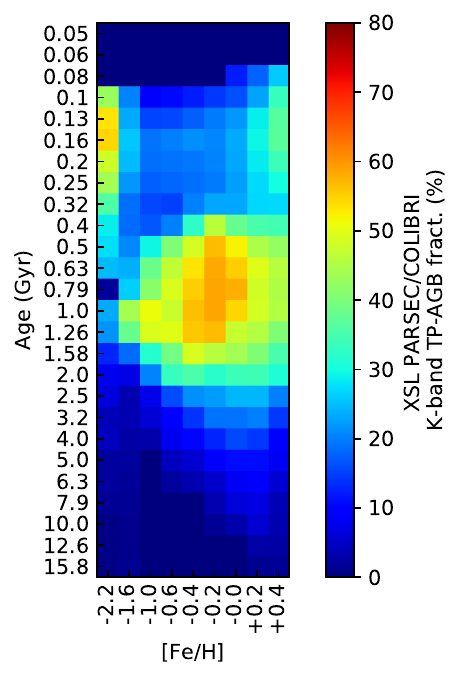}
    \end{subfigure}%
    ~ 
     \begin{subfigure}[]
        \centering
        \includegraphics[width = 0.22\linewidth]{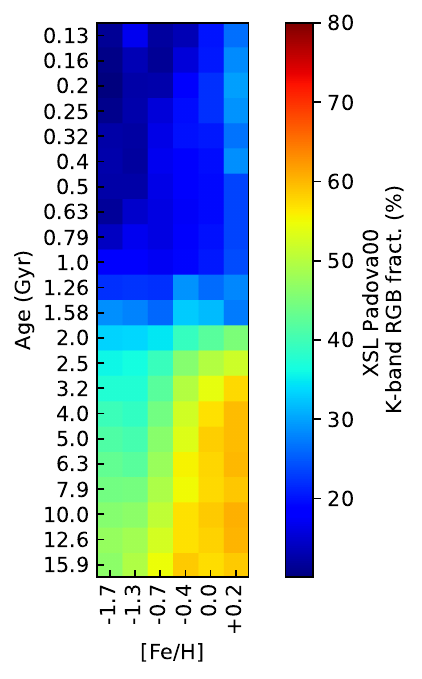}
    \end{subfigure}%
    ~ 
    \begin{subfigure}[]
        \centering
        \includegraphics[width =0.22 \linewidth]{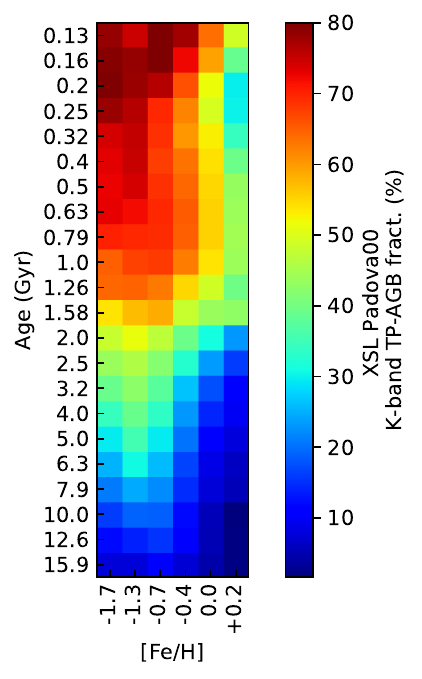}
    \end{subfigure}%
    ~ 
 
    \caption{The contribution of RGB stars and \mbox{TP-AGB} stars to the total $K$-band luminosity of the XSL models. Panels (a) and (b): the contribution of RGB and \mbox{TP-AGB} stars, respectively, in the PARSEC/COLIBRI models. Panels (c) and (d): the contribution of RGB and \mbox{TP-AGB} stars, respectively, in the Padova00 models.}
    \label{fig:Cooliantfracs}
\end{figure*}

We show the contribution of RGB stars and \mbox{TP-AGB} stars to the total K-band luminosity of the XSL models in Fig.~\ref{fig:Cooliantfracs}. In the XSL PARSEC/COLIBRI SSP models, the RGB contribution changes from low in young populations to high in old populations, with a strong transition around 2\,Gyr. A high contribution of \mbox{TP-AGB} stars to the $K$-band flux extends roughly from 0.5 to 1.6\,Gyr, contributing 40\% or more of the flux in the $K$-band at these ages, peaking around 0.8\,Gyr and $\mathrm{[Fe/H]}=-0.2$\,dex, contributing 55\%--60\% of the $K$-band flux in this population. Mainly \mbox{C-rich} \mbox{TP-AGB} stars contribute to this peak. C-star $H$-band signatures can be recognized in Fig. \ref{fig:fullSSPs}a for the 1~Gyr solar metallicity XSL PARSEC/COLIBRI SSP models. However, we do not see the `AGB boosting' peak in our 1.58~Gyr and $\mathrm{[Fe/H]}=-0.4$\,dex SSP models. For younger and older ages, the predicted \mbox{TP-AGB} contribution is almost entirely due to \mbox{O-rich} stars and increases with metallicity, together with the \mbox{O-rich} \mbox{TP-AGB} lifetimes. There is another peak in the \mbox{TP-AGB} contribution in very young and very metal-poor SSP models, where the flux contribution from other stars is lower. 

These behaviours are expected. The \mbox{TP-AGB} phase for low and intermediate mass stars ($M = 2$--$7\,M_\odot$) culminates in stellar populations of ages between 0.5 and 2\,Gyr. These stars emit mainly in the NIR spectral range, given their low temperatures. The SSP models calculated in \citet{Pastorelli2020} predict a \mbox{TP-AGB} contribution peak at around 1\,Gyr (roughly between 0.3 and 2\,Gyr) that does not exceed 55\% in the $K$-band luminosity. In comparison, the peak is as high as 80\% in the $K$-band at $\mathrm{[Fe/H]}=-0.3$\,dex for the M05 models.

On the other hand, the XSL Padova00 models have a completely different \mbox{TP-AGB} fraction behaviour with SSP parameters: the younger and more metal-poor models have higher \mbox{TP-AGB} fraction. For example, a 0.2\,Gyr, $\mathrm{[Fe/H]}=-1.7$\,dex model has 80\% of its $K$-band flux coming from \mbox{TP-AGB} stars. This is why we discourage the usage of XSL Padova00 models outside of the narrow safe zone suggested in Sect.~\ref{sect:safety}.

\subsection{Colour--temperature relations}
\label{sect:ctrelations}
\begin{figure}
    \centering
    \includegraphics[width = 0.5\textwidth]{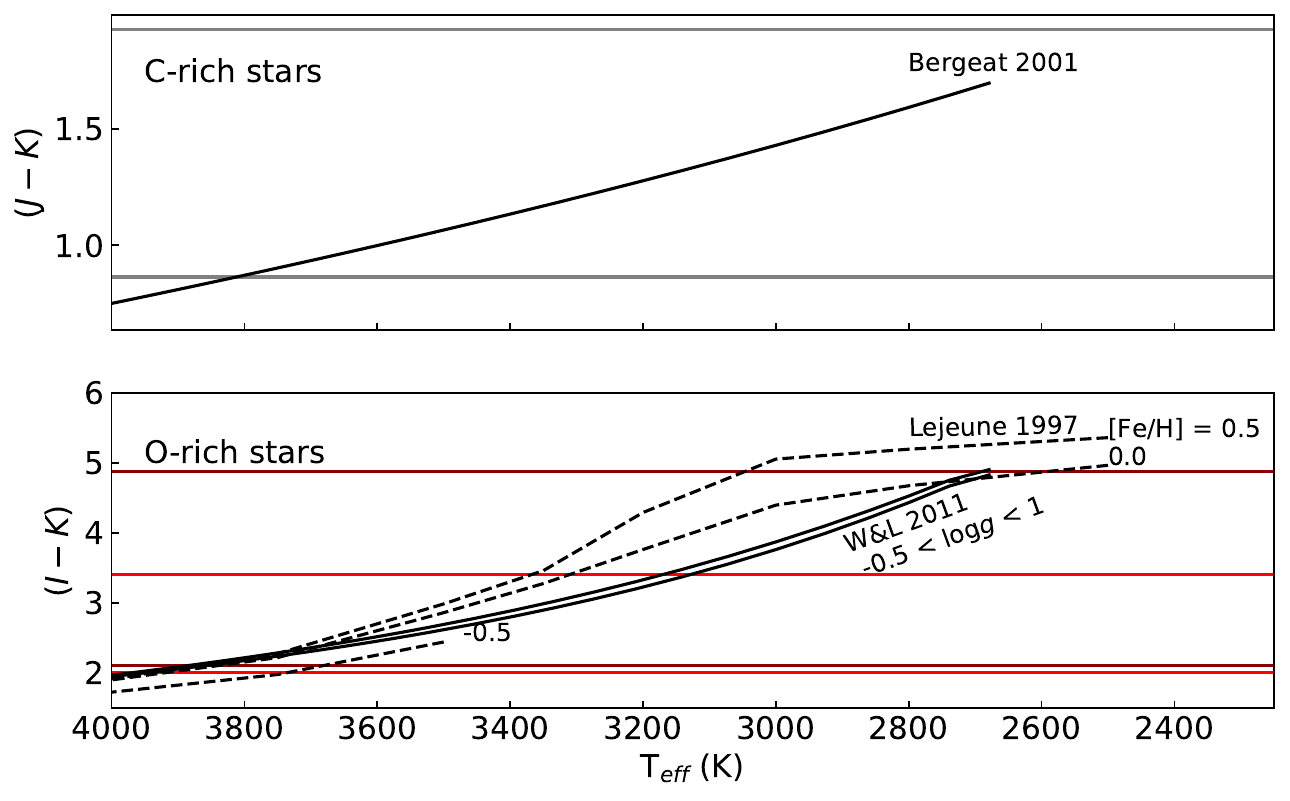}
    \caption{The colour--temperature relations for \mbox{C-rich} and \mbox{O-rich} stars. The grey horizontal lines (upper panel) represent the colours of the bluest and reddest average spectra on the \mbox{C-rich} \mbox{TP-AGB} sequence. The red and the dark red horizontal lines represent the bluest and reddest spectra from the \mbox{O-rich} `static' and \mbox{O-rich} \mbox{TP-AGB} sequences respectively.}
    \label{fig:colour_param_relation}
\end{figure}

We have used colour--temperature relations of \citet{Worthey2011} and \citet{Bergeat2001}, shown in Fig.~\ref{fig:colour_param_relation}, to assign an average spectrum of a \mbox{O-rich} static/variable or \mbox{C-rich} star to a point on an isochrone when generating an SSP model. This allows us to bypass stellar parameter estimation for the complex XSL stars which make up these average spectra. This assignment comes with some caveats. 

\begin{figure}
    \centering
    \includegraphics[width = 0.5\textwidth]{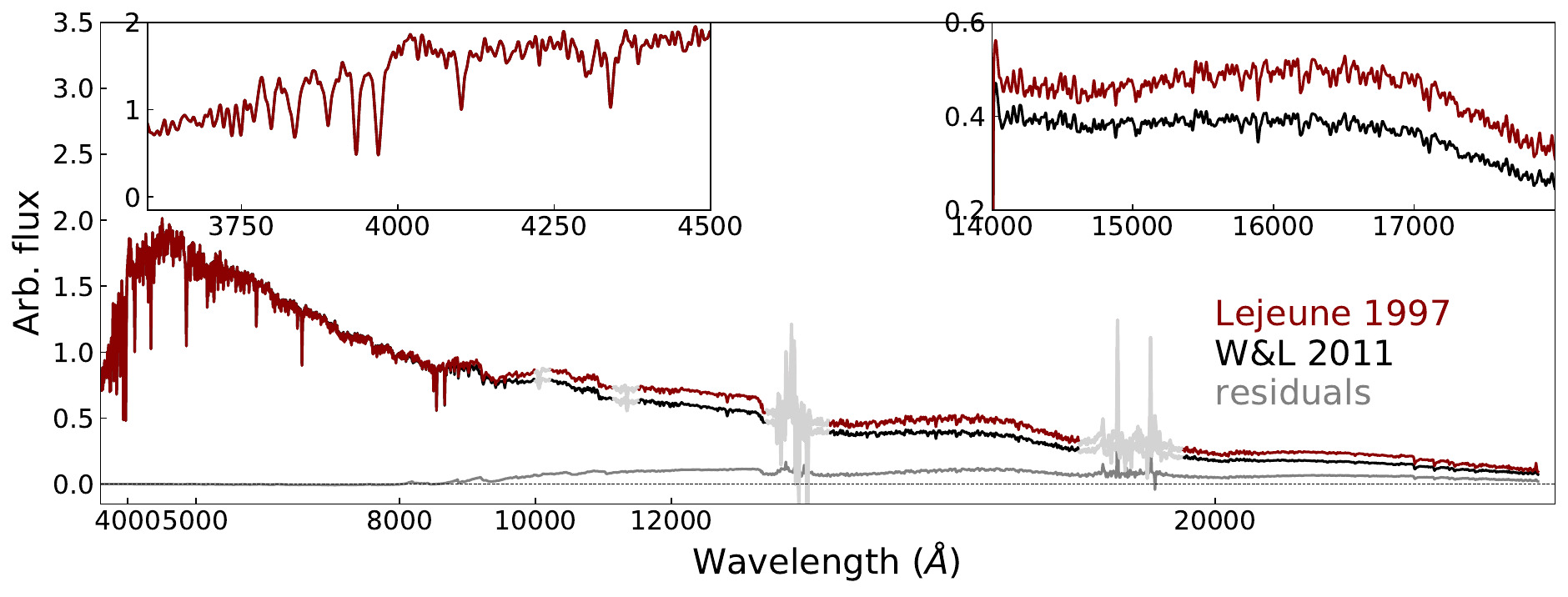}
    \caption{A 1\,Gyr old, supersolar-metallicity XSL SSP generated using the \citet{Lejeune1997} (red) and \citet{Worthey2011} (black) colour--temperature relations.}
    \label{fig:W_L_exapleSSP}
\end{figure}

We have made a version of the models using the colour--temperature relation of \citet{Lejeune1997} to compare with the \citet{Worthey2011} relation used in our default models. Figure~\ref{fig:W_L_exapleSSP} shows that the \citet{Lejeune1997} relation undesirably enhances the NIR fluxes and introduces stronger \mbox{O-rich} \mbox{TP-AGB} features (such as the $H$-band H$^-$/H$_2$O feature) to the models. 

We also note here that the static sequence does not have spectra that appropriately represent RGB or E-AGB stars with colour-inferred temperatures less than 3100\,K. This affects the older metal rich models the most, where the tip of the RGB dominates the NIR light. This is the reason why the supersolar-metallicity XSL SSP models cannot reach NIR colours as red as E-MILES models in the colour comparisons in Sects.~\ref{sect:generalbehaviour} and \ref{sect:comacolours}. The E-MILES models might achieve these colours by effectively mixing the redder spectra of \mbox{TP-AGB} with the spectra of RGB stars (within the local interpolator scheme) at the tip of the RGB in these populations, as those models do not distinguish between RGB and post-RGB (E-AGB and \mbox{TP-AGB}) stars. Our empirical separation into static and O-rich \mbox{TP-AGB} stars based in Fig. \ref{fig:IK_Hbump}, might exclude from our static sequence a few spectra of very cool stars that would in fact be acceptable representations of the coolest RGB stars. However, it remains difficult to match these cool spectra with synthetic ones (see \citet{Lancon2019,Lancon2021}) and hence to separate effects of temperature, metallicity, circumstellar extinction and variability. A star-by-star study that would incorporate variability and mid-infrared information where available, may help improve this separation in future versions.

Furthermore, the XSL does not have spectra that appropriately represent O-rich TP-AGB stars with colour-inferred temperatures less than 2700\,K and C-rich TP-AGB stars with colour-inferred temperatures less than 1700\,K. The \citet{Worthey2011} colour-temperature relation for O-rich giants does not go to lower temperatures. Also, none of the colour-temperature relations discussed in \citet{Worthey2011} go to lower temperatures. By contrast, the PARSEC/COLIBRI models include extremely cool TP-AGB stars (e.g the 1~Gyr model in Fig. \ref{fig:limits_HRD}). The reddest average spectrum of the O-rich or C-rich TP-AGB sequence will represent these stars in our models.

\subsection{Metallicity effects}
The metallicities of the majority of the stars from which the `static', \mbox{O-rich} and \mbox{C-rich} \mbox{TP-AGB} sequences were constructed, are unknown or not accurately known. These stars come from a variety of environments - the solar neighbourhood, star clusters, the Galactic bulge and the Magellanic Clouds. Furthermore, the range of ages in those stars is likely to correspond to a range of metallicities, as determined by the chemical evolution of these environments. Hence, we have combined together spectra with various metallicities. We note that the SEDs of \mbox{C-rich} \mbox{TP-AGB}s are not as sensitive to metallicity as to the effective temperature, or the C/O ratio \citep{Lancon2002}, so we do not consider these stars in this discussion.

The combination of O-rich TP-AGB stars with different metallicities will have three separate consequences on the resulting SSP modelling. Firstly, the relation between effective temperature and spectrophotometric properties changes due to molecular opacities. Lower metallicity cool giants have bluer spectra with weaker molecular bands \citep{Hauschildt1999,Lancon2002}. This would mean one set of `static' and \mbox{O-rich} \mbox{TP-AGB} averages would not be enough and we would need to create `static' and \mbox{O-rich} \mbox{TP-AGB} sequences for different metallicity bins separately. Considering that each average on the `static' and \mbox{O-rich} sequence consists of a handful of spectra, further division would be impossible. Furthermore, neither the \citet{Bergeat2001} or \citet{Worthey2011} relations we use in the modelling are metallicity dependent. 

Secondly, the AGB evolutionary tracks shift to lower effective temperatures. Again, signatures of cooler giant stars will be less pronounced at lower metallicities. We do take this effect into account, as the lower metallicity isochrones we use here shift into cooler temperatures and we select giants with lower colour-temperature from the `static' and \mbox{O-rich} and \mbox{TP-AGB} sequences.

The third effect is the metallicity dependence of the mass loss. It affects the AGB lifetimes and the efficiency of the production of \mbox{C-rich} \mbox{TP-AGB} stars. This is addressed in the stellar evolution calculations, that we use as an input \citep[e.g.][and other papers from the PARSEC/COLIBRI group]{Pastorelli2020}.

\section{Applicability of the XSL stellar population models}
\label{sect:safety}

\begin{figure}
    \centering
    \includegraphics[width = 0.5\textwidth]{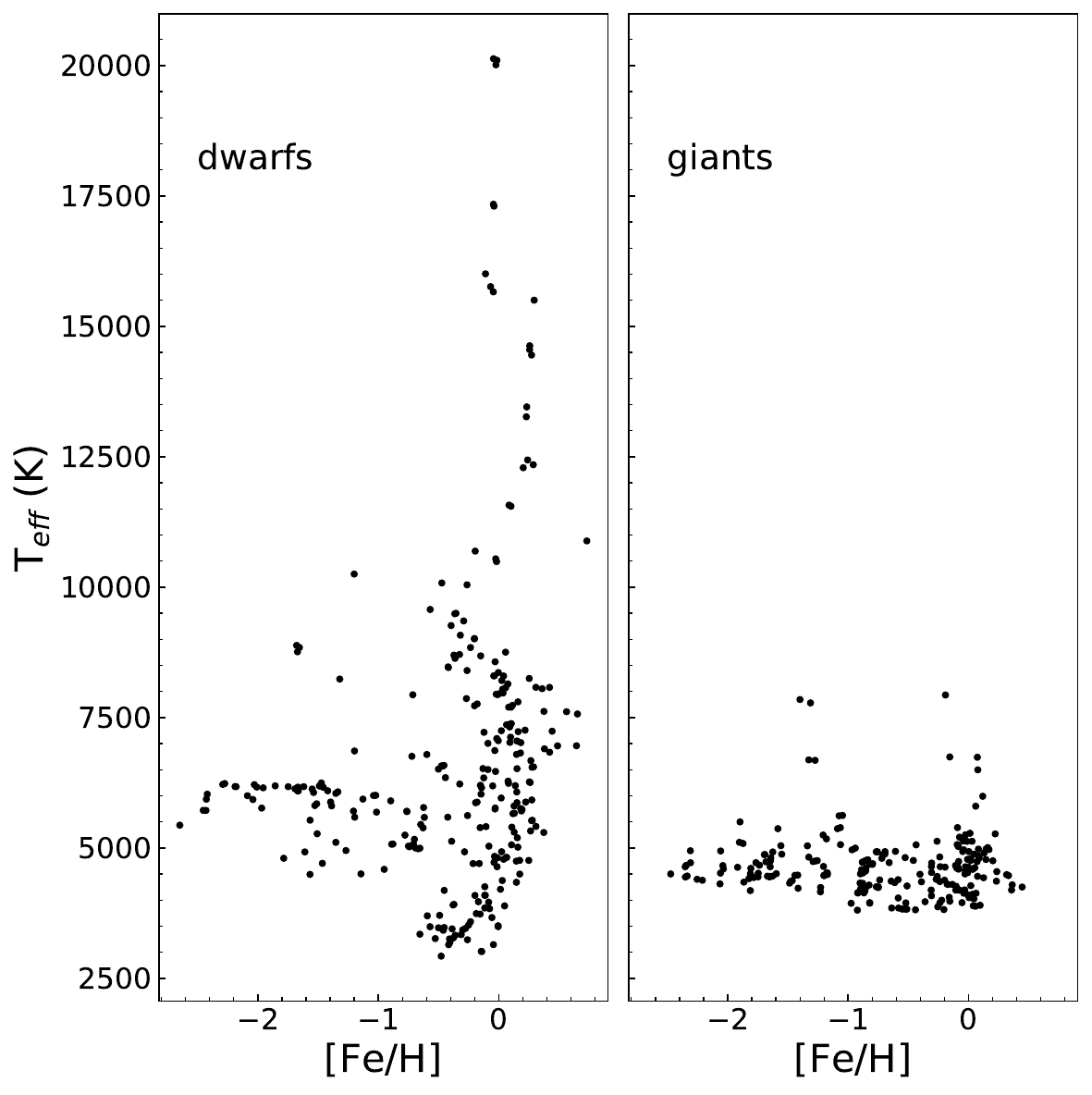}
    \caption{The fundamental parameter coverage of the XSL stars, for dwarfs (left-hand panel) and giants (right-hand panel). Dwarfs and giants are separated at $\log g=3$. The stars which are used to construct static, variable (\mbox{O-rich} \mbox{TP-AGB}) and carbon (\mbox{C-rich} \mbox{TP-AGB}) star sequences are not included, as we ignore the information about their metallicity in the models (Sect.~\ref{Sect:coolevolvedstars}).}
    \label{fig:fundamentalplane}
\end{figure}

The coverage of the spectral library, the capabilities of interpolator(s) and the selected isochrones determine the stellar population models which we can create. Figure~\ref{fig:fundamentalplane} shows the parameter coverage of the XSL stars for dwarfs and giants (separated at $\log g=3$). At solar metallicity, all types of stars are well represented, and  the coverage is good even for the lower metallicities. Reliable models can be computed down to $[\mathrm{Fe/H}] = -2.2$\,dex, for very old populations, due to lack of hot metal poor ($6500 < T_\mathrm{eff}< 10000$\,K) stars, and down to $[\mathrm{Fe/H}] = -1.6$\,dex for intermediate stellar population models. The coverage of metal-rich dwarf and giant stars allows us to safely compute stellar population models up to $[\mathrm{Fe/H}] = +0.20$\,dex. 

\begin{figure}
        \centering
        \includegraphics[width = \linewidth]{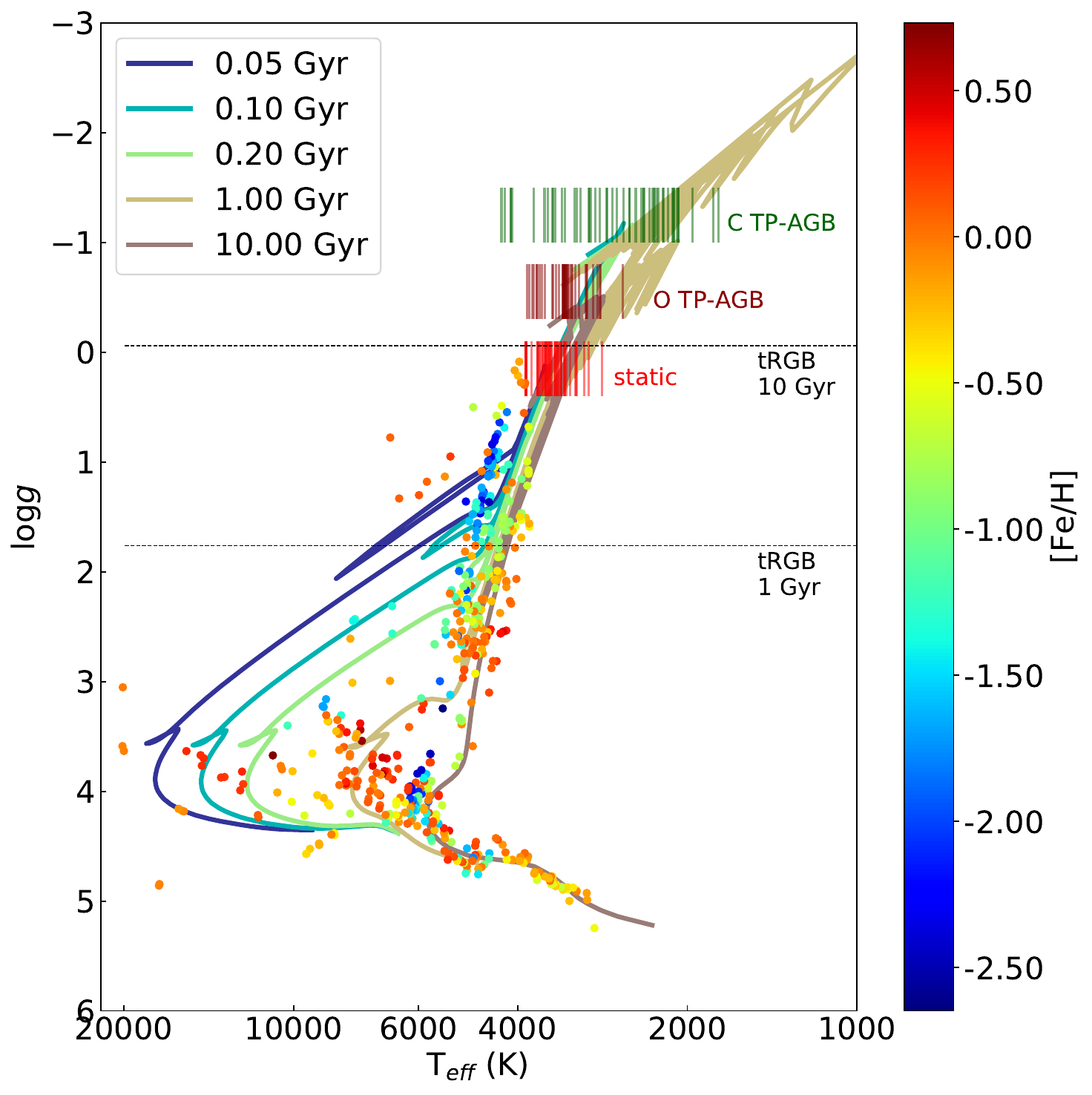}
    \caption{Selected PARSEC/COLIBRI solar metallicity isochrones illustrating the coverage of isochrones by the XSL DR3. We have pointed out the locations of the tip of the RGB for the 1~Gyr and 10~Gyr isochrones with black horizontal lines. We have used the colour-temperature relations used in this study to add the giants into this Figure to illustrate how well the cool RGB and AGB stars are represented (at arbitrary $\log g$ values). Because luminous cool stars are particularly important contributors to the red and near-infrared light of galaxies, XSL was designed to contain a large number of such objects. We are limited by the lack of hot stars to be able to create XSL SSP models younger than 50\,Myr.}
     \label{fig:limits_HRD}
\end{figure}

An HR diagram, such as those shown in Fig.~\ref{fig:limits_HRD} with 50, 80, 100, 150, 200\,Myr, 1\,Gyr and 10\,Gyr solar-metallicity PARSEC/COLIBRI isochrones, gives another perspective. We are limited by the lack of hot stars in XSL for very young ages even at solar metallicity. Furthermore, we have removed supergiants. As the hottest turn-off star determines the shape of the optical population model and the supergiants dominate the NIR light of young populations, we can create stellar population models of 50\,Myr and older. However, the lack of blue loop stars limits the models to ages older than roughly 80\,Myr at solar metallicity or older than roughly 100\,Myr at subsolar metallicities. 

Because luminous cool stars are particularly important contributors to the red and near-infrared light of galaxies, XSL was designed to contain a large number of such objects. The PARSEC/COLIBRI models in Fig.~\ref{fig:limits_HRD} include TP-AGB stars cooler than any XSL TP-AGB star, given the colour-temperature relations of \citet{Worthey2011}. We believe the XSL sample has allowed for significant progress in inclusion of TP-AGB stars into SSP models. That is why we do not base our judgement of the age and metallicity limits of the models on the XSL coverage of this extreme region of the HRD. However, the limits of the XSL Padova00 models are justified based on the discussion in Sect. \ref{sect:RGBAGBlight} regarding the handling of \mbox{TP-AGB} stars.

\begin{figure}
    \centering
    \includegraphics[width = 0.30\textwidth]{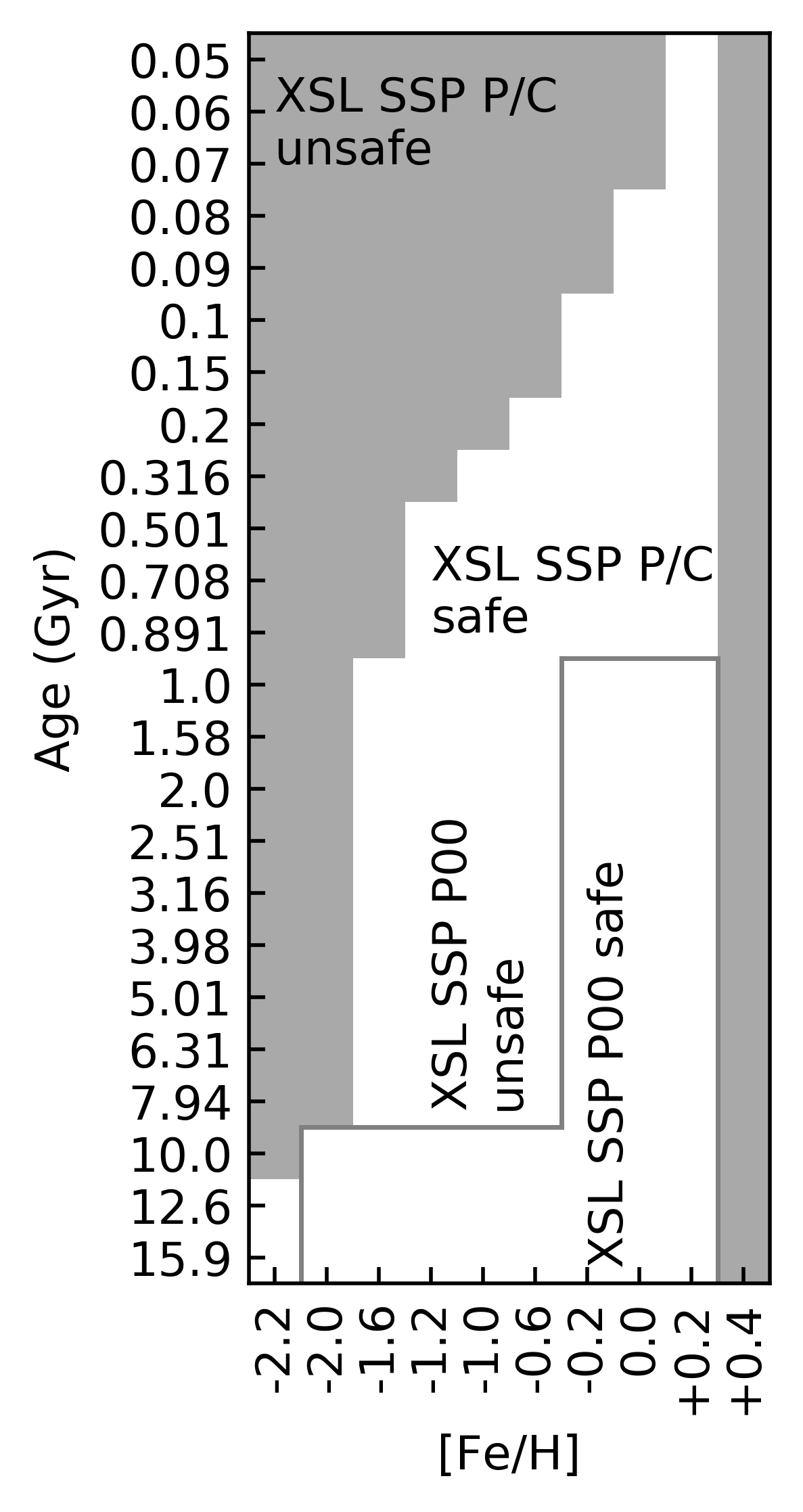}
    \caption{Safe zones of XSL simple stellar population models. `P/C': PARSEC/COLIBRI-based models (white/gray -- safe/unsafe); `P00': Padova00-based models (gray outline -- safe/unsafe).}
    \label{fig:safezones}
\end{figure}

Figure~\ref{fig:safezones} shows the age and metallicity limits of the XSL simple stellar population models, based on the age and metallicity coverage of the isochrones and the temperatures and metallicity of the XSL stars included in the model. 

Stellar population modelling smears out the individual issues of the constituent stellar spectra to some extent. Nevertheless, we warn that certain areas in the SSP model spectrum may be influenced by the dichroic contamination or residuals from telluric absorption correction present in many of the XSL DR3 spectra (Verro 2021a). Table \ref{tab:dichroic_bands} defines the problematic spectral regions. 

XSL SSP models are based on an empirical stellar library which, like any empirical stellar library, has limited coverage of the HR diagram. Here we have given rough limits to which extent we believe the XSL SSP models to be reliable to use. We note that \citet{Coelho2020} described extensively the effect of the coverage of the stellar library on the SSP model predictions. They found that predicted colours are more affected by the coverage effect than the choice of a synthetic versus empirical library. Derived galaxy ages can be underestimated when stellar population synthesis models with limited parameter coverage are used. On the other hand, metallicities are robust against limited HR diagram coverage but are underestimated when using synthetic libraries. 

\section{Conclusions}
We present the XSL simple stellar population models, which are based on 639 stellar spectra from  XSL DR3. These simple stellar population models have various improvements compared to other models available. XSL SSP models cover a wide wavelength range, from the NUV (350\,nm) to the NIR (2480\.nm); have moderate-high resolution throughout the wavelength range, with original $\sigma = 13$/$11$/$16\,\mathrm{km\,s}^{-1}$ in the UVB/VIS/NIR arms of X-shooter; are constructed from stars for which the spectra have been observed simultaneously at all wavelengths, and extend over a metallicity range of $-2.2 < \mathrm{[Fe/H]}<+0.2$\,dex and an age range of $0.05 < t_\mathrm{SSP}<16$\,Gyr.

To construct these models we have used recent PARSEC/COLIBRI stellar evolutionary tracks, which include the \mbox{TP-AGB} phases that control the evolution of NIR colours. Particular care was taken to include the RGB, E-AGB and \mbox{TP-AGB} stars in the stellar population models, with the use of average spectra of static giants, variable \mbox{O-rich} \mbox{TP-AGB} stars, and \mbox{C-rich} \mbox{TP-AGB} stars. Instead of relying on stellar parameter estimation of these stars, we used established \mbox{colour--temperature(--metallicity)} relations of \citet{Worthey2011} (for \mbox{O-rich} cool giants) and \citet{Bergeat2001} (\mbox{C-rich} cool giants). We also provide XSL SSP models constructed with older Padova00 stellar evolutionary tracks, but we discourage the use of these models outside the narrow safe zone defined in Sect. \ref{sect:safety}. 

We have gone through an extensive characterization of the stellar population models. On one hand, we have compared colours and absorption line indexes with existing stellar population models (E-MILES, C18, M09). On the other hand, we have compared our model predictions with colours of \citet{Eisenhardt2007} ETGs from Coma cluster and spectral features of SBG and ETG from \citet{Riffel2019} and find encouraging agreement with the observations. The XSL SSP models can reproduce the optical colours of ETGs in Coma cluster, comparable to the success of the E-MILES and C18 models. Differences between models are largest at NIR supersolar metallicities of old populations, which can be due to the inclusion of cool giant stars using separate giant sequences and the colour–temperature relation. The NIR colours are constrained by the reddest `static' and `variable' giant template. Offsets in $(I-J)$ might also come from the inaccurate merging of the VIS and NIR arms of spectra of cool giants. While the behaviour of optical absorption-line indices is similar between E-MILES and C18, there are discrepancies between models for NIR indices. The XSL models improve the range of predicted values for many NIR indices, such as MgI1.7. Careful separation of XSL RGB, E-AGB and \mbox{TP-AGB} stars, and including them into the XSL SSP models will allow us to analyse NIR indices more systematically in the future.

The extended wavelength coverage, and high resolution of the new XSL-based stellar population models will help us to bridge the optical and the near-IR studies of intermediate and old stellar populations and clarify the role of evolved cool stars in stellar population synthesis. 

\begin{acknowledgements}
AA acknowledges funding from the European Research Council (ERC) under the European Unions Horizon 2020 research and innovation programme (grant agreement No. 834148).
RFP acknowledges financial support from the European Union’s Horizon 2020 research and innovation programme under the Marie Sk\l{}odowska-Curie grant agreement No. 721463 to the SUNDIAL ITN network. A.V. and J.F-B acknowledge support through the RAVET project by the grant PID2019-107427GB-C32 from the Spanish Ministry of Science, Innovation and Universities (MCIU), and through the IAC project TRACES which is partially supported through the state budget and the regional budget of the Consejer\'ia de Econom\'ia, Industria, Comercio y Conocimiento of the Canary Islands Autonomous Community. PC acknowledges support from Conselho Nacional de Desenvolvimento Cient\'ifico e Tecnol\'ogico (CNPq) under grant 310041/2018-0 and from Funda\c{c}\~{a}o de Amparo \`{a} Pesquisa do Estado de S\~{a}o Paulo (FAPESP) process number 2018/05392-8. PSB acknowledges the financial support from the Spanish National Plan for Scientific and Technical Research and Innovation, through the grant PID2019-107427GB-C31. L.M. thanks FAPESP (grant 2018/26381-4) and CNPQ (grant 306359/2018-9) for partial funding of this research. 
\end{acknowledgements}
%-------------------------------------------------------------------

\bibliographystyle{aa} % style aa.bst
\typeout{} 
\bibliography{refer} % your references Yourfile.bib
\clearpage

\begin{appendix} %First online appendix

\section{The weighted median residuals around four spectral line indices.}

\label{app:fourlines}
\begin{figure}[!h]
        \centering
        \includegraphics[width =2 \linewidth]{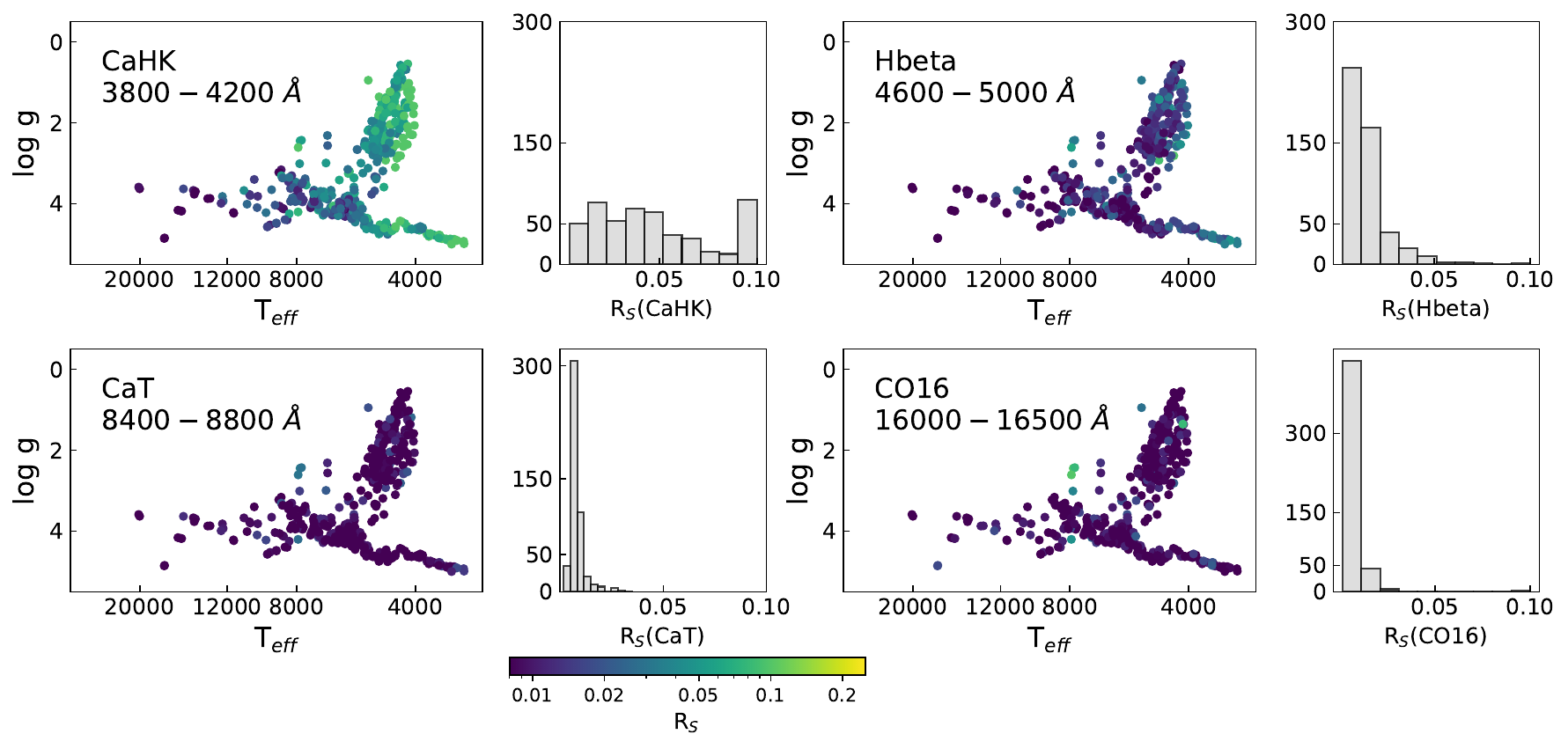}
    \caption{The weighted median residual $R_\mathrm{S}$ between the original spectrum and the interpolated spectrum for four spectral line indices (CaHK, Hbeta, CaT and CO1.6), as a function of position in the HR diagram. We have normalised the spectra over the wavelengths of interest before calculating the line-level median residuals. The colour-bar is logarithmic. Histograms show the
distributions of $R_\mathrm{S}$ calculated within these spectral ranges at the full XSL resolution. For ease of visualisation, spectra with $R_\mathrm{S} > 0.1$ are placed into the $R_\mathrm{S} = 0.1$ bin in the histograms.}
     \label{fig:fourlines_HRD}
\end{figure}
\FloatBarrier

\clearpage
\onecolumn
\section{Static giants}
\begin{table*}[!h]
    \caption[]{Selected static giants and removed supergiants}
    \centering
    \resizebox{0.7\linewidth}{!}{
    \begin{tabular}{|l|l|c|c|c|c|l|}
        \hline
         XSL ID & name & $(I-K)$ & H$^-$ / H$_2$O & H2O & CN & comment\\
         \hline
         \multicolumn{7}{|c|}{Bin 1}\\
         \hline
         X0402 &	CL* NGC 6522 ARP 1073 & 2.01 & -0.12 & 0.04 & 2.99 & \\
         X0804	&	CL* NGC 1978 LE 09	&	2.01 & -0.15 &	0.05 & 3.02 & \\
         X0805	&	CL* NGC 1978 LE 09	& 2.01 &-0.15 & 0.08 &	2.25 & \\
         X0806	&	CL* NGC 1978 LE 09	& 2.02 & -0.15 & 0.08 &	1.88 & \\
         X0364 & CL* NGC 6121 LEE 4611 & 2.04 & -0.16 &	0.06 &	0.33 & \\
         X0844 & BD-16 1934 & 2.05 &-0.13 &	0.06 & 4.67 & \\
         \hline
         \multicolumn{7}{|c|}{Bin 2}\\
         \hline
         X0852 & BD-16 1934 & 2.07 &-0.12 & 0.06 & 4.62 & \\
         X0328 & HD 79349 & 2.07 & -0.14 &	0.07 &	-0.19 & \\
         X0323 & HD 79349 &	2.07 & -0.11 & 0.06 & 1.12 & \\
         X0908 & CL* NGC 6121 LEE 4611	& 2.08 & -0.15 & 0.07 & 2.59 & \\
         X0756 & CL* NGC 288 OCH 531 & 2.08 & -0.19 & 0.11 & 0.22 &  \\
         X0909 & CL* NGC 6121 LEE 4613 & 2.09 & -0.16 & 0.07 & 2.46 & \\
         \hline
         \multicolumn{7}{|c|}{Bin 3}\\
         \hline
         X0869 & HD 69701 &  2.25 & -0.13 &	0.07 &	3.75 & \\
         X0845 & HD 69701 & 2.26 & -0.123 & 0.07 &	4.34 & \\
         X0495 & HD 212516 & 2.26 & -0.16 & 0.07 &  3.14 & \\
         X0767 & CL* NGC 288 OCH 531 & 	2.29 & -0.20 & 	0.11 &  0.20 &  \\ 
         X0801 & SHV 0529355-694037	&  2.29 & -0.14 & 0.05 & 2.98 & \\ 
         \hline
         \multicolumn{7}{|c|}{Bin 4}\\
         \hline
         X0551 & SHV 0525012-694829 & 2.33 & -0.19 & 0.11 & 0.84 & \\
         X0798 & SHV 0527122-695006	& 2.34 & -0.15 & 0.07 &	4.91 &   \\
         X0811 & SHV 0531398-701050	& 2.40 & -0.16 & 0.07 & 2.32 &  \\
         X0799 & SHV 0529355-694037	& 2.40 & -0.15 & 0.08 & 1.39 & \\
         \hline
         \multicolumn{7}{|c|}{Bin 5}\\
         \hline
         X0509 & SHV 0520036-692817	& 2.43 & -0.24 & 0.16 & 0.22 & \\
         X0800 & SHV 0529355-694037	& 2.44 & -0.15 & 0.076 & 1.95 & \\
         X0030 & ISO-MCMS J005314.8-730601 & 2.52 & -0.18 &	0.12 & 4.44 &\\
         \hline
         \multicolumn{7}{|c|}{Bin 6}\\
         \hline
         %X0335 & BS 4517 & 2.54 & -0.20 & 0.04 & -10.94 & looks different Hband \\
         X0783 & SHV 0448341-691510	& 2.54 & -0.18 &0.11 & -1.81 &  \\
         X0099 & SHV 0526364-693639 & 2.55 & -0.17 & 0.10 & -1.34 &\\
         X0587 & IRAS 10151-6008? & 2.56 & -0.11 & 0.09 & 4.02 & \\
         \hline
         \multicolumn{7}{|c|}{Bin 7}\\
         \hline
         X0785 & SHV 0448341-691510	& 2.66 & -0.17 & 0.14 &	-1.40 & \\
         X0819 & SHV 0535237-700720 & 2.67 & -0.17 & 0.09 & 2.59 & \\
         \hline
         \multicolumn{7}{|c|}{Bin 8}\\
         \hline
         X0807 & SHV 0530380-702618 & 2.72 & -0.24 & 0.15 & 2.13 & \\
         X0117 & SHV 0543367-695800	& 2.79 & -0.27 & 0.20 & -0.27 & \\
         X0531 & SHV 0518331-685102 & 2.86 & -0.23 & 0.11 &	4.06 & strong VO1.1 band \\ 
         X0517 & SHV 0515313-694303	& 2.89 & -0.21 & 0.13 &	4.29 & strong VO1.1 band \\ 

         \hline
         \multicolumn{7}{|c|}{Bin 9}\\
         \hline
         X0592 & SHV 0520342-693911 & 3.07 & -0.18 & 0.13 & -0.54 & \\
         X0172 & [B86] 133 & 3.14 &	-0.15 &	0.08 & 3.08 & \\
         X0254 & OGLEII DIA BUL-SC01 1821 &	3.31 & -0.19 & 0.10 & 0.65 & \\
         \hline
         \multicolumn{7}{|c|}{Bin 10}\\
         \hline
         X0153 & BMB 245 & 3.41 & -0.15 & 0.08 & 3.08 & \\
         X0815 & SHV 0533130-702409	 & 3.43 & -0.20 & 0.16 & 3.40 & strong VO1.1 band\\
         X0257 & BMB 13 & 3.47 & -0.17 & 0.10 &	0.77 & \\
         X0246 & OGLEII DIA BUL-SC03 1890 &	3.65 & -0.25 & 0.18 & 0.06 &\\
         X0399 & V5475 Sgr & 3.75 & -0.12 & 0.12 & 0.58 &\\
         \hline
         \multicolumn{7}{|c|}{Removed supergiants}\\
         \hline
         X0850  & IRAS 06404+0311  & 2.00  & -0.13  &	0.07  & 13.17  & \\
         X0265	& [M2002] LMC 162635 &  2.02  & -0.12  & 0.05  & 11.98 & \\
         X0849  & IRAS 06404+0311  & 2.03  & -0.13  & 0.07  & 12.58  & \\
         X0411 & CL* NGC 121 T V1 	& 2.18	& -0.21	& 0.15 & 7.79 &  \\
         X0786 & SV* HV 2555 & 2.36 & -0.21 & 0.17 & 12.79 &  \\
         X0005 & [M2002] SMC 46662 & 2.42 & -0.18 &	0.14 & 7.12 & \\
         X0021 & [M2002] SMC 83593 & 2.48 & -0.24 & 0.17 & 6.86 & \\
         X0118 & [M2002] LMC 143035 & 2.47 & -0.22 & 0.13 & 7.90 & \\
         X0120 & [M2002] LMC 150040	& 2.66 & -0.18 & 0.14 &	2.87 & \\
         X0266 & [M2002] LMC 168757	& 2.69 & -0.21 & 0.14 & 7.01 &\\
         X0420 & SV* HV 11223 &	2.89 & -0.18 & 0.12 & 8.60 & \\
         X0205 & SV* HV  2255 & 3.03 & -0.17 & 0.10 & 10.10 & \\
         X0260 & SV* HV  2255 &	3.06 & -0.17 & 0.11 & 10.69 & \\
         \hline
         
    \end{tabular}
    }
    \label{tab:static_giants}
\end{table*}

\onecolumn
\begin{figure}
     \centering
     \begin{subfigure}
         
         \includegraphics[width=0.33\linewidth]{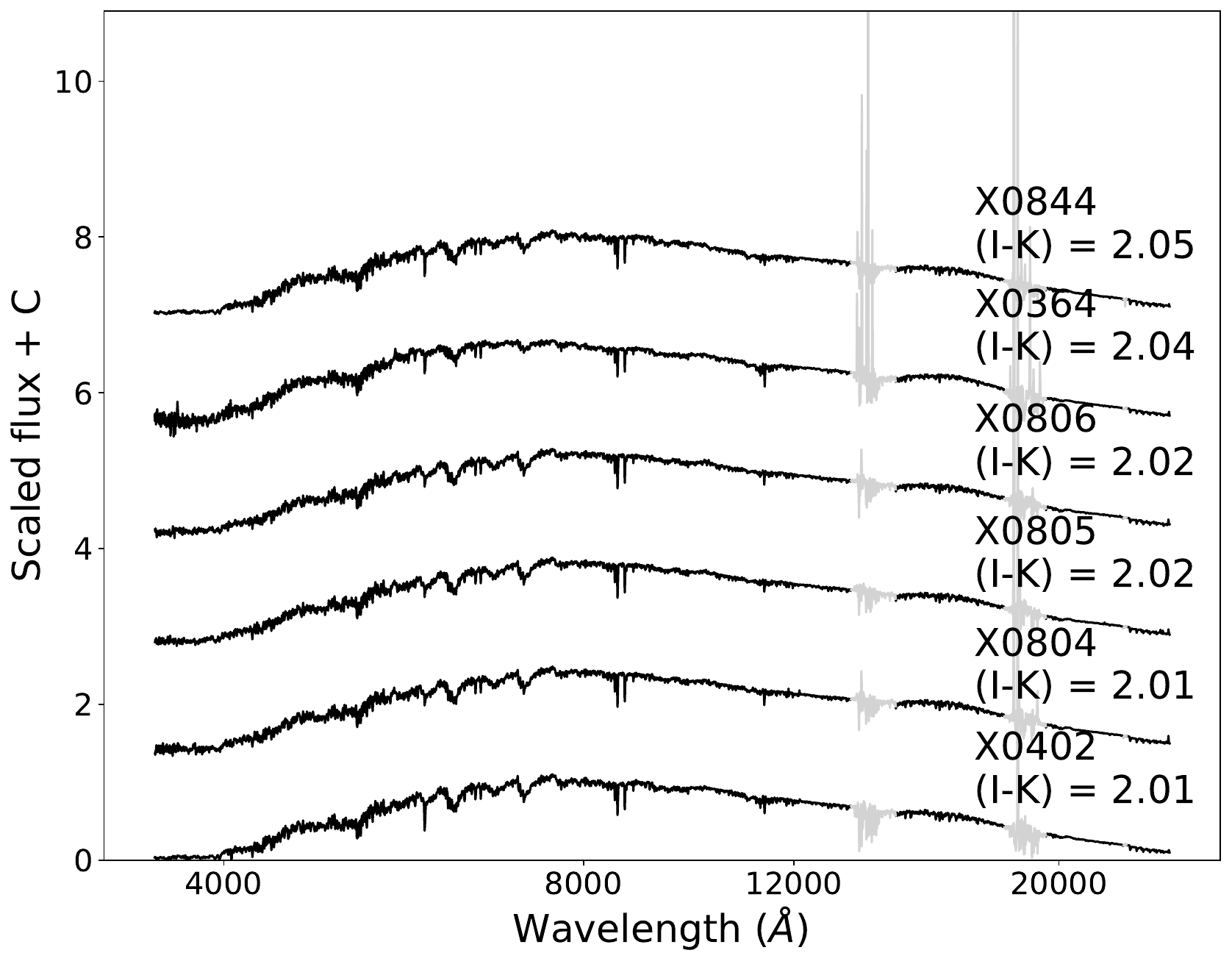}
         %\caption{Bin 1}
     \end{subfigure}
     \begin{subfigure}
         
         \includegraphics[width=0.33\linewidth]{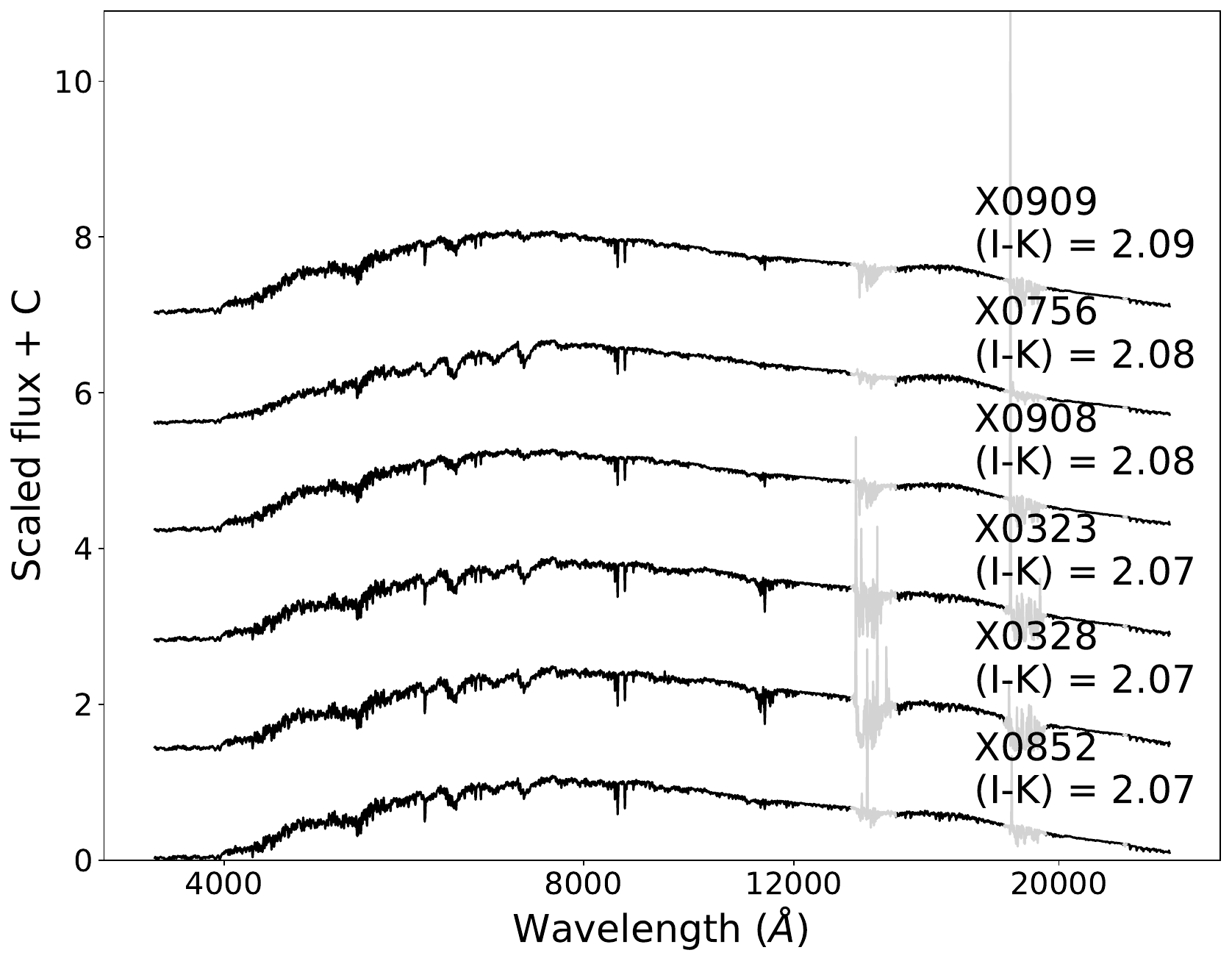}
     \end{subfigure}
     \begin{subfigure}
         
         \includegraphics[width=0.33\linewidth]{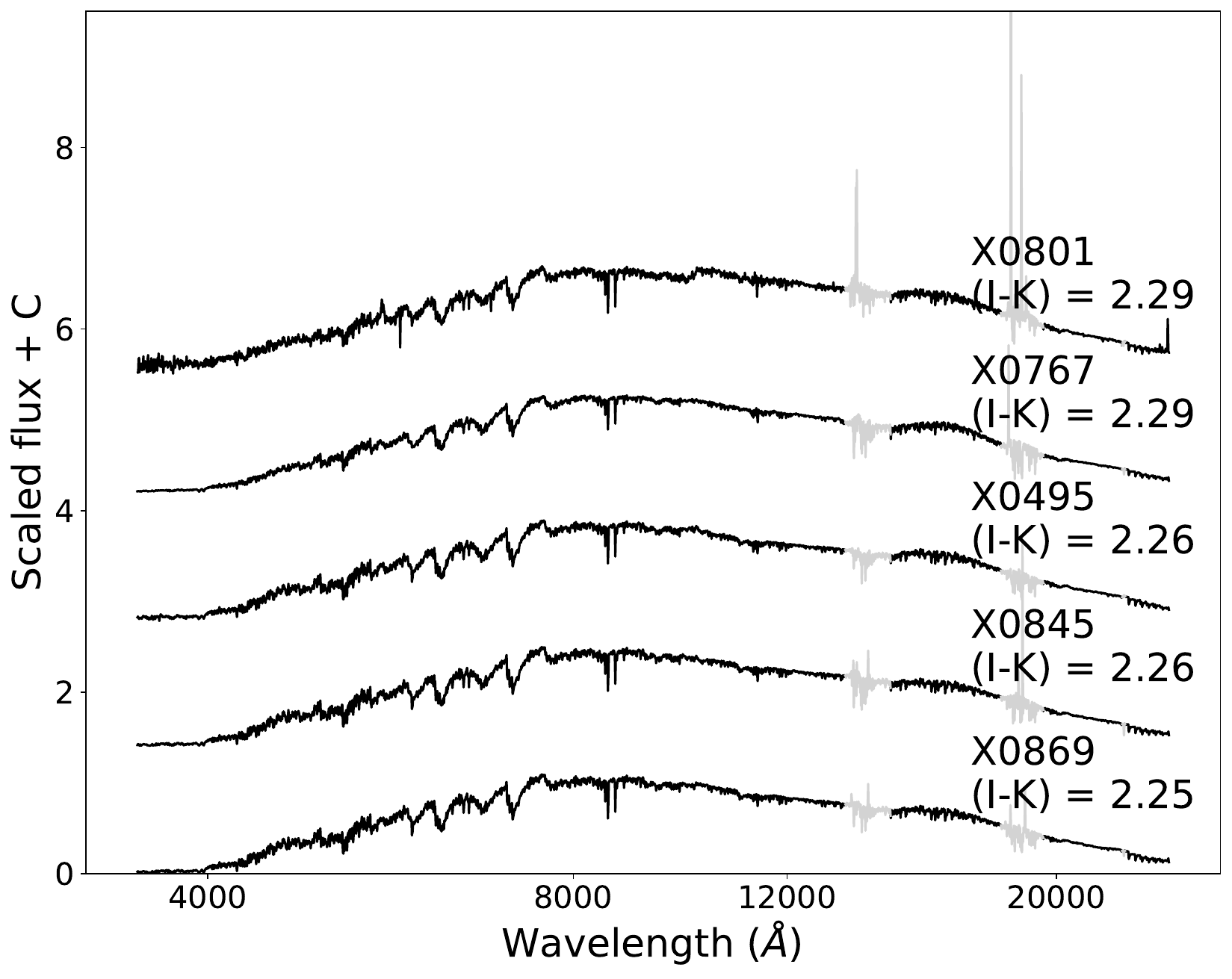}
     \end{subfigure}
     \begin{subfigure}
         
         \includegraphics[width=0.33\linewidth]{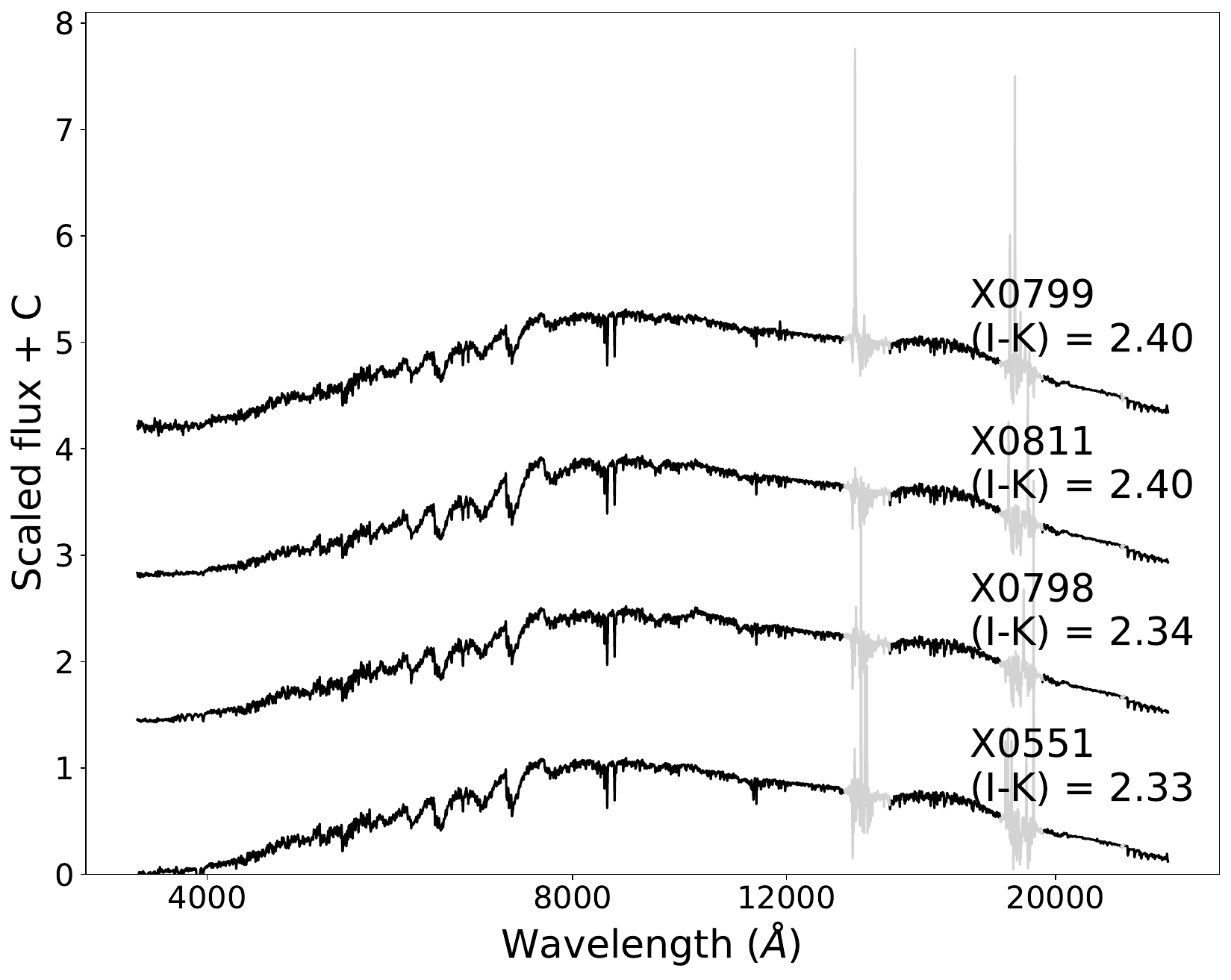}
     \end{subfigure}
     \begin{subfigure}
         
         \includegraphics[width=0.33\linewidth]{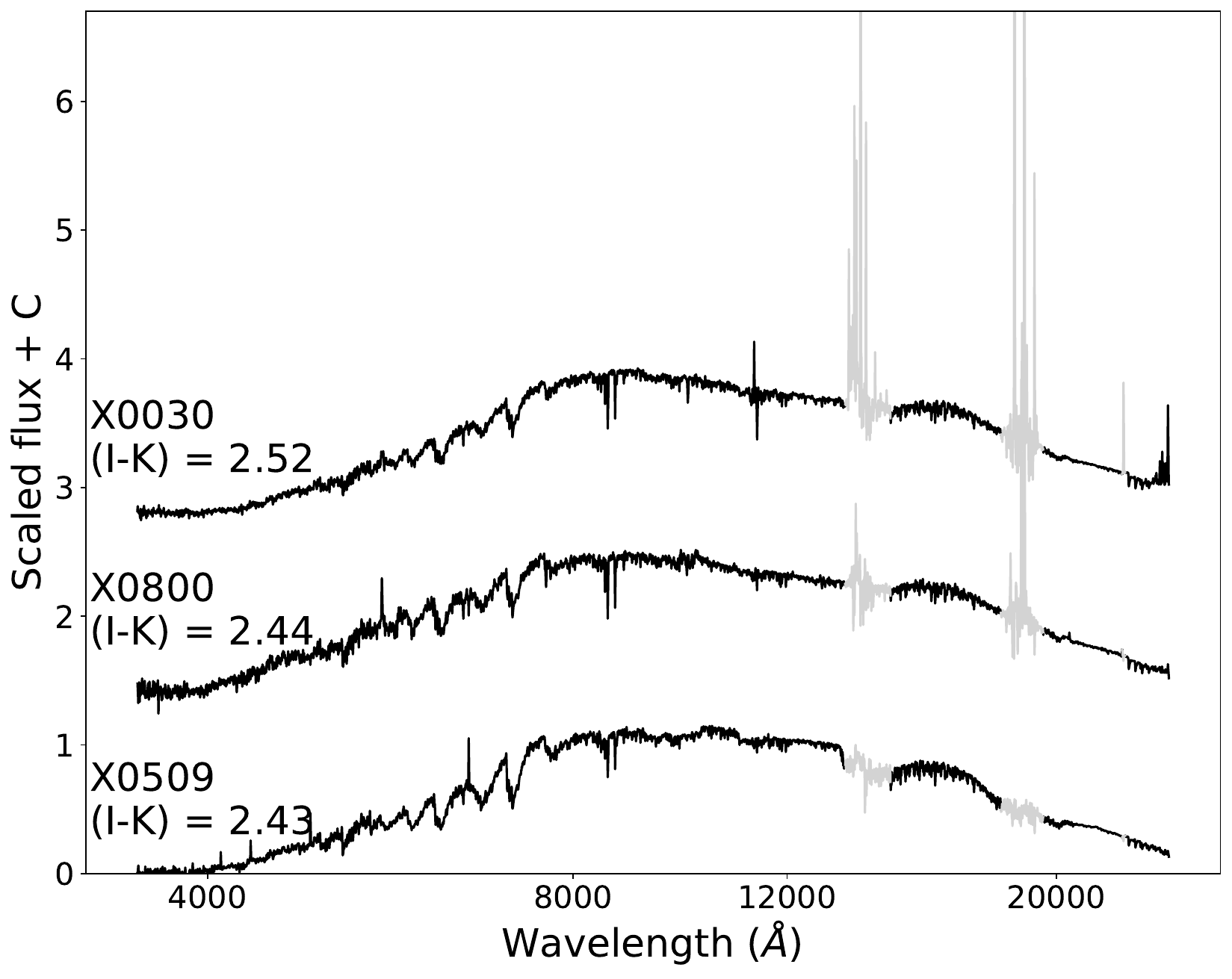}
     \end{subfigure}
     \begin{subfigure}
         
         \includegraphics[width=0.33\linewidth]{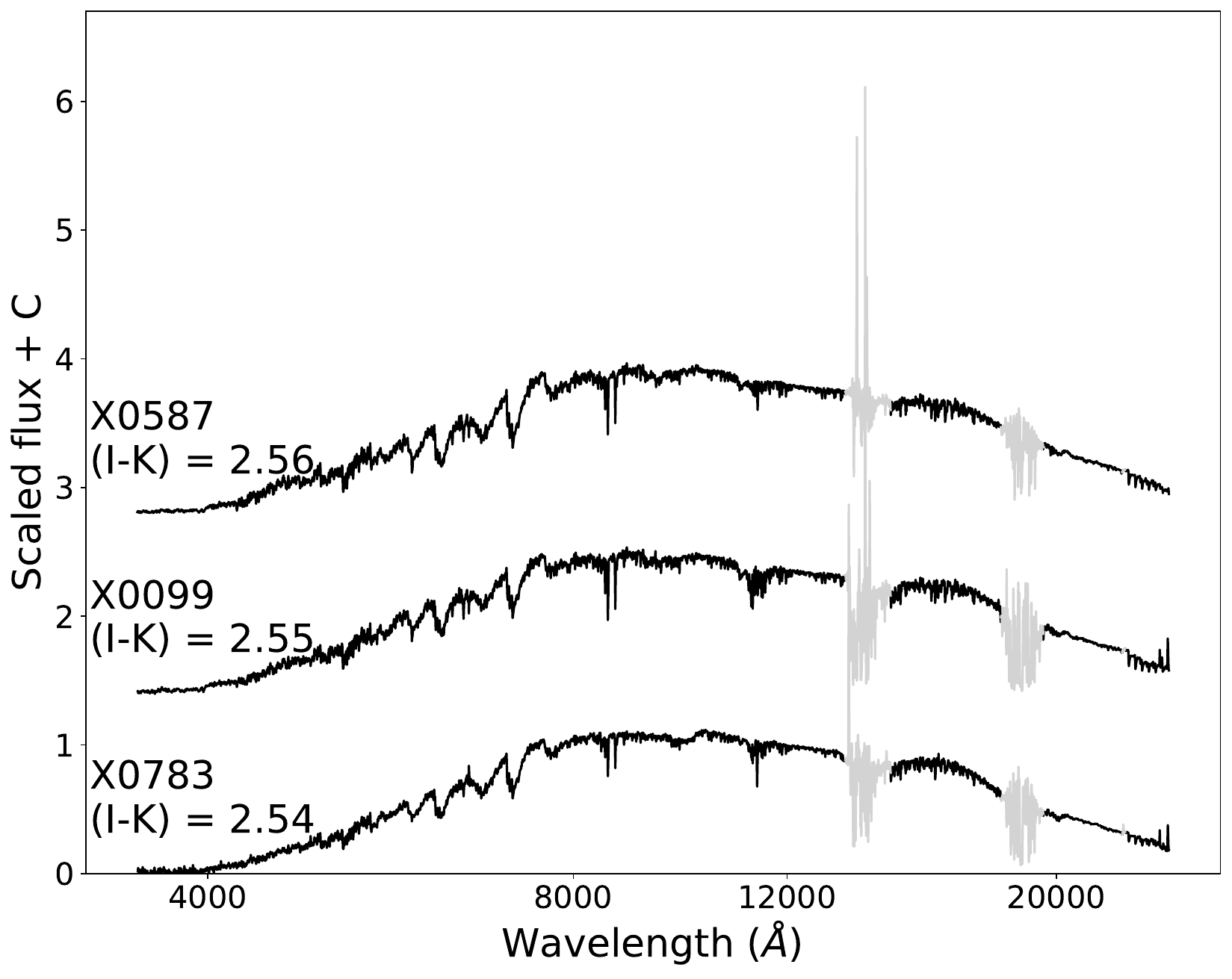}
     \end{subfigure}
     \begin{subfigure}
         
         \includegraphics[width=0.33\linewidth]{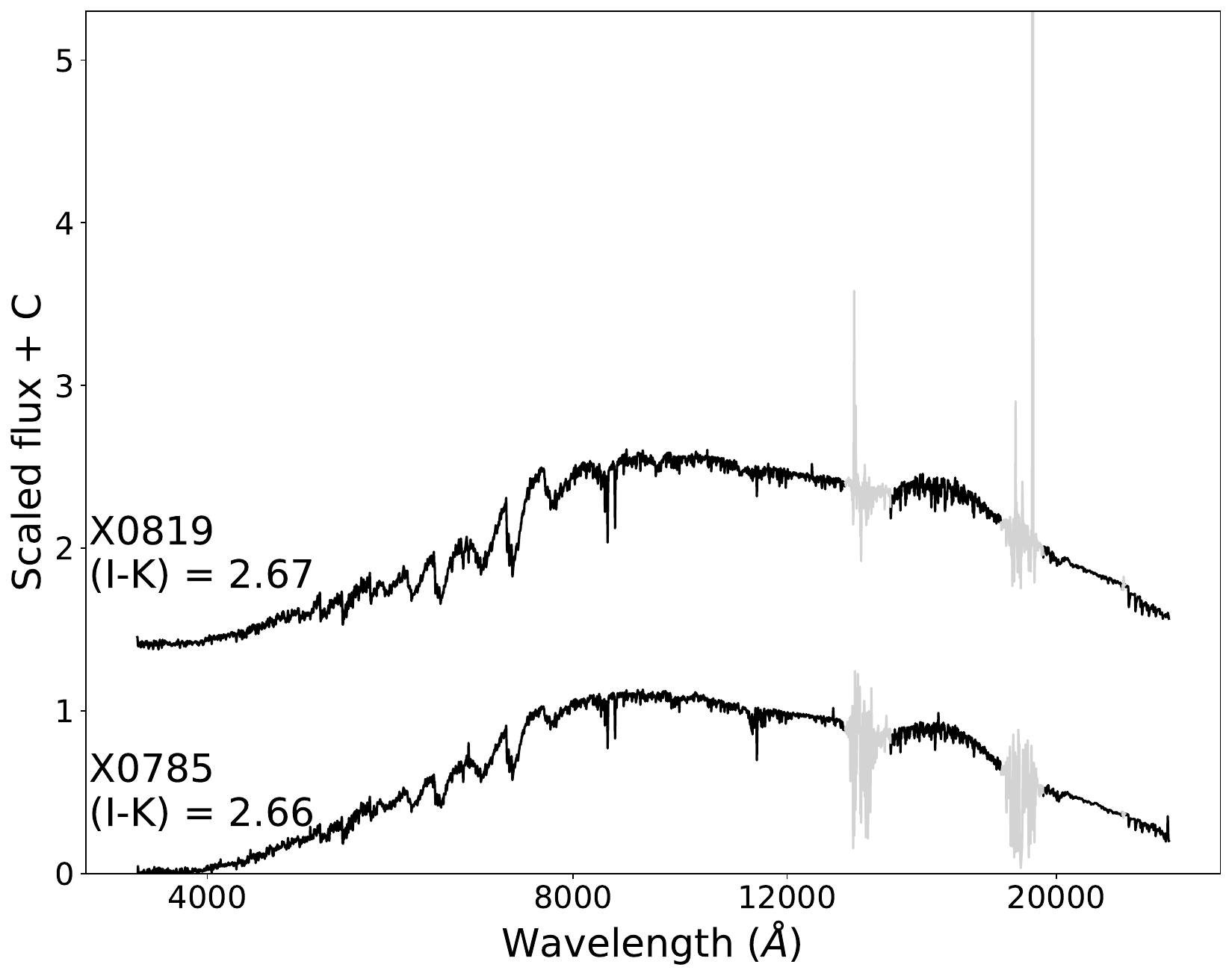}
     \end{subfigure}
     \begin{subfigure}
         
         \includegraphics[width=0.33\linewidth]{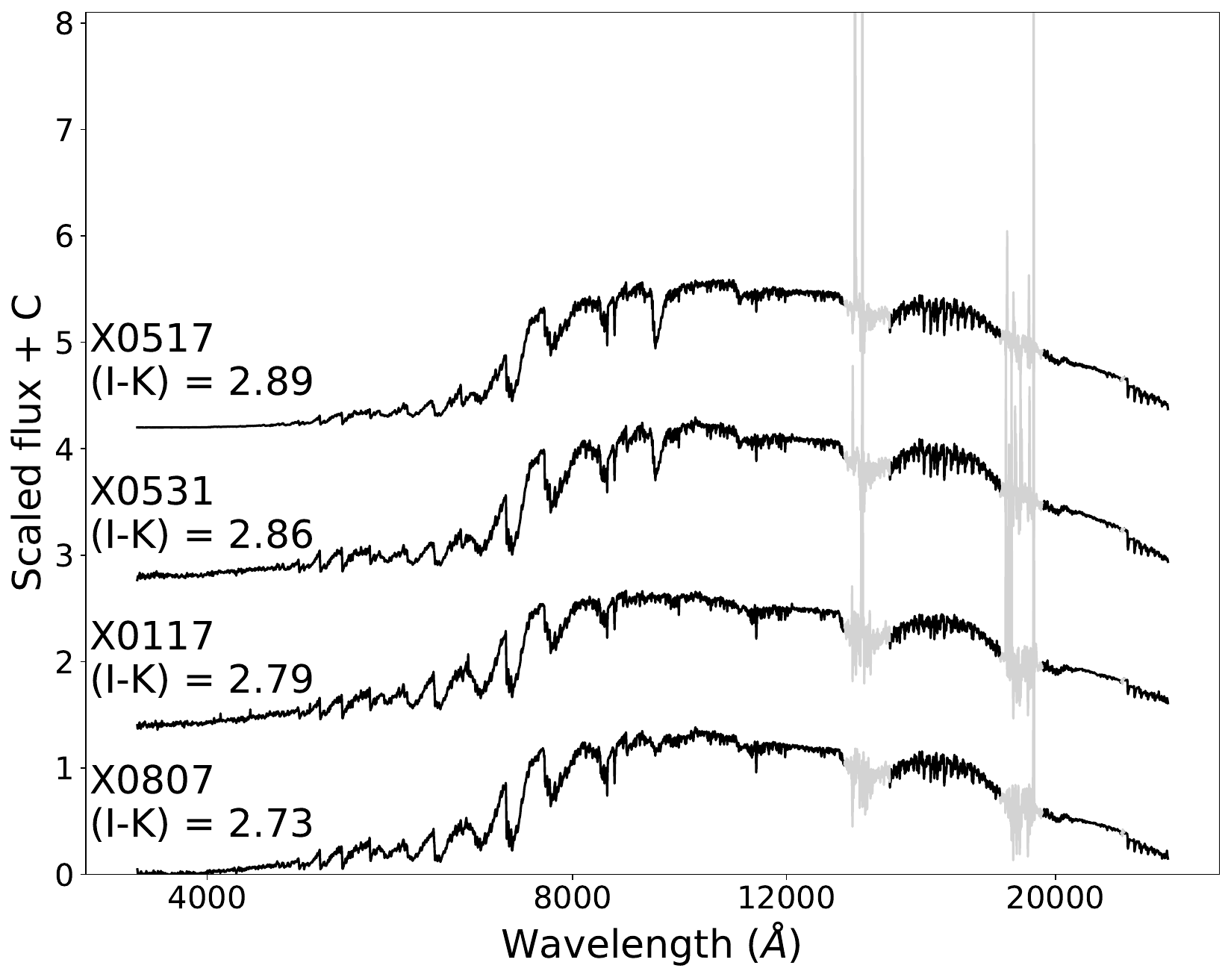}
     \end{subfigure}
     \begin{subfigure}
         
         \includegraphics[width=0.33\linewidth]{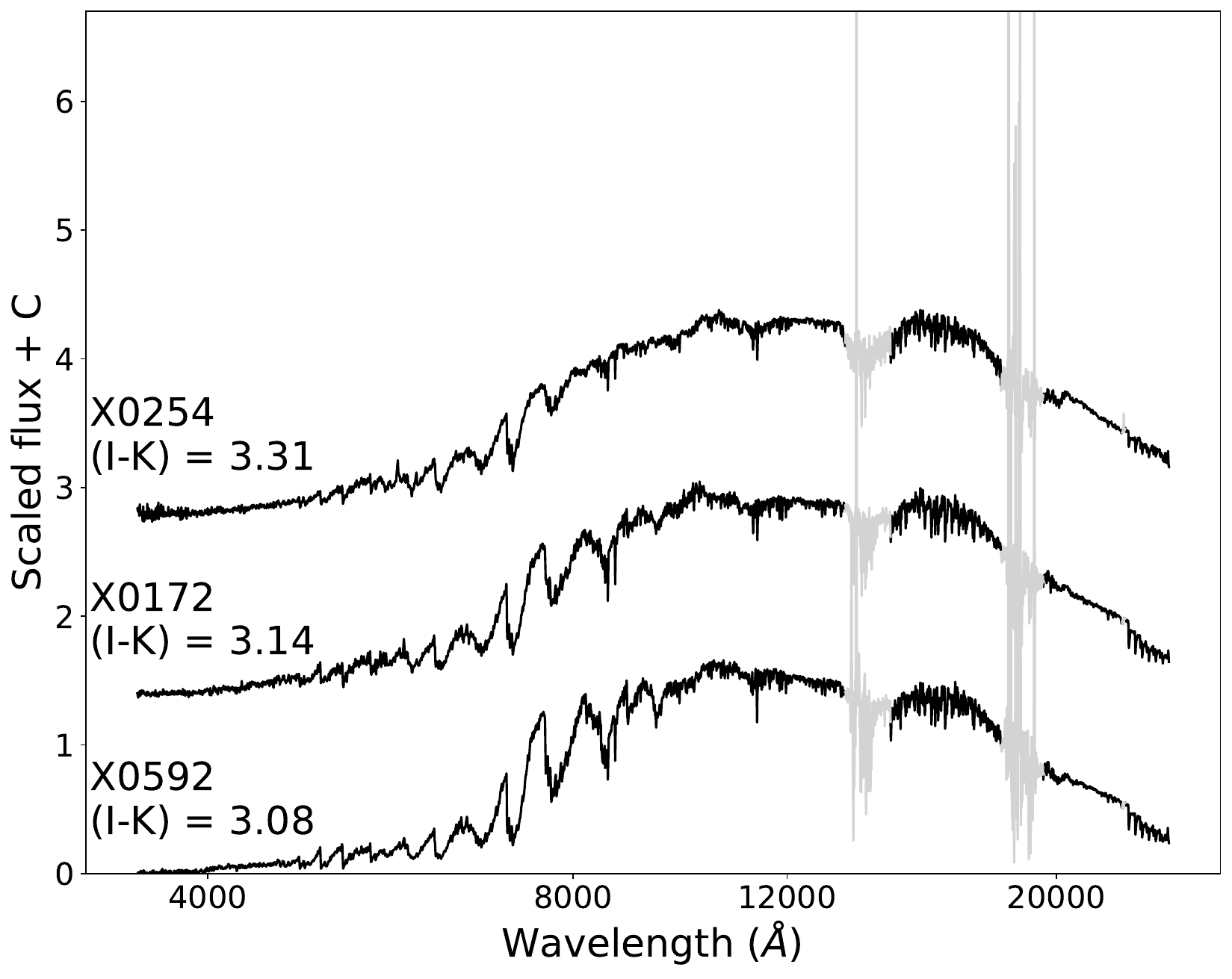}
     \end{subfigure}
     \begin{subfigure}
         
         \includegraphics[width=0.33\linewidth]{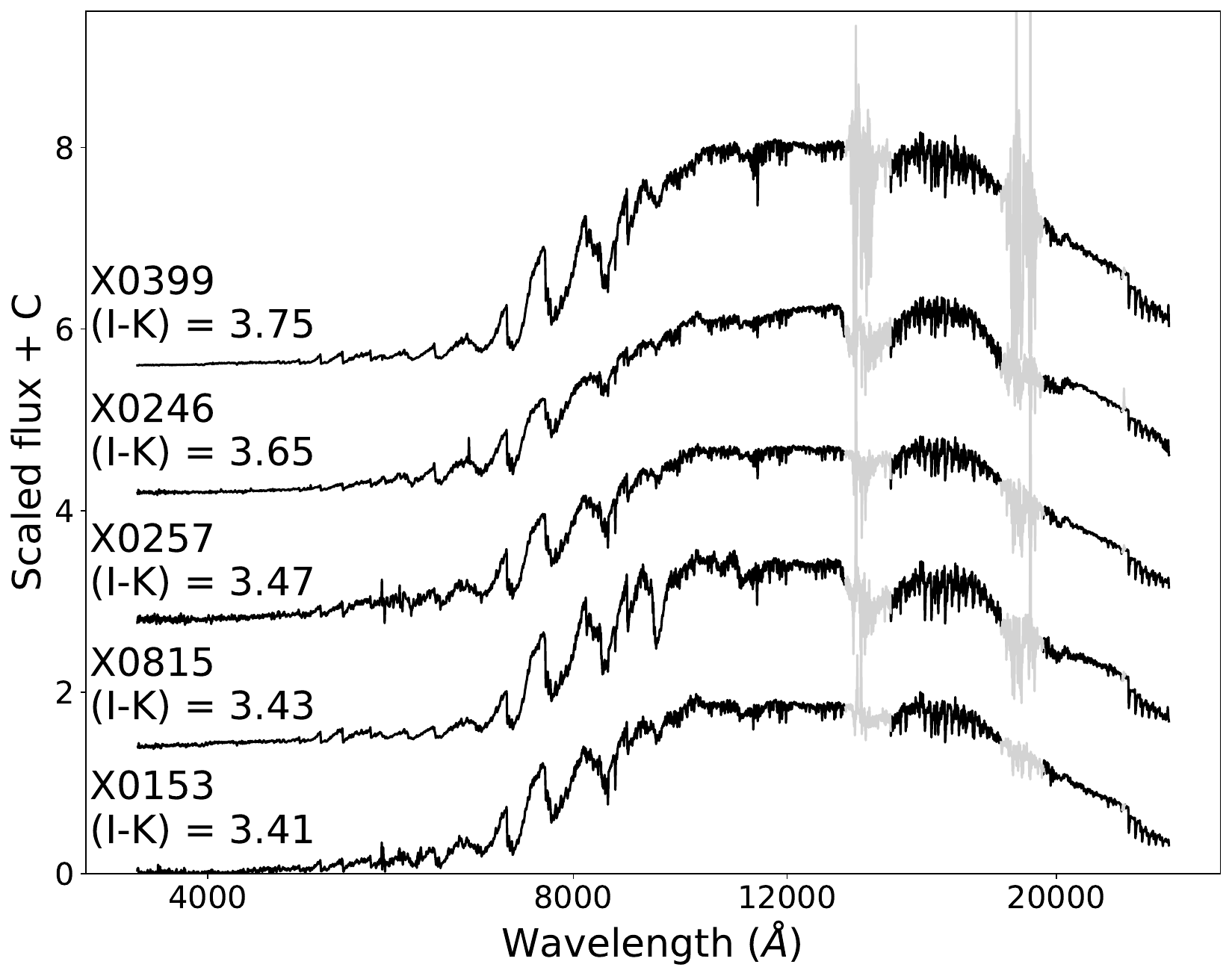}
     \end{subfigure}
    \caption{XSL spectra of \mbox{O-rich}, cool static giant stars from which the static sequence is constructed.}

    \label{fig:staticstars}
\end{figure}
\clearpage
%%%%%%%%%%%%%%%%%%%%%%%%%%%%%%%%%%%%%%%%%%%%%
\section{\mbox{O-rich} \mbox{TP-AGB} star average bins}

\begin{table*}[!h]
    \caption[]{Selected \mbox{O-rich} \mbox{TP-AGB} stars and supergiants removed from the library}
    \centering
    \small
    \begin{tabular}{|l|l|c|c|c|c|}
        \hline
         XSL ID & name & $(I-K)$ & H$^-$ / H$_2$O & H2O & CN \\
        \hline
         \multicolumn{6}{|c|}{Bin 1}\\
         \hline
         X0527 & U Psc & 2.11 & -0.42 & 0.34 & -1.86  \\
         X0642 & SY Pav & 2.15 & -0.29 & 0.19 &-3.68  \\
         X0644 & SY Pav & 2.20 & -0.29 & 0.19 &	-3.56  \\
         X0487 & V335 Aql &	2.22 & -0.35 & 0.24 & 0.31  \\
         \hline
         \multicolumn{6}{|c|}{Bin 2}\\
         \hline
         X0428 & RY CrA & 2.27 & -0.34 & 0.25 & -2.95   \\
         X0672 & BH Tel	& 2.16 & -0.42 & 0.33 &	-0.86  \\
         X0489 & XZ Her & 2.30 & -0.44 & 0.28 &	-3.80  \\
         X0690 & X Lib & 2.36 & -0.30 & 0.23 & 2.75 \\
         X0689 & X Lib & 2.42 &	-0.30 &	0.23 & 2.04   \\
         \hline
         \multicolumn{6}{|c|}{Bin 3}\\
         \hline
         X0905 & V Crv & 2.59 & -0.45 &	0.35 & 1.59 \\
         X0911 & SY Pav	& 2.60 & -0.56 & 0.41 & -5.43  \\
         X0910 & SY Pav & 2.70 & -0.56 & 0.43 & -5.04  \\
         X0638 & FR Her & 2.81 & -0.36 & 0.28 & 1.65  \\
         X0054 & SHV 0515461-691822 & 3.00 & -0.65 & 0.50 & -5.25  \\
         X0511 & SV* HV 1963 & 3.01 & -0.30 & 0.21 & 1.84  \\
         X0134 & U Crt & 3.00 & -0.39 & 0.30 & -1.81  \\
         
         \hline
         \multicolumn{6}{|c|}{Bin 4}\\
         \hline
         X0037 & SHV 0549503-704331 & 3.09 & -0.41 & 0.28 &	-3.67   \\
         X0888 & V354 Cen & 3.10 & -0.61 & 0.50 & -8.14  \\
         X0149 & CM Car	 & 3.13 & -0.54 & 0.38 & -4.57  \\
         X0237 & SHV 0510004-692755	& 3.10 & -0.42 & 0.34 & -3.00   \\
         \hline
         \multicolumn{6}{|c|}{Bin 5}\\
         \hline
         X0050 & HV 2360 & 3.19 & -0.50 & 0.42 & -4.13  \\
         X0557 & AL Mon	& 3.19 & -0.29 & 0.20 & -0.14   \\
         X0532 & SHV 0518571-690729 & 3.22 & -0.34 & 0.20 & -2.62   \\
         X0492 & DG Peg & 3.31 & -0.52 & 0.40 & -7.00  \\
         \hline
         \multicolumn{6}{|c|}{Bin 6}\\
         \hline
         X0675 & V874 Aql & 3.33 & -0.31 &	0.24 & 1.09  \\
         X0251 & OGLEII DIA BUL-SC04 9008 &	3.44 & -0.30 & 0.19 &-2.35   \\
         X0397 & RR Ara & 3.69 & -0.56 & 0.54 &	-5.43  \\
         X0647 & V348 Sco & 3.70 & -0.37 & 0.32 & -5.19  \\
         X0242 & IRAS 14303-1042 & 3.71 & -0.55 & 0.44 & -9.53 \\
         \hline
         \multicolumn{6}{|c|}{Bin 7}\\
         \hline 
         X0160 & OGLEII DIA BUL-SC03 3941 & 3.88 & -0.35 & 0.25 & 1.37  \\
         X0154 & BMB 286 & 	4.03 & -0.38 & 0.29 & -4.40  \\
         X0296 & OGLEII DIA BUL-SC13 0324 &	4.23 & -0.54 & 0.44 & -7.89  \\
         X0253 & OGLEII DIA BUL-SC22 1319 & 4.32 & -0.60 & 0.55 & -2.75 \\
         \hline
         \multicolumn{6}{|c|}{Bin 8}\\
         \hline
         X0020 & ISO-MCMS J005714.4-730121 & 4.88 & -0.67 & 0.66 & -19.87  \\

         \hline
         \multicolumn{6}{|c|}{Bin 9}\\
         \hline
         X0145 & OGLEII DIA BUL-SC41 3443 &  5.76 & -0.37 & 0.43 & -7.11  \\
         \hline
         \multicolumn{6}{|c|}{Removed supergiants}\\
         \hline
         X0761 & Y Sge & 2.98 & -0.33 & 0.26 & 7.72  \\
         X0119 & [M2002] LMC 148035	& 3.01 & -0.30 & 0.24 & 1.07  \\ 
         X0004 & [M2002] SMC 55188 & 3.11 &	-0.34 &	0.27 &	4.54  \\
         X0148 & [M2002] LMC 170452	& 3.51 & -0.37 & 0.25 &	6.58  \\

         \hline
         
    \end{tabular}
    \label{tab:variable_giants}
\end{table*}

\onecolumn
\begin{figure}
     \centering
     \begin{subfigure}
         
         \includegraphics[width=0.33\linewidth]{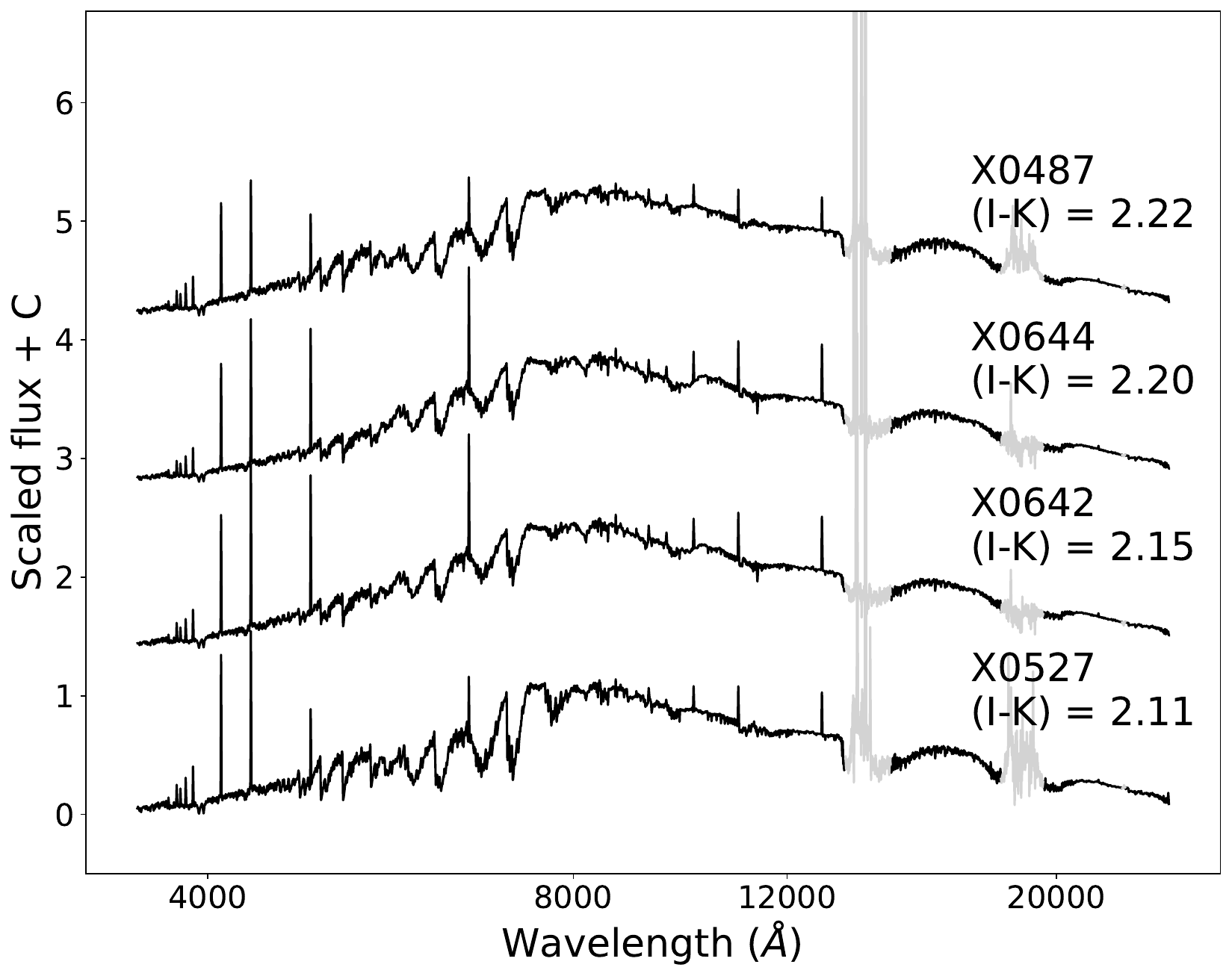}
         %\caption{Bin 1}
     \end{subfigure}
     \begin{subfigure}
     
         \includegraphics[width=0.33\linewidth]{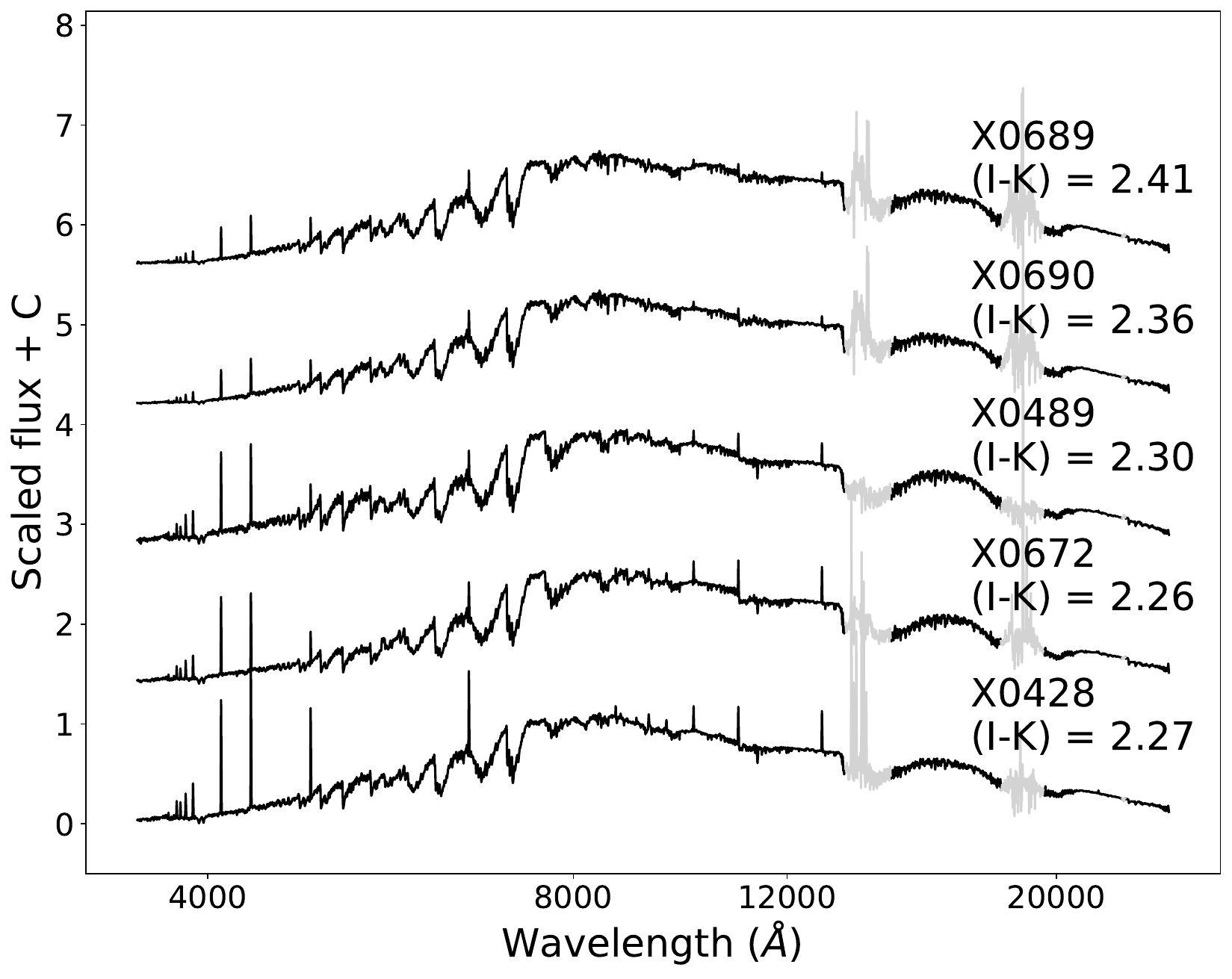}
     \end{subfigure}
     \begin{subfigure}
     
         \includegraphics[width=0.33\linewidth]{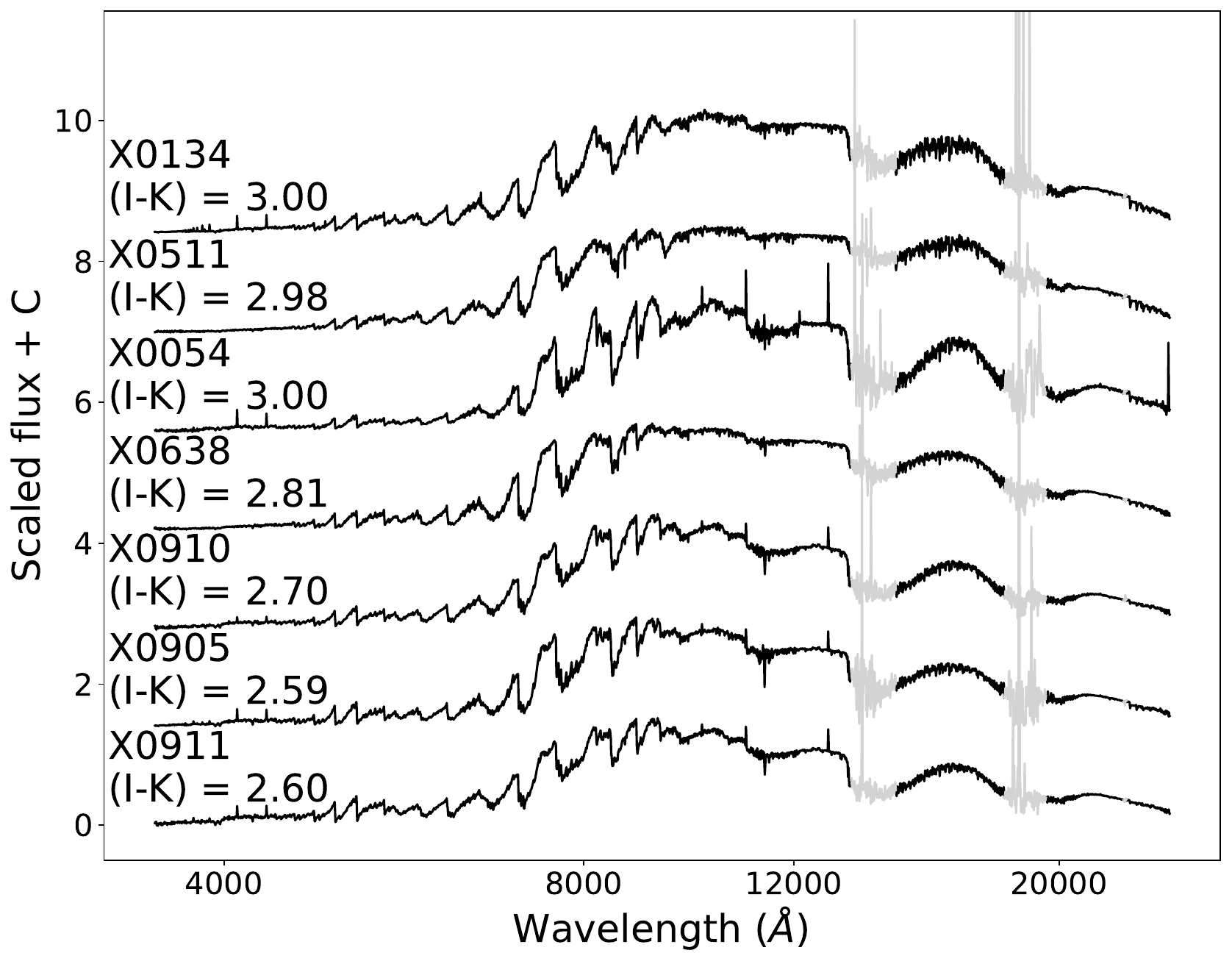}
     \end{subfigure}
     \begin{subfigure}
     
         \includegraphics[width=0.33\linewidth]{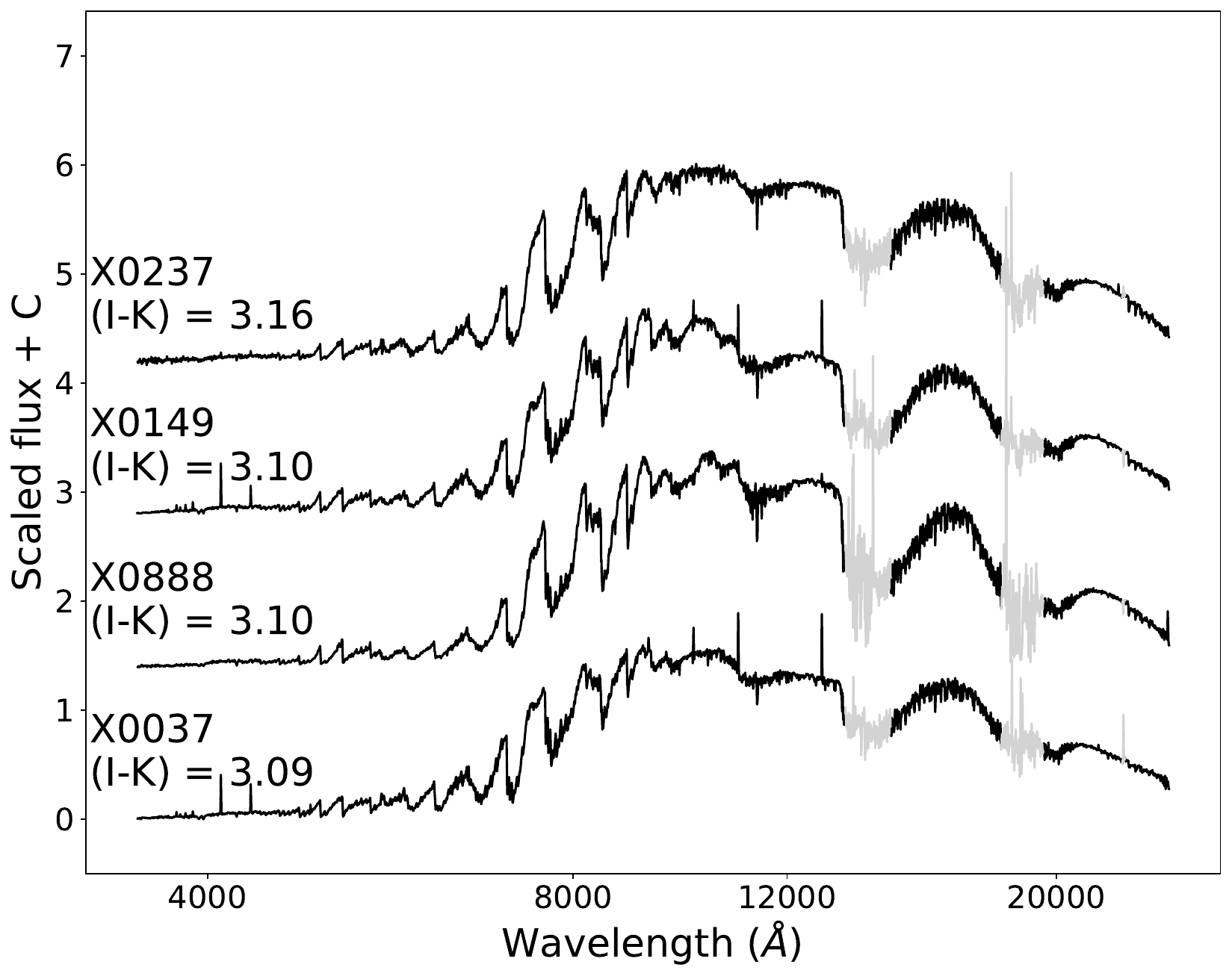}
     \end{subfigure}
     \begin{subfigure}
     
         \includegraphics[width=0.33\linewidth]{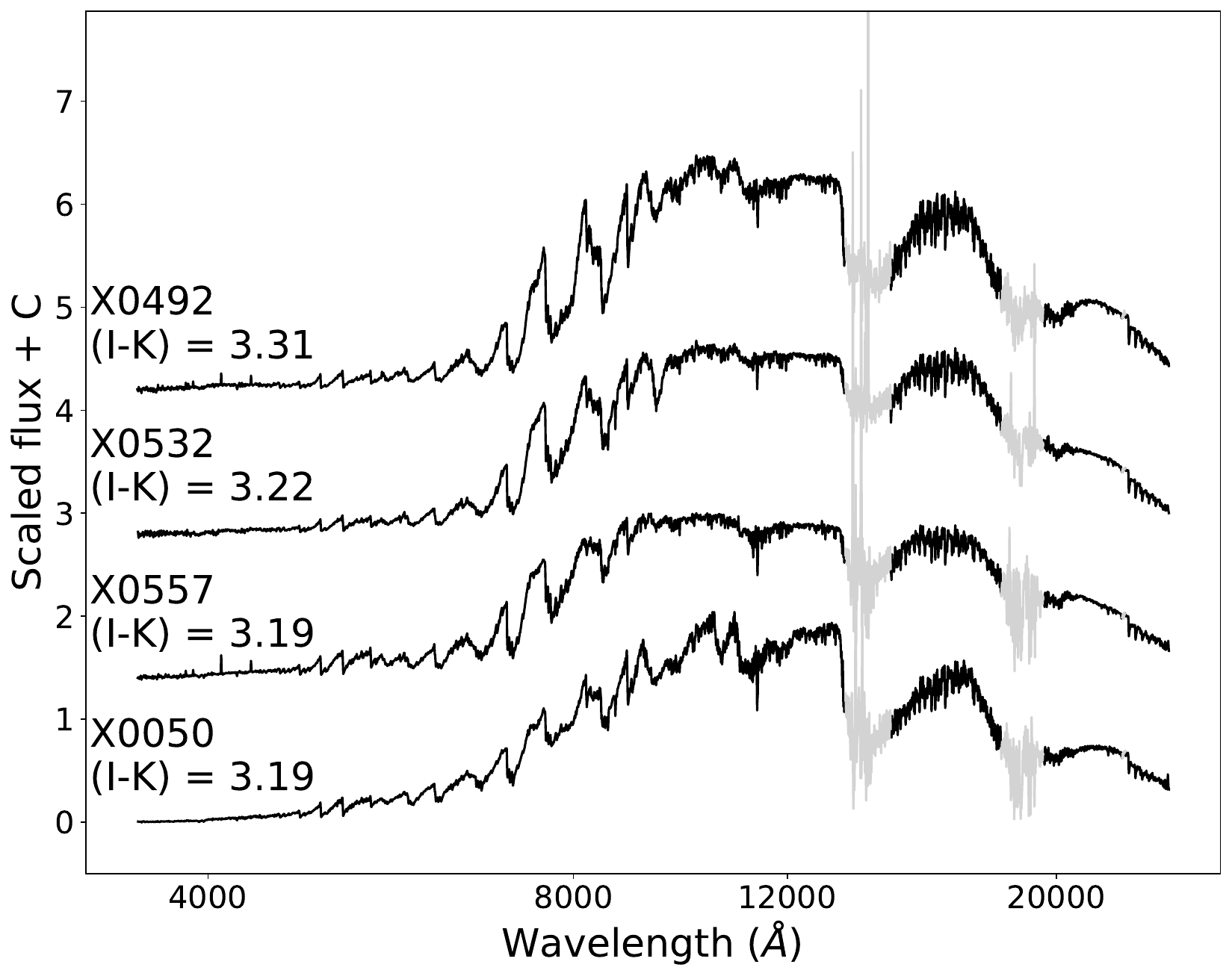}
     \end{subfigure}
     \begin{subfigure}
     
         \includegraphics[width=0.33\linewidth]{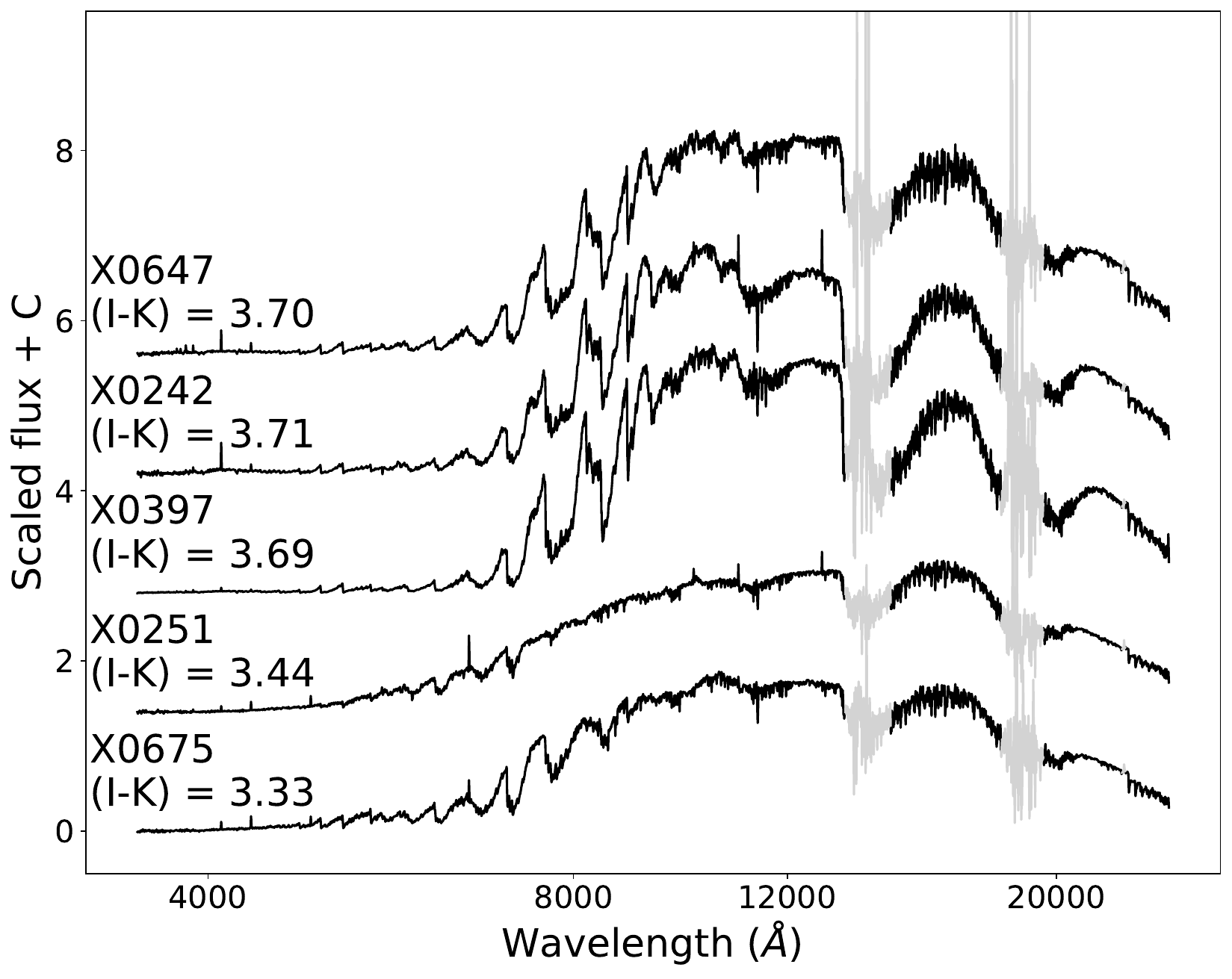}
     \end{subfigure}
     \begin{subfigure}
     
         \includegraphics[width=0.35\linewidth]{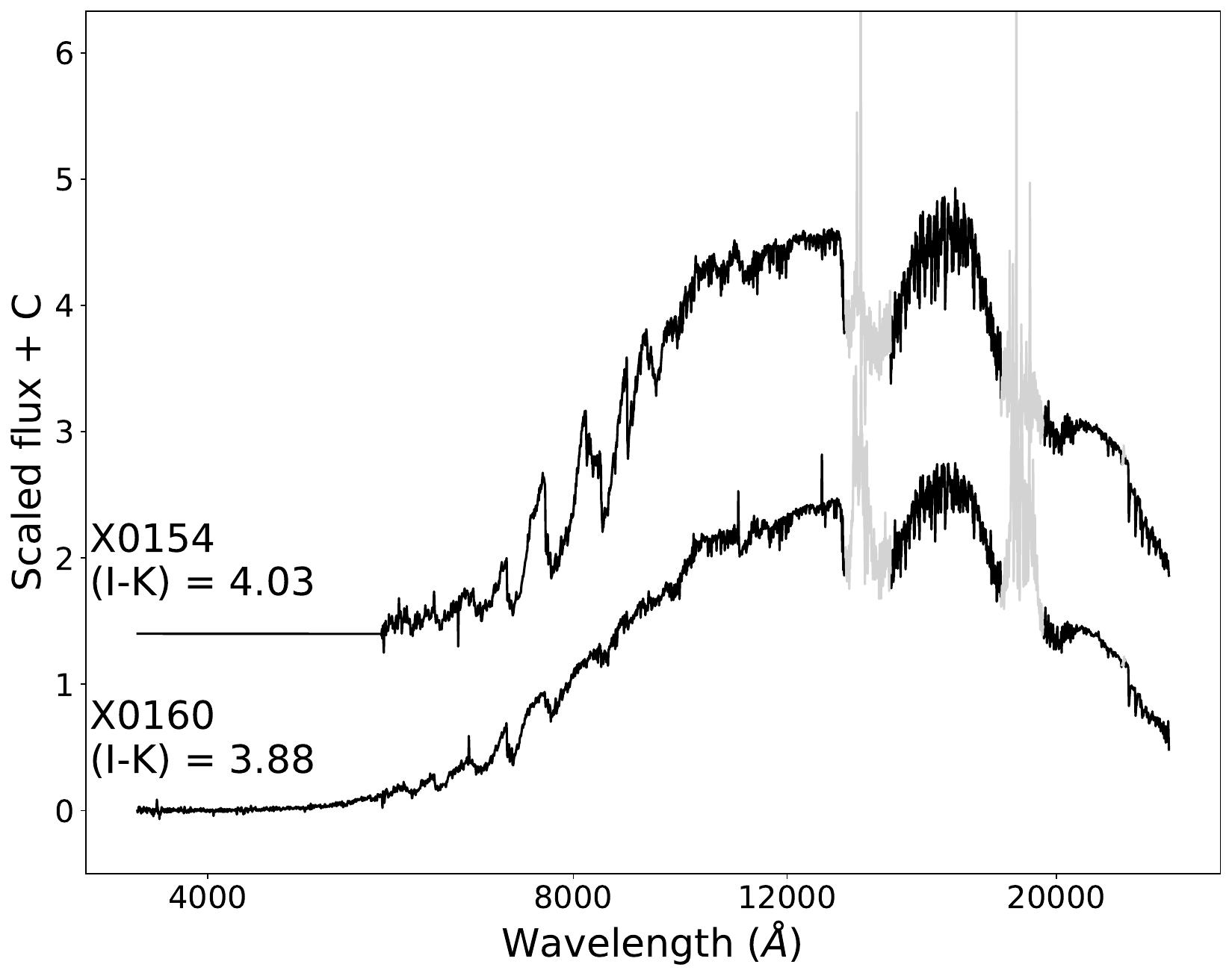}
     \end{subfigure}
     \begin{subfigure}
     
         \includegraphics[width=0.33\linewidth]{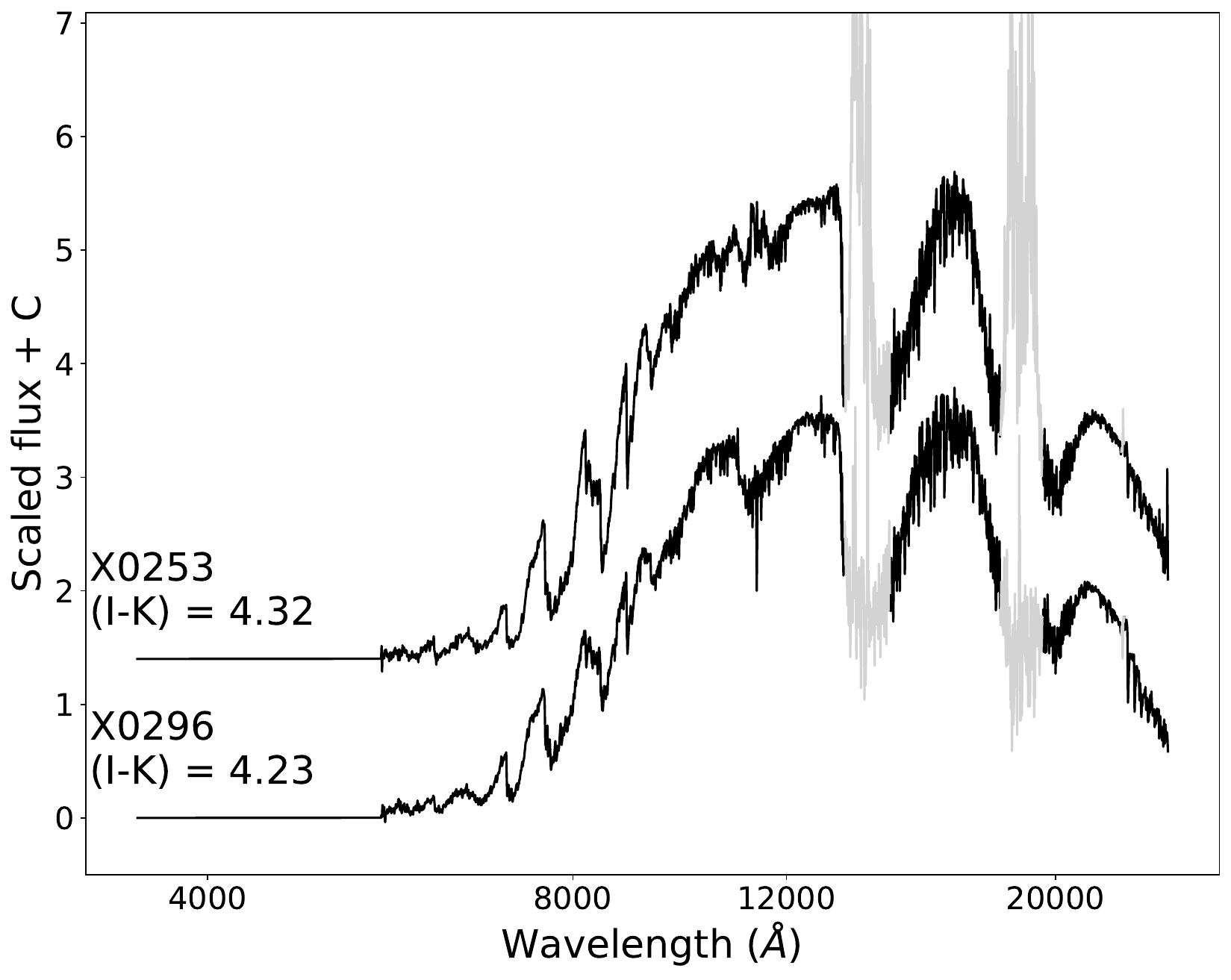}
     \end{subfigure}
     \begin{subfigure}
     
         \includegraphics[width=0.33\linewidth]{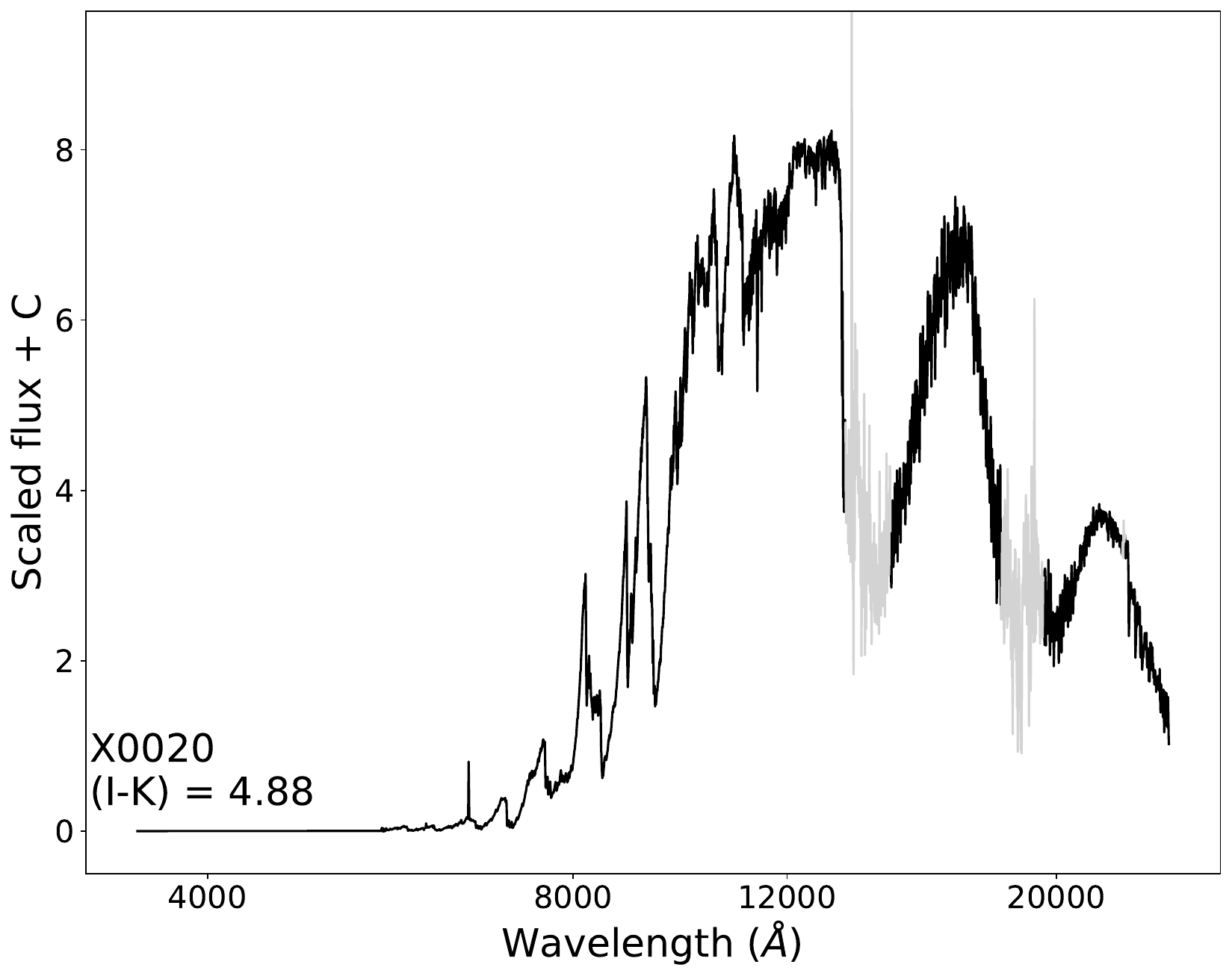}
     \end{subfigure}
     \begin{subfigure}
     
         \includegraphics[width=0.33\linewidth]{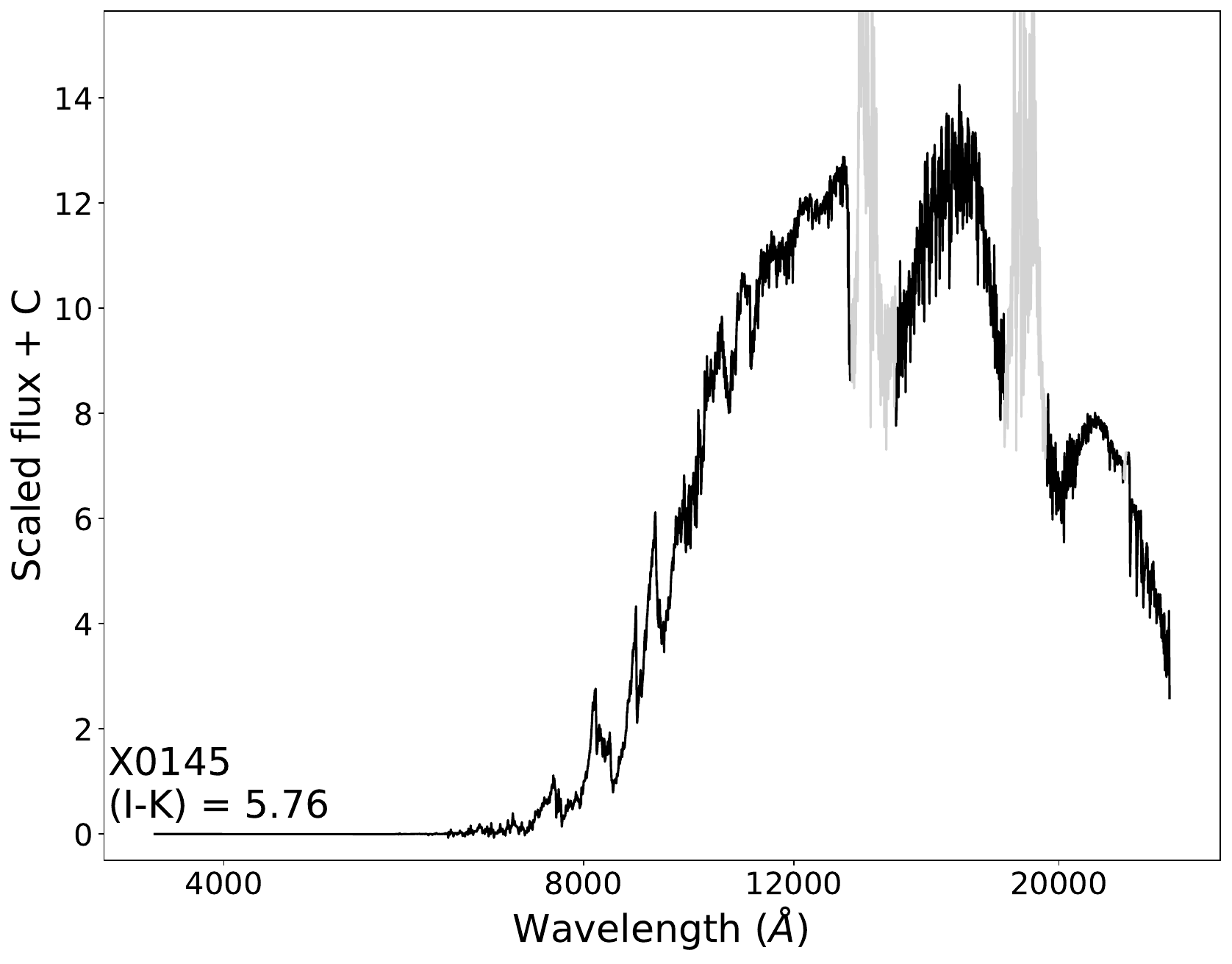}
     \end{subfigure}
     \caption{XSL spectra of \mbox{O-rich} \mbox{TP-AGB} stars from which the variable \mbox{O-rich} \mbox{TP-AGB} sequence is constructed.}
    \label{fig:AGBstars}
\end{figure}

\FloatBarrier
\clearpage
%%%%%%%%%%%%%%%%%%%%%%%%%%%%%%%%%%%%%%%
\section{Carbon star average bins}
\begin{table}[!h]
    \caption[]{Selected \mbox{C-rich} \mbox{TP-AGB} stars}
    \centering
    \begin{tabular}{|c|l|c|}
        \hline
         XSL ID & name & (R-H) \\
        \hline
         \multicolumn{3}{|c|}{Bin 1}\\
         \hline 
            X0530 & SHV 0518161-683543& 2.61  \\
            X0385 & CL* NGC 121 T V8 & 2.67  \\
            X0809 & SHV 0542111-683837 & 2.68  \\
            X0821 & SHV 0534578-702532 & 2.81  \\
            X0810 & SHV 0542111-683837 & 2.82  \\
            \hline
            \multicolumn{3}{|c|}{Bin 2}\\
            \hline 
            X0822 & SHV 0534578-702532 & 2.94  \\
            X0823 & SHV 0534578-702532 & 2.99  \\
            X0519 & SHV 0517337-725738 & 3.00   \\
            \hline
            \multicolumn{3}{|c|}{Bin 3}\\
            \hline 
            X0017 & ISO-MCMS J005716.5-731052 & 3.22  \\
            X0018 & ISO-MCMS J010031.5-730724 & 3.37  \\
            X0001 & ISO-MCMS J004900.4-732224 & 3.41  \\
            X0034 & ISO-MCMS J005307.8-730747 & 3.41  \\
            X0325 & HD 70138 & 3.44  \\
            X0002 & ISO-MCMS J004932.4-731753 & 3.48  \\
            \hline
            \multicolumn{3}{|c|}{Bin 4}\\
            \hline 
            X0013 & ISO-MCMS J005712.2-730704 & 3.64  \\
            X0515 & SHV 0500412-684054 & 3.65  \\
            X0040 & ISO-MCMS J005644.8-731436 & 3.89  \\
            X0012 & ISO-MCMS J005700.7-730751 & 3.94  \\
            X0609 & [W65] c2 & 4.10  \\
            X0512 & CL* NGC 419 LE 27 & 4.12  \\
            \hline
            \multicolumn{3}{|c|}{Bin 5}\\
            \hline 
            X0864 & [W71b] 008-03 & 4.16  \\
            X0534 & SHV 0520427-693637 & 4.17  \\
            X0860 & IRAS 09484-6242 & 4.24  \\
            X0038 & ISO-MCMS J005422.8-730105 & 4.29  \\
            X0591 & [W65] c2  & 4.30  \\
            X0505 & CL* NGC 371 LE 31 & 4.32  \\
            \hline
            \multicolumn{3}{|c|}{Bin 6}\\
            \hline 
            X0513 & CL* NGC 419 LE 35 & 4.46  \\
            X0047 & SHV 0504353-712622 & 4.54  \\
            X0803 & SHV 0529222-684846 & 4.58  \\
            X0039 & ISO-MCMS J005531.0-731018 & 4.89  \\
         \hline
    \end{tabular}
    \label{tab:carbon_stars}
\end{table}

\onecolumn
\begin{figure}
     \centering
     \begin{subfigure}

         \includegraphics[width=0.35\linewidth]{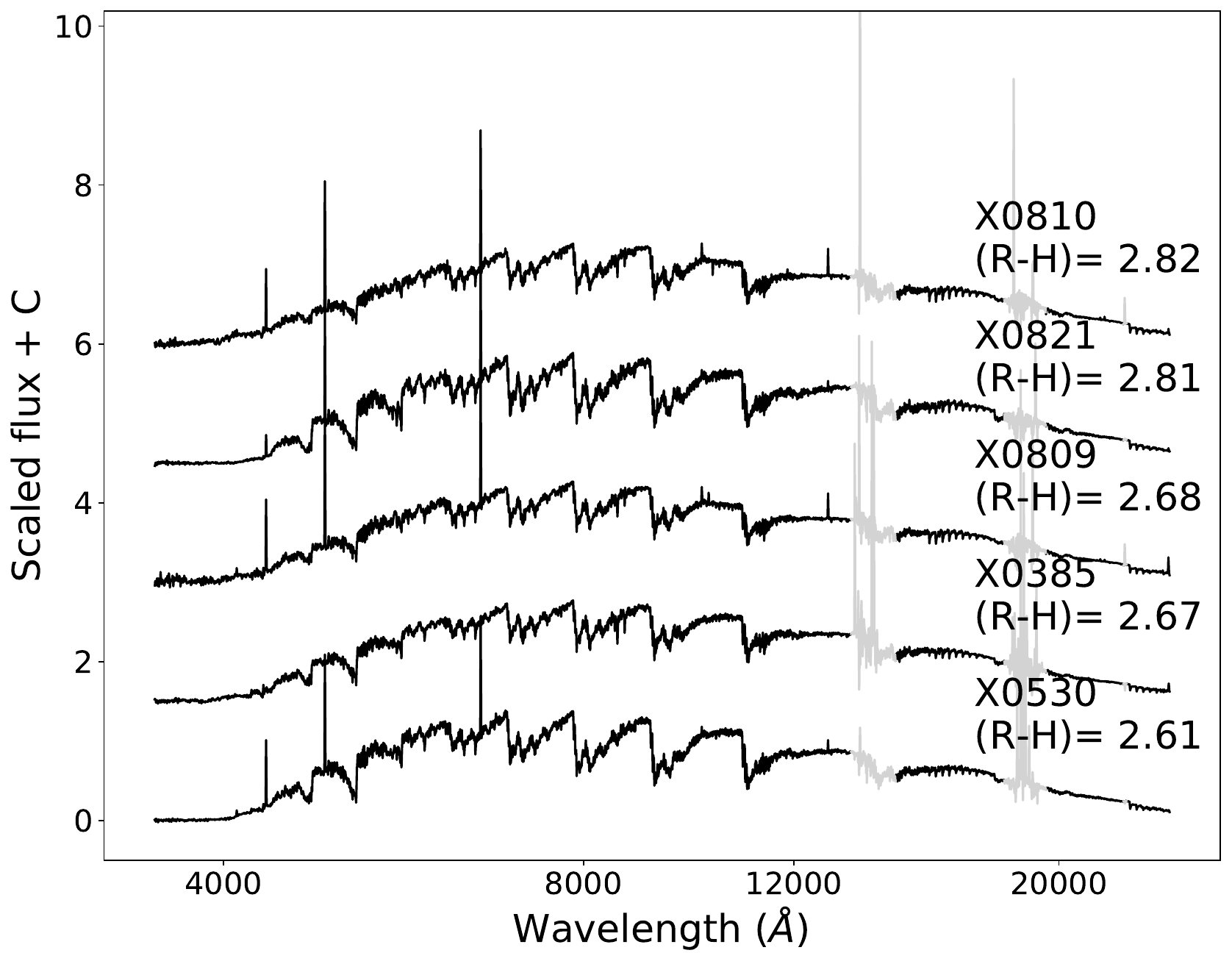}
     \end{subfigure}
     \begin{subfigure}

         \includegraphics[width=0.35\linewidth]{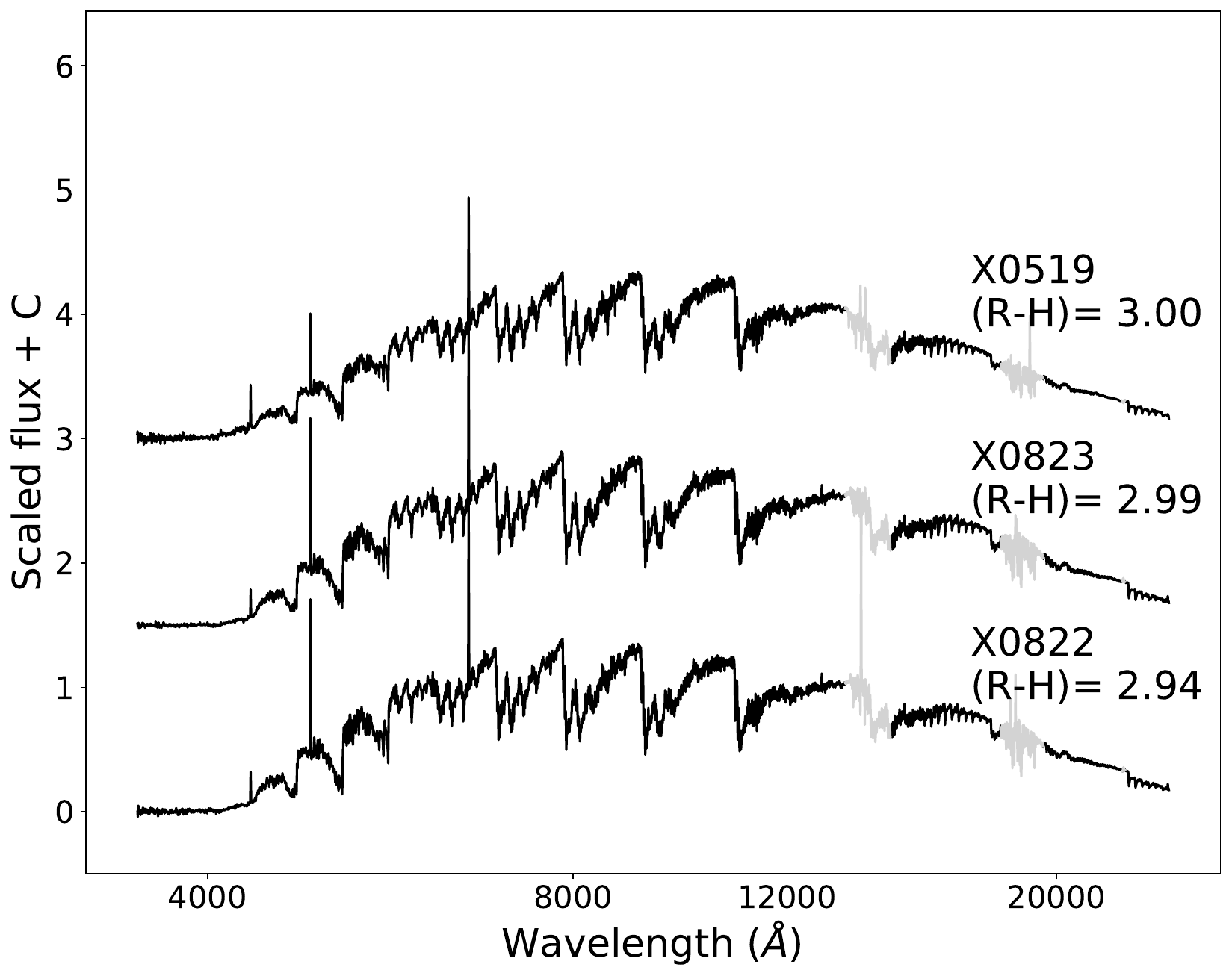}
     \end{subfigure}
     \begin{subfigure}

         \includegraphics[width=0.35\linewidth]{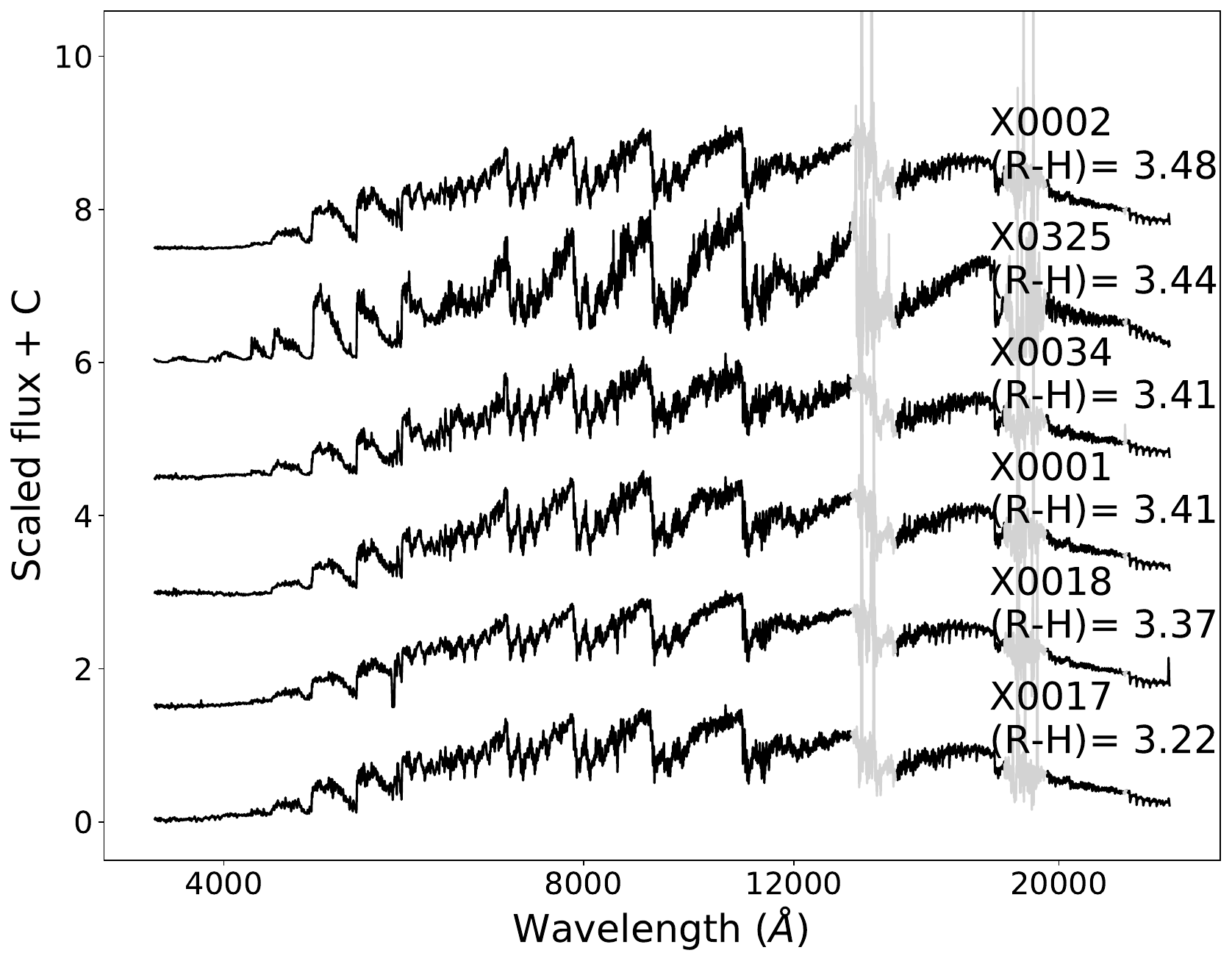}
     \end{subfigure}
     \begin{subfigure}

         \includegraphics[width=0.35\linewidth]{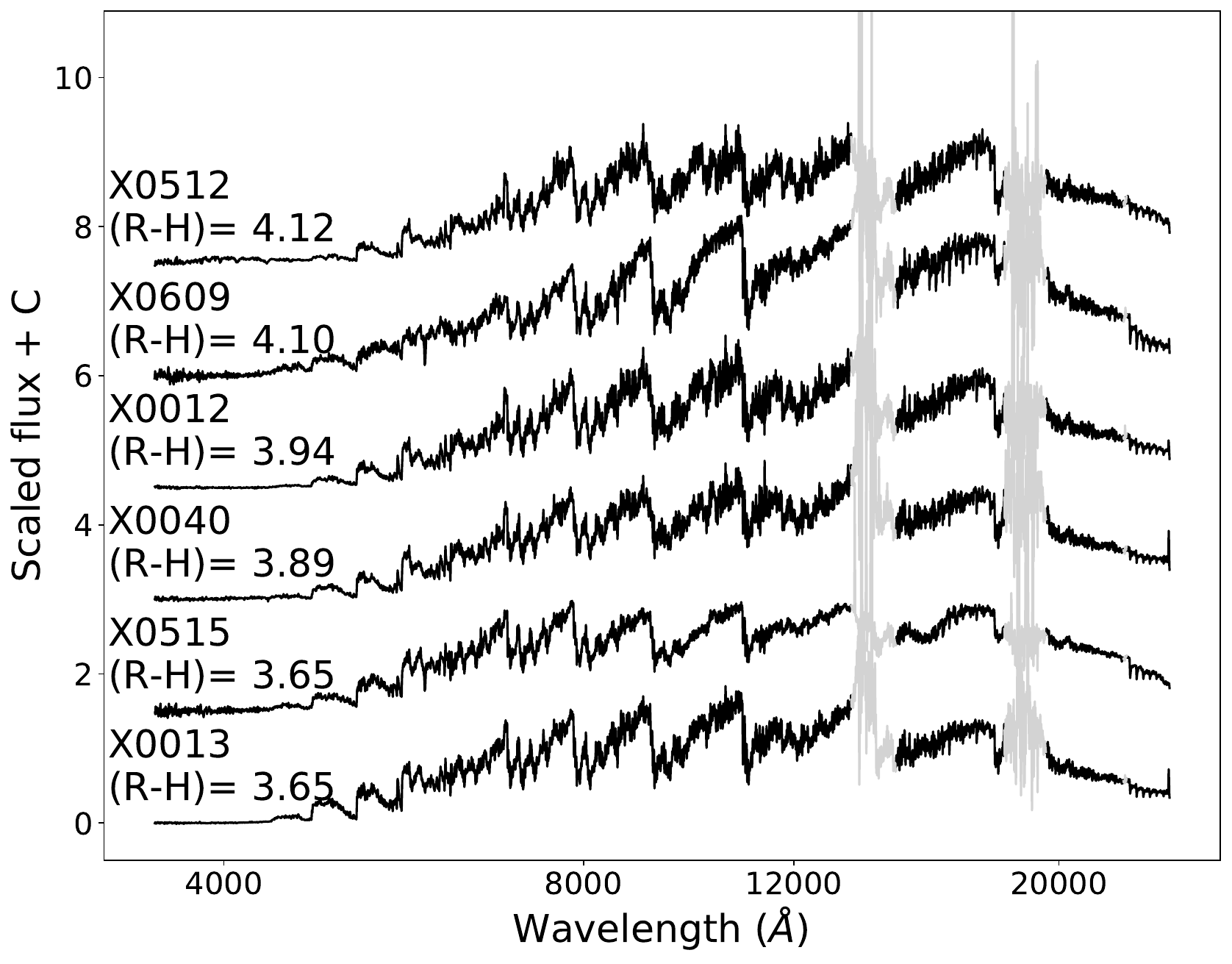}
     \end{subfigure}
     \begin{subfigure}

         \includegraphics[width=0.35\linewidth]{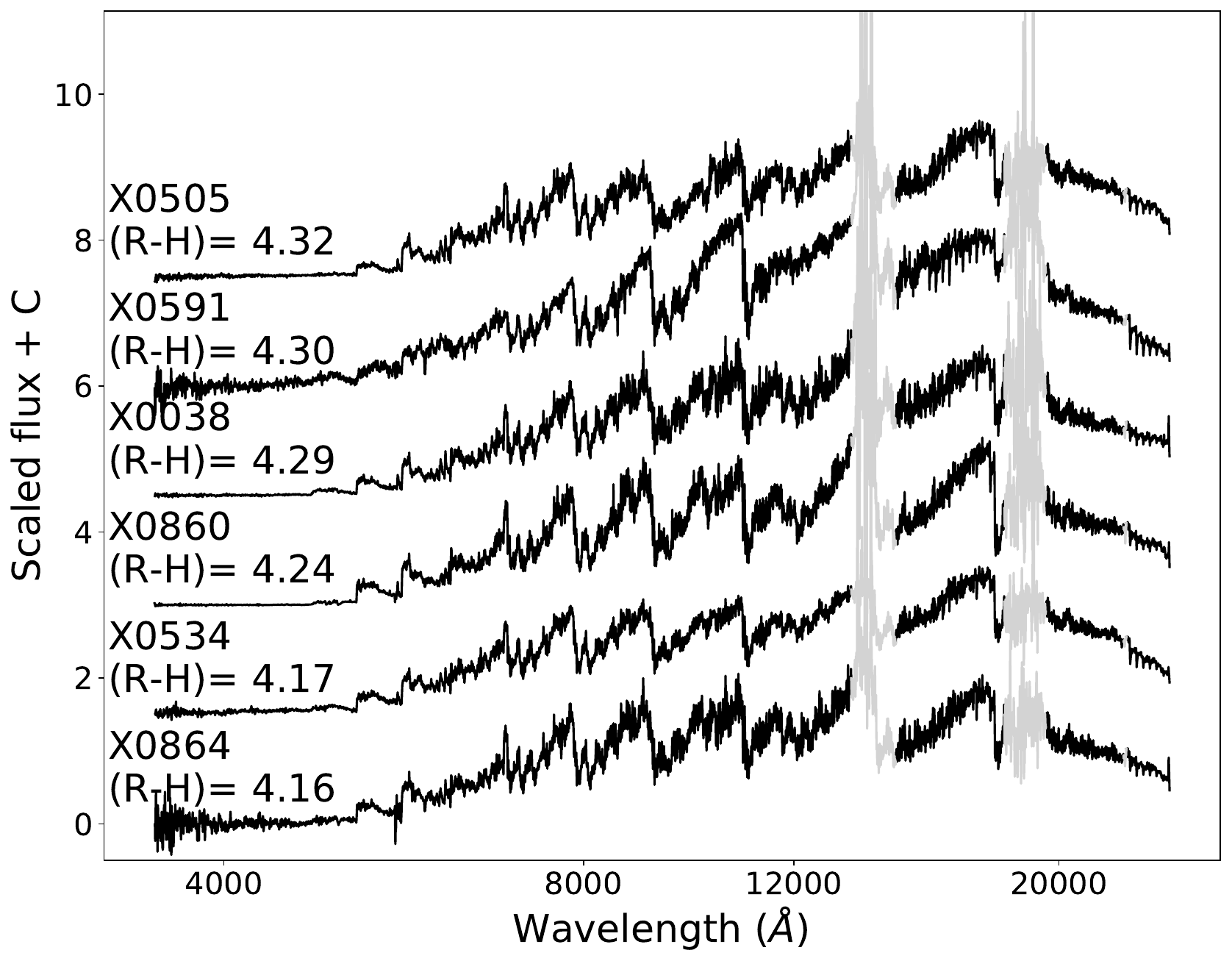}
     \end{subfigure}
     \begin{subfigure}

         \includegraphics[width=0.35\linewidth]{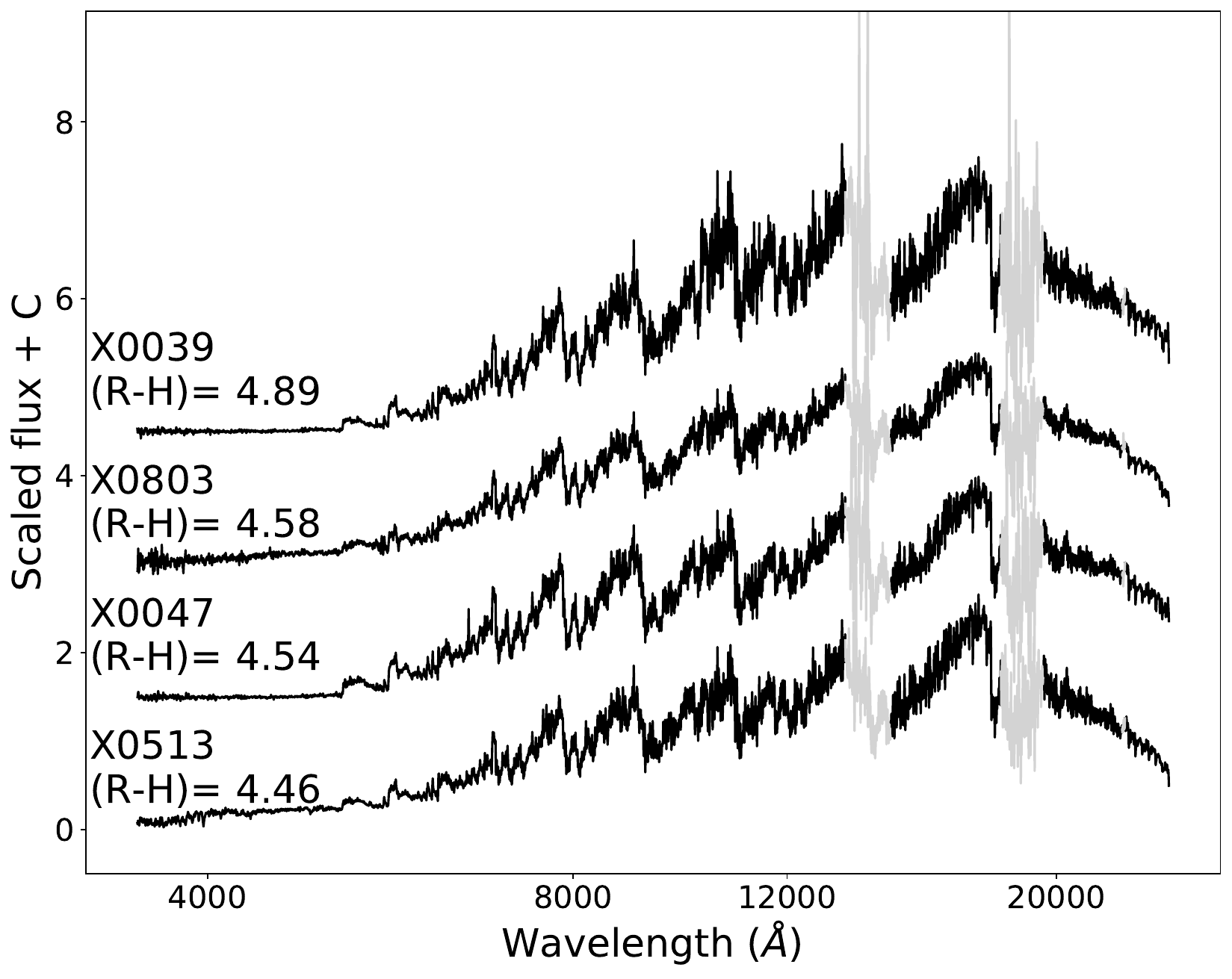}
     \end{subfigure}\
        \caption{Spectra of XSL \mbox{C-rich} \mbox{TP-AGB} stars from which the \mbox{C-rich} \mbox{TP-AGB} sequence was constructed.}
    \label{fig:Cstar_bins}
\end{figure}
\FloatBarrier
\clearpage
\begin{multicols}{2}
\section{XSL PARSEC/COLIBRI and E-MILES BaSTI absorption line index grids}
\end{multicols}
\begin{figure*}[!h]
    \centering
    \begin{subfigure}[]
        \centering
        \includegraphics[width = 0.3\linewidth]{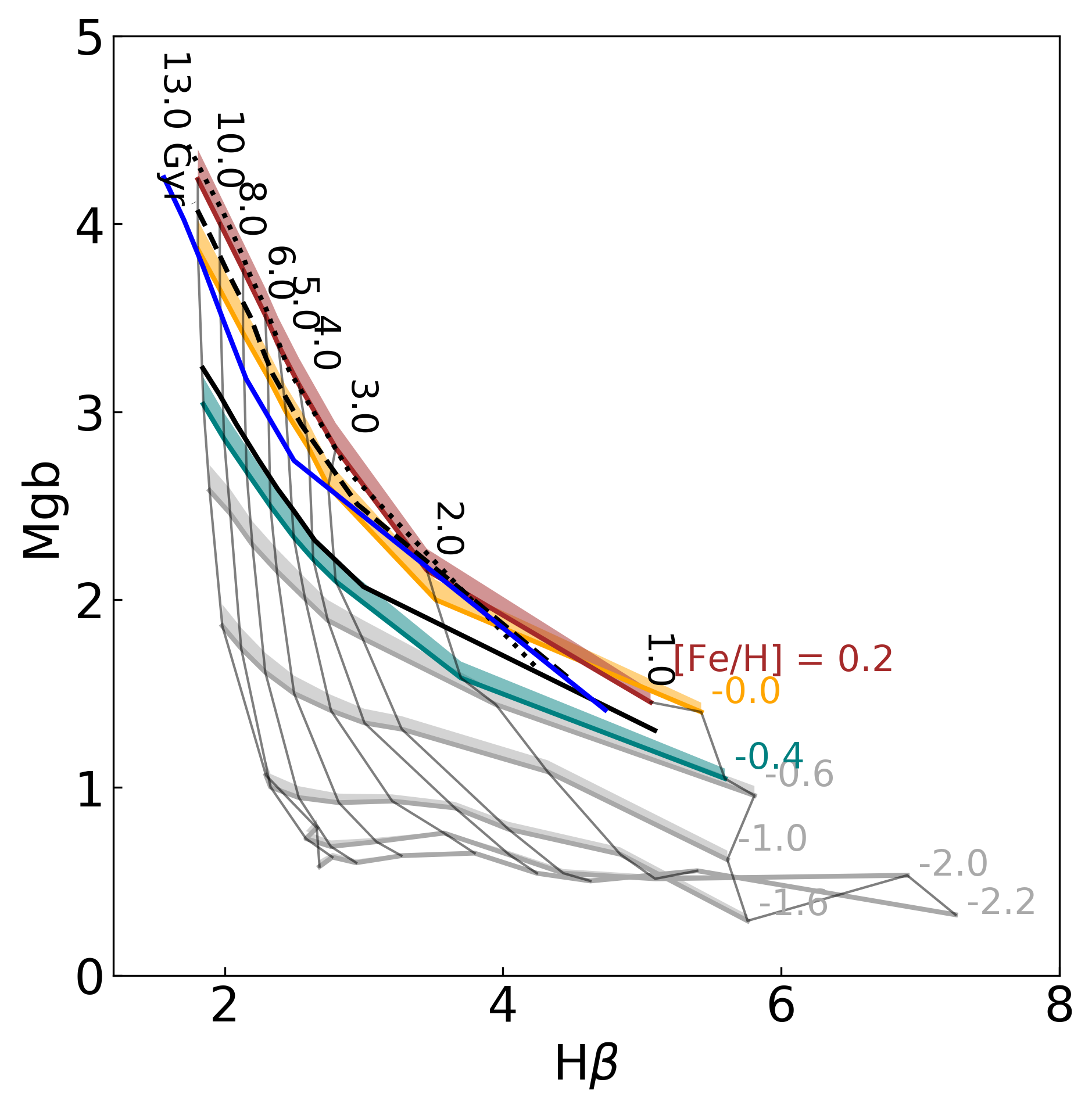}
    \end{subfigure}%
    ~ 
    \begin{subfigure}[]
        \centering
        \includegraphics[width = 0.3\linewidth]{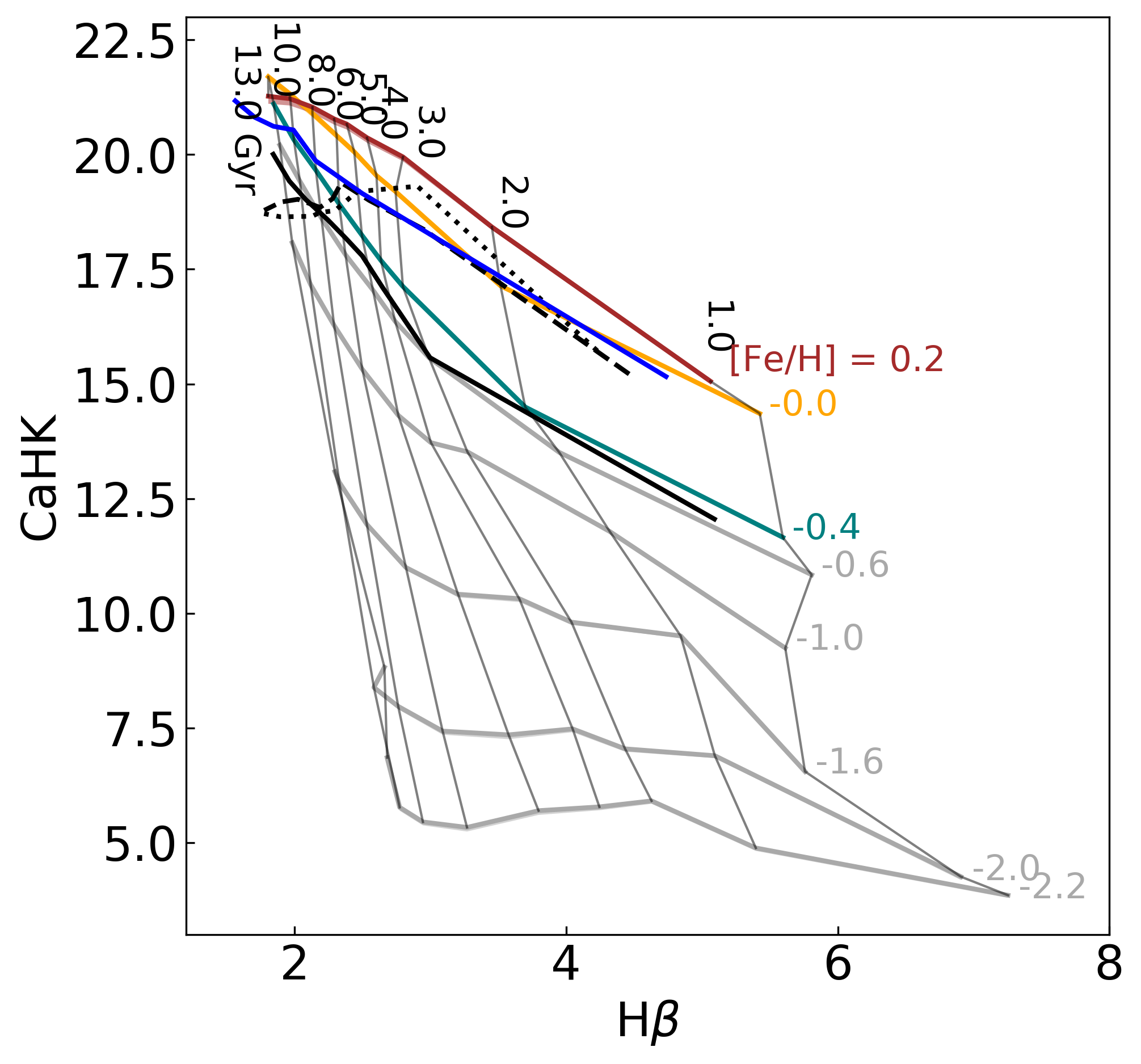}
    \end{subfigure}%
    ~ 
    \begin{subfigure}[]
        \centering
        \includegraphics[width = 0.3\linewidth]{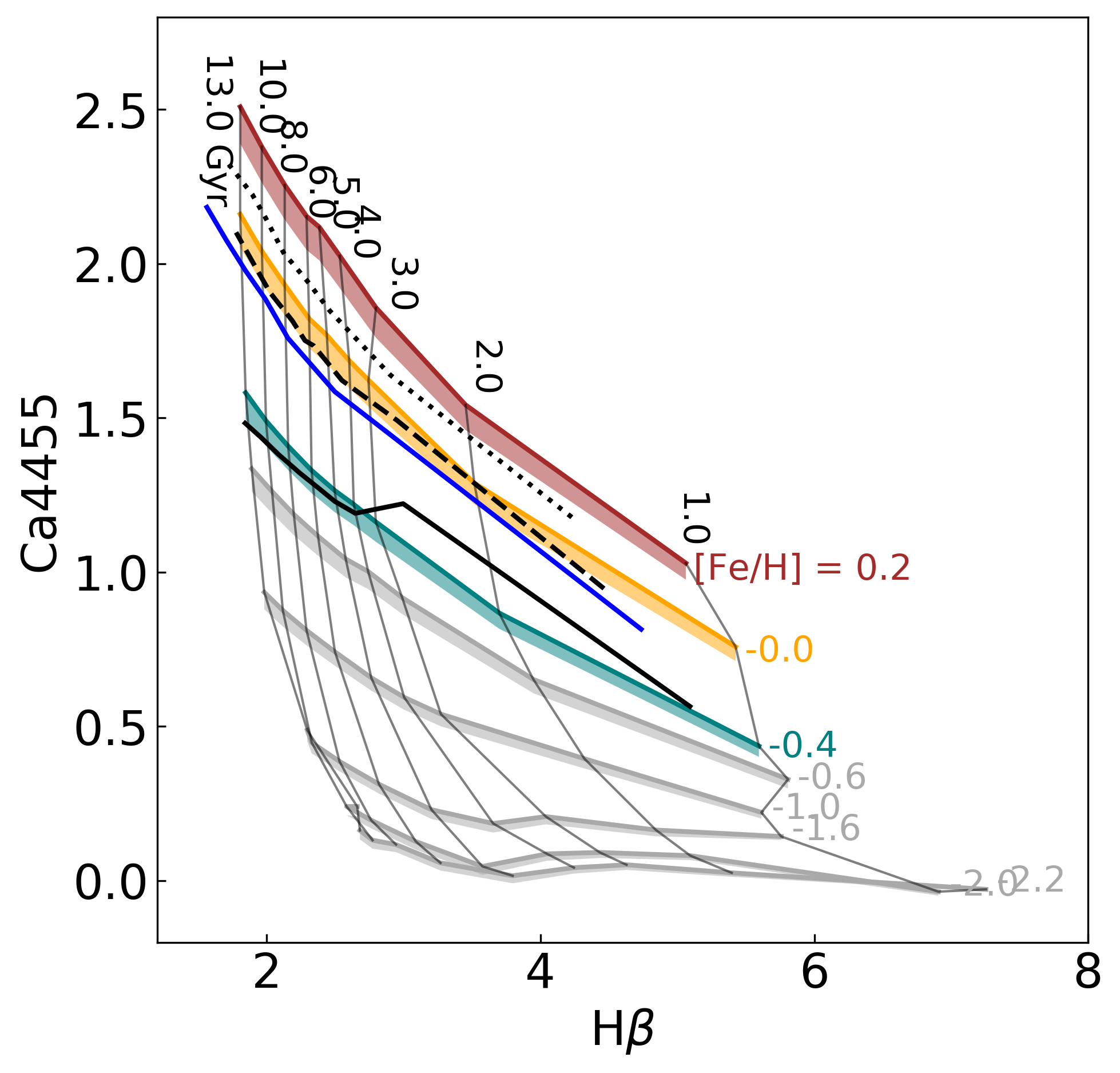}
    \end{subfigure}
    ~ 
    \begin{subfigure}[]
        \centering
        \includegraphics[width = 0.3\linewidth]{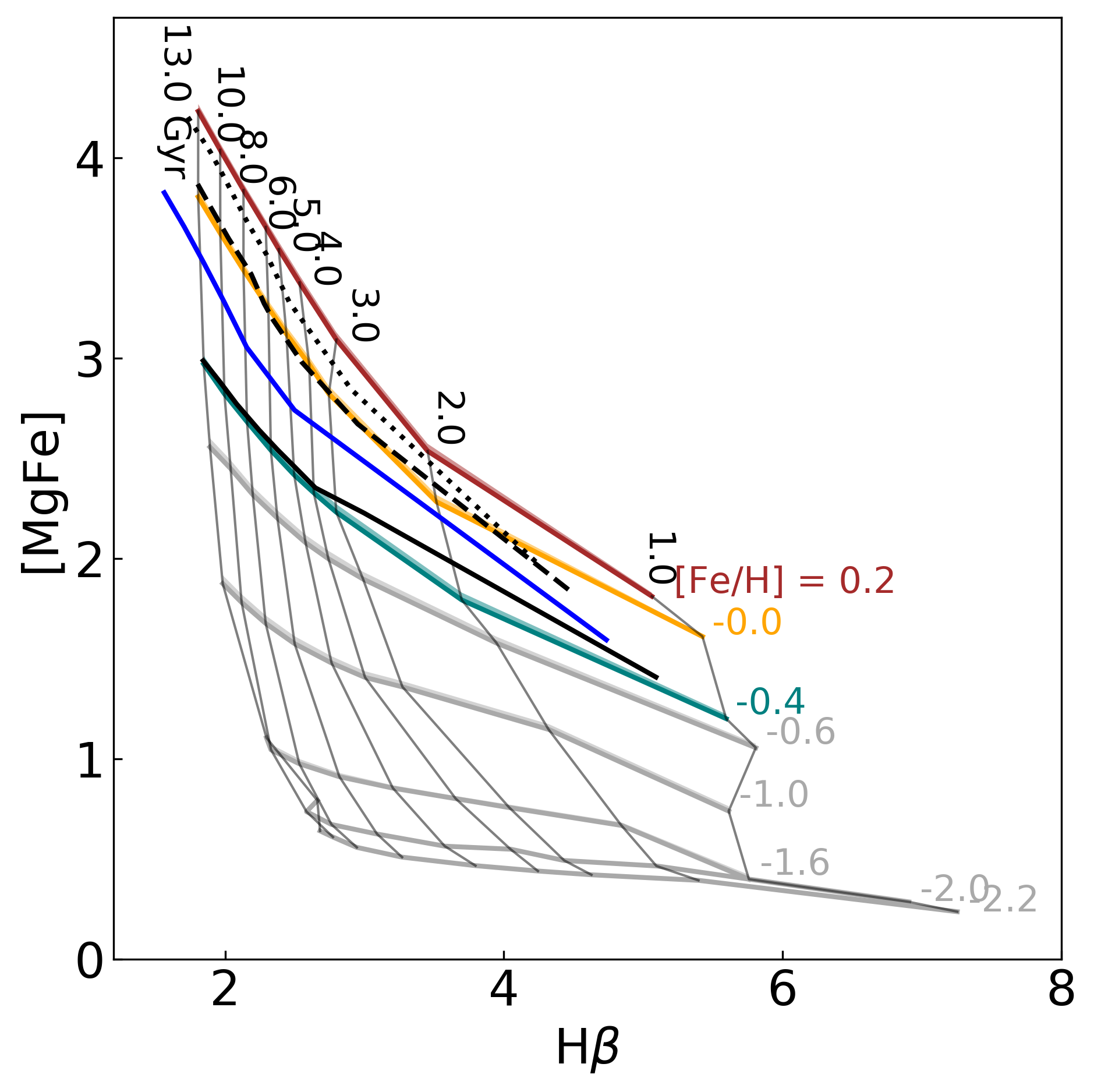}
    \end{subfigure}
    ~ 
    \begin{subfigure}[]
        \centering
        \includegraphics[width = 0.3\linewidth]{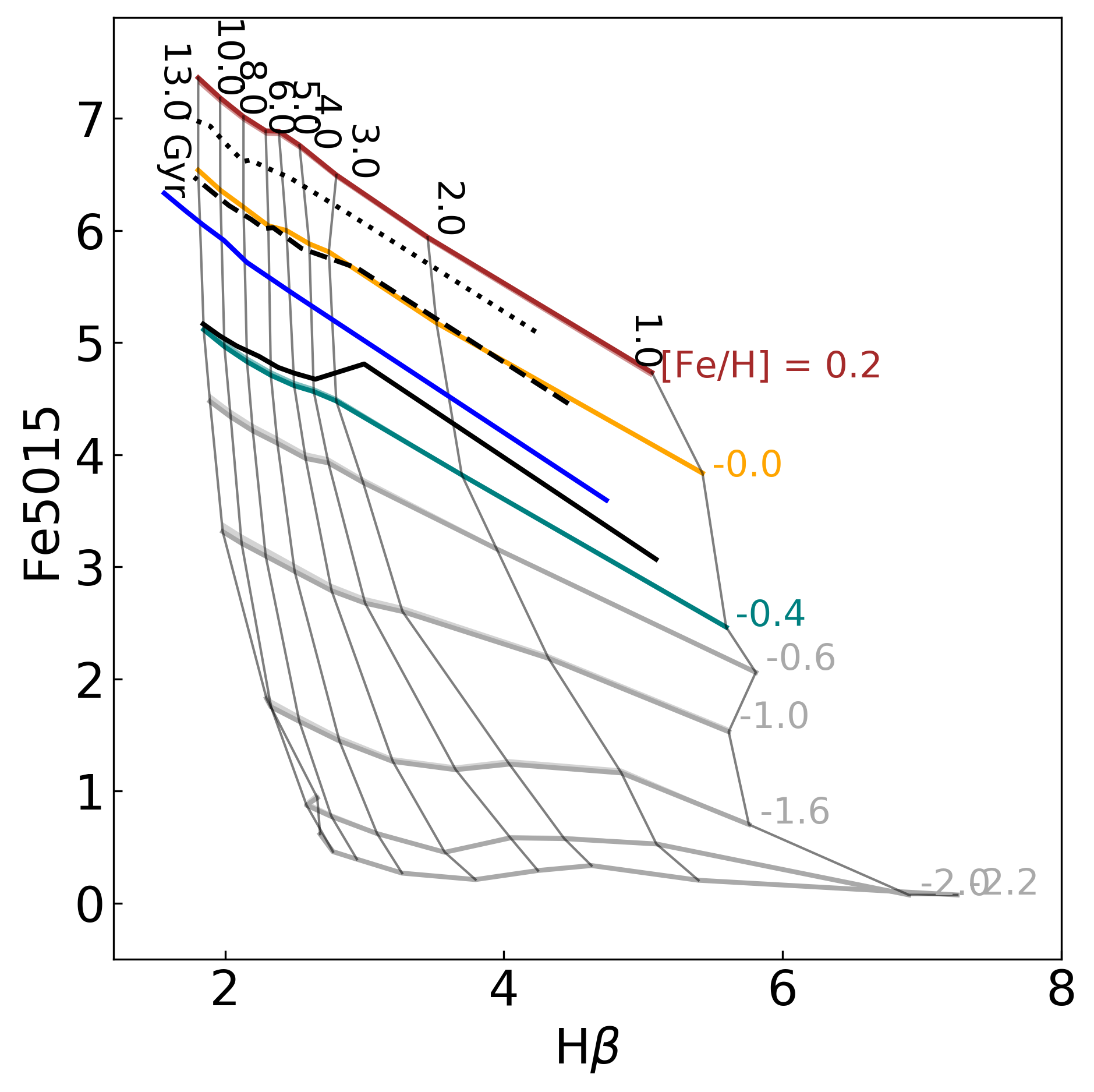}
    \end{subfigure}
    ~ 
     \begin{subfigure}[]
        \centering
        \includegraphics[width = 0.3\linewidth]{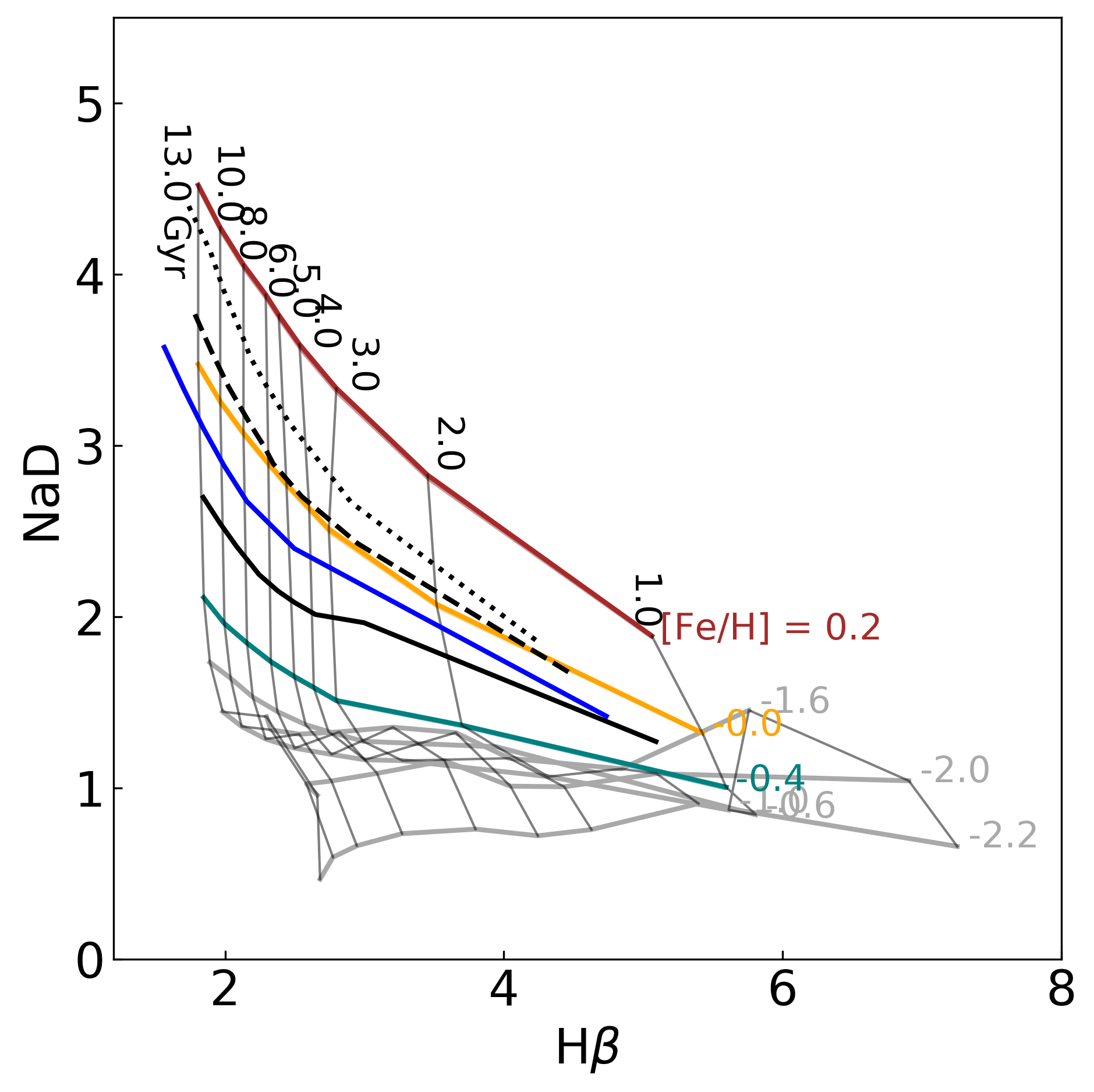}
    \end{subfigure}
    ~ 

    \caption{Comparison of the behaviour of the Mgb, CaHK, Ca4455, [MgFe], Fe5015, and NaD absorption-line indices as a function of the H$\beta$ index. The shaded areas represent XSL PARSEC/COLIBRI models with varying spectral resolution from $\sigma=13\,\mathrm{km\,s}^{-1}$ (the XSL native resolution) to $\sigma=60\,\mathrm{km\,s}^{-1}$ (the minimum E-MILES resolution in the optical). Black lines represent E-MILES BaSTI model predictions with dotted, dashed and solid line representing $\mathrm{[Fe/H]}=+0.2$, $0.0$, and $-0.4$\, dex, respectively, measured at the original E-MILES resolution. The blue solid line represents the predictions of the C18 solar-metallicity models.}
    \label{fig:indices_comparison_PC}
    \end{figure*}
    
\clearpage    
\section{$M_*/L$ ratio model comparisons}
\label{MLratiomodelcomparison}  

\begin{figure}[!h]
    \centering
    \includegraphics[width = 0.74\textwidth]{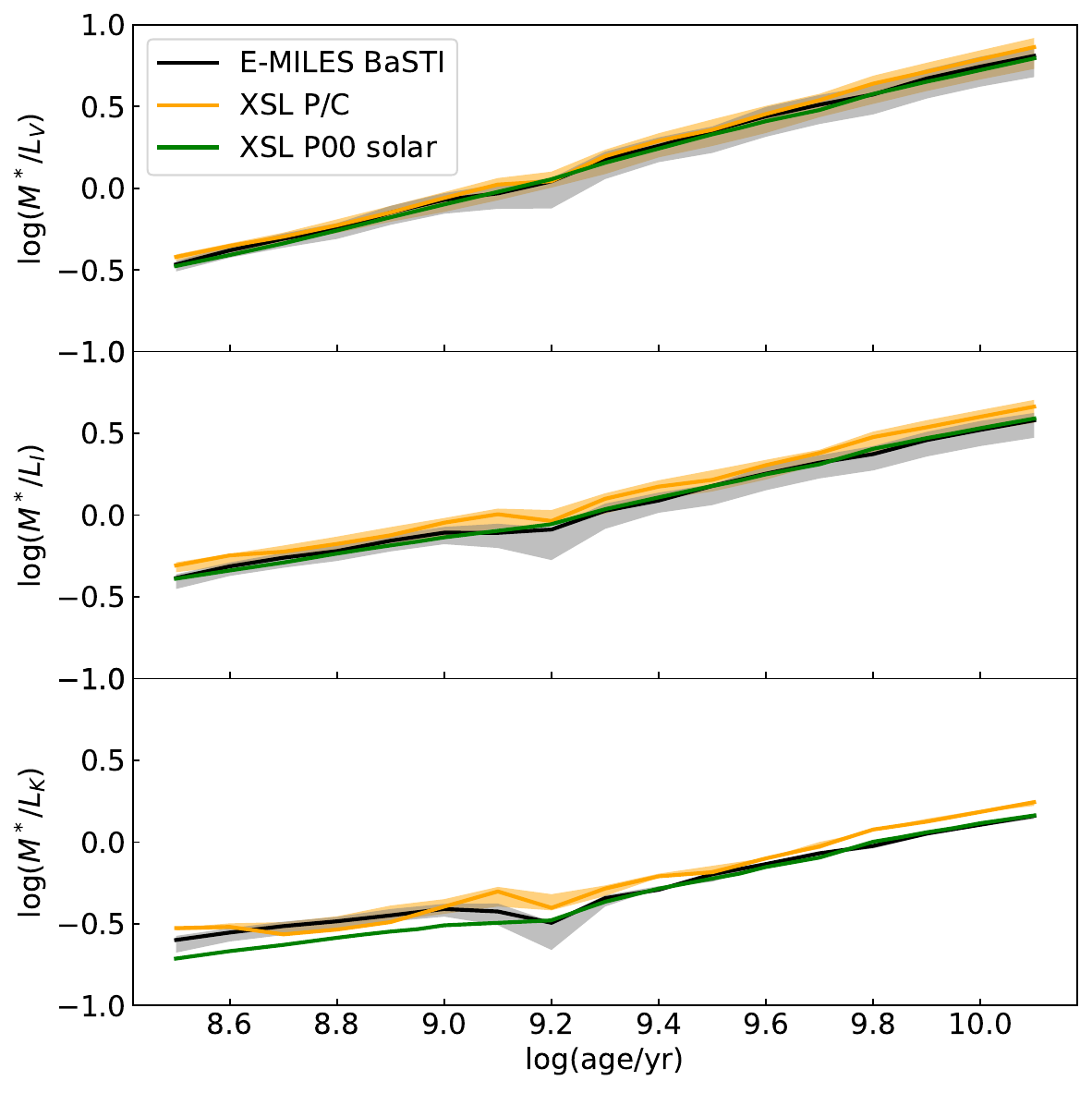}
    \caption{$M_*/L$ ratio derived from XSL PARSEC/COLIBRI (P/C) Salpeter models (orange), XSL Padova00 (P00) Salpeter models (green) and E-MILES BaSTI Salpeter models (black). The shaded areas represent the spread in $M_*/L$ arising from the metallicity spread from \mbox{$\mathrm{[Fe/H]} = -0.4$\,dex} (lower $\log M_*/L$) to \mbox{$\mathrm{[Fe/H]} = +0.2$\,dex} (higher $\log M_*/L$), centered on the solar metallicity values.}
    \label{fig:ML_comp_emiles}
\end{figure}
\begin{multicols}{2}
Figure \ref{fig:ML_comp_emiles} shows $\log(M_*/L)$ in the $V$-, $I$-, and $K$-bands derived from the XSL PARSEC/COLIBRI Salpeter models, XSL Padova00 Salpeter models and E-MILES BaSTI Salpeter models. The $M_*/L$ ratios for E-MILES models are taken from the model prediction tables provided by the MILES collaboration on their web page\footnote{The E-MILES  $M_*/L$ ratio predictions in the Johnson/Cousin filters (Vega system) can be found at \url{http://research.iac.es/proyecto/miles/pages/photometric-predictions-based-on-e-miles-seds.php}} \citep{MIUSCAT,Ricciardelli2012,UVMILES}. 

The predicted $M_*/L$ ratios differ among SSP models (assuming they use the same IMF). This discrepancy reflects, in part, a different accounting of stellar remnants and possible stellar mass span of the IMF. However, the biggest differences originate from the modelling of the advanced evolutionary stages of stars and found in the NIR. The XSL PARSEC/COLIBRI models include more stellar evolutionary phases, namely the TP-AGB phase, than the Padova00 models. These NIR-bright stars give their contribution to the light budget of a population, increasing its NIR luminosity and lowering its $\log(M_*/L_\mathrm{NIR})$. 

This is clear from Fig. \ref{fig:ML_comp_emiles}, where the XSL PARSEC/COLIBRI models show a dip at $\log(M_*/L_K)$  0.5-1 Gyr (coinciding with the SSP model ages where TP-AGB start becoming dominant, see Fig. \ref{fig:Cooliantfracs}), but XSL Padova00 and E-MILES BaSTI models do not. XSL Padova00 models monotonically decrease $\log(M_*/L)$ towards younger ages with the increase of the $K$-band flux fraction of TP-AGB stars, seen in Fig. \ref{fig:Cooliantfracs}. 

Over all ages, the XSL PARSEC/COLIBRI models have higher $\log(M_*/L)$ compared to XSL Padova00 models, due to having more mass in existing stars. The TP-AGB stars after the first thermal pulse are non-existent in Padova00 isochrones (replaced by remnants), but PARSEC/COLIBRI models still have them. The $K$-band luminosities of older XSL PARSEC/COLIBRI models are also lower, which is responsible for the increased $M_*/L_{K}$ compared to XSL Padova00 models. Hence, the $M_*/L_\mathrm{NIR}$ ratio depends on the isochrones used as well as on the coverage of the spectral library and the modelling methods.
\end{multicols}
\clearpage
\section{Examples of XSL PARSEC/COLIBRI SSP models}

\label{fig:SSP_examples}
\begin{figure*}[!h]
    \centering
    \begin{subfigure}

        \includegraphics[width = 0.35\linewidth]{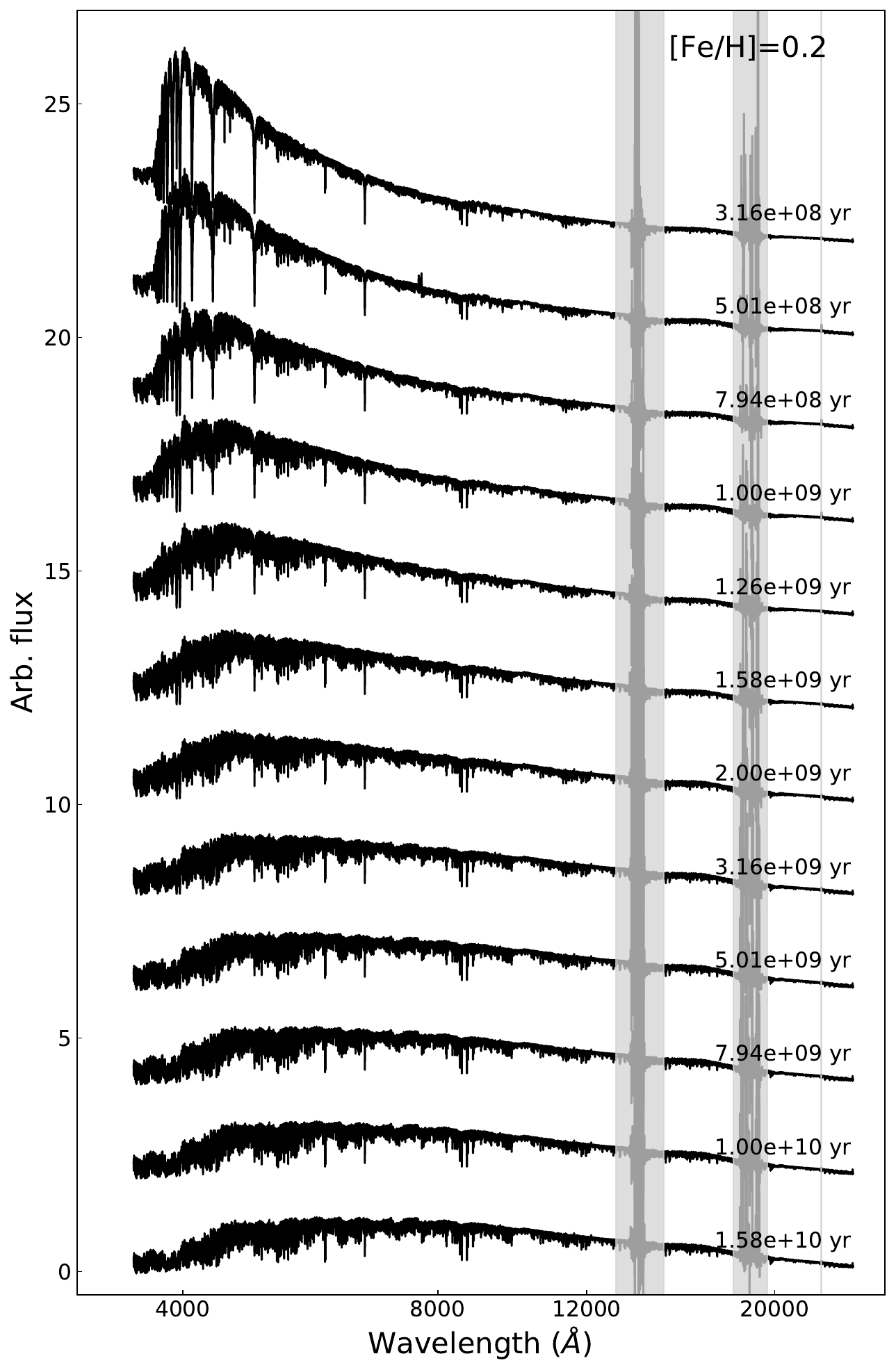}
    \end{subfigure}%
    \begin{subfigure}

        \includegraphics[width = 0.35\linewidth]{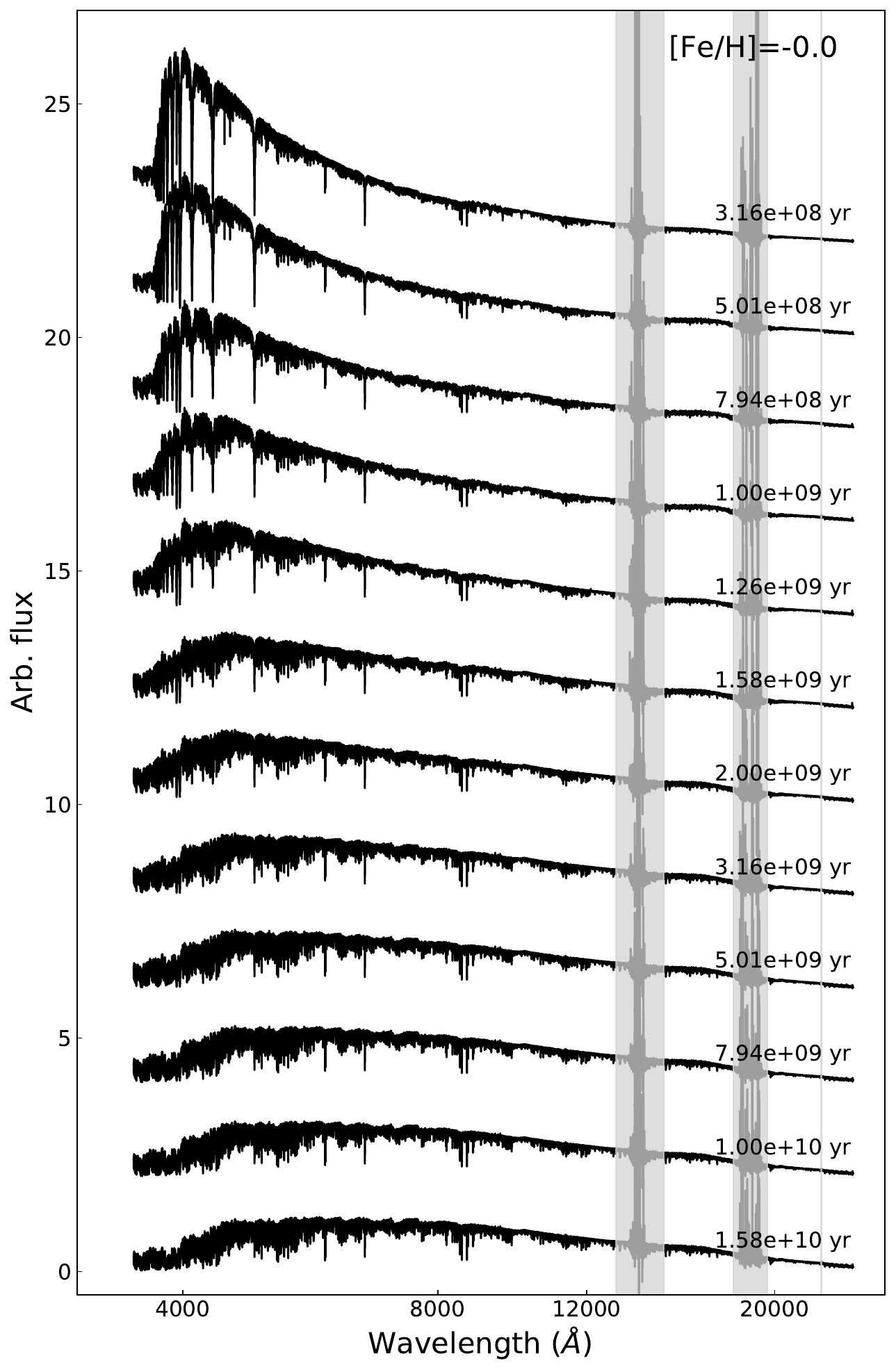}
    \end{subfigure}%
    \begin{subfigure}

        \includegraphics[width = 0.35\linewidth]{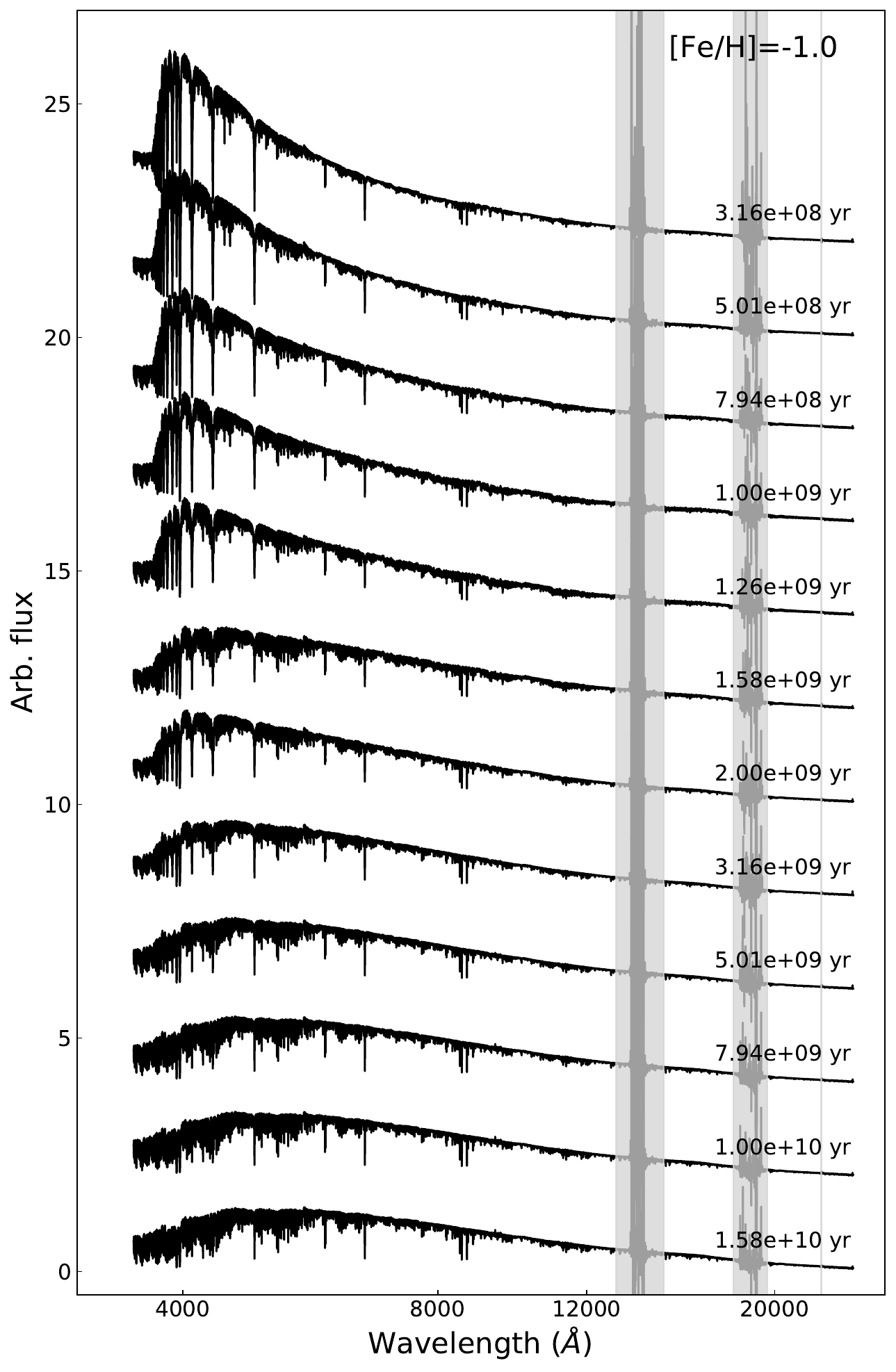}
    \end{subfigure}
    \begin{subfigure}

        \includegraphics[width = 0.35\linewidth]{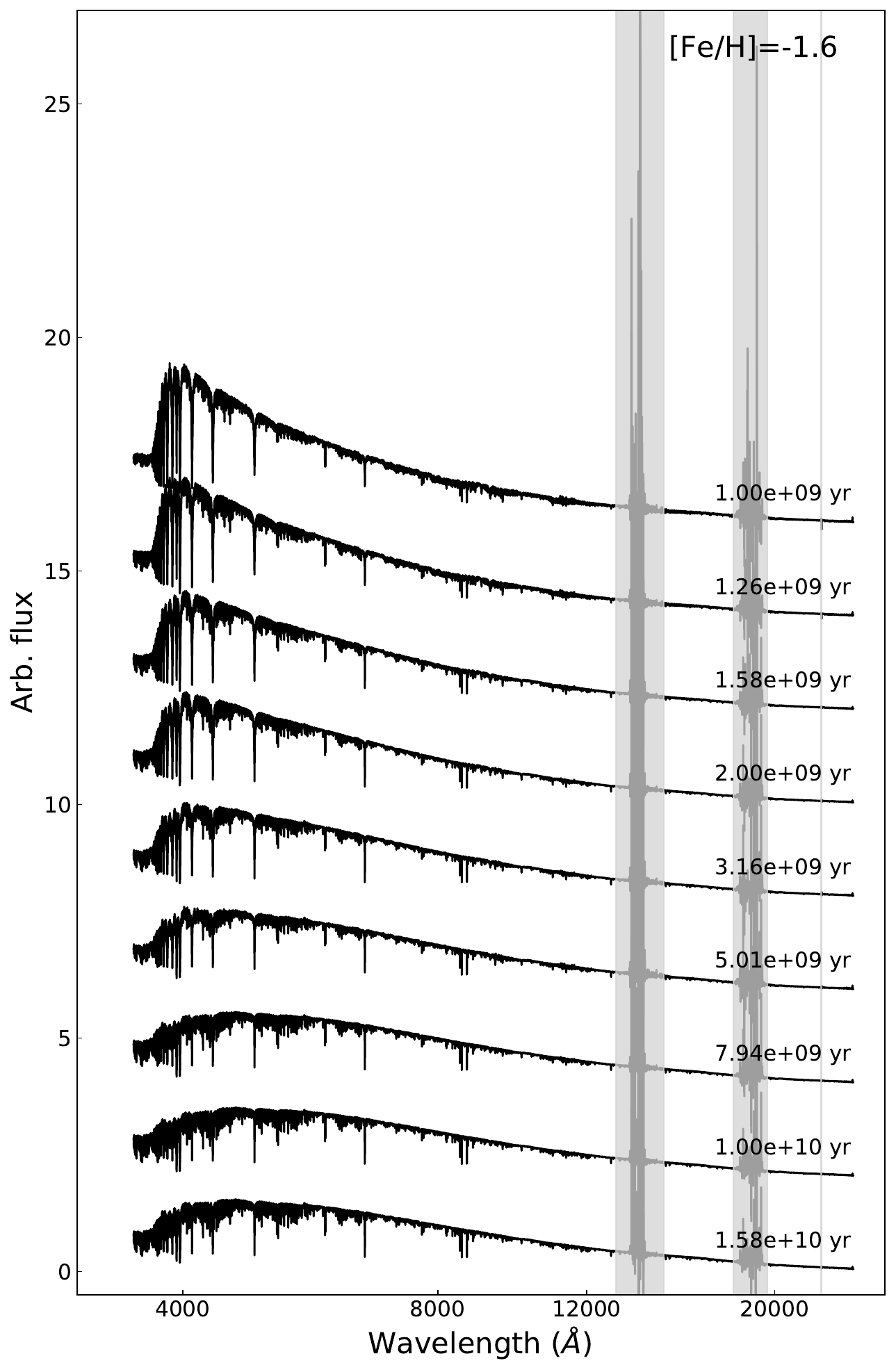}
    \end{subfigure}

\caption{Examples of XSL PARSEC/COLIBRI Salpeter SSP models, for four metallicities: $\mathrm{[Fe/H]}=+0.2$, $0.0$, $-1.0$ and $-1.6$\, dex. We note that for $\mathrm{[Fe/H]}=-1.6$\, dex younger than 1~Gyr models are not shown, because they are outside the safe zones of XSL SSP models. Spectra are show in original resolution. The telluric absorption is marked in gray.}
    
    \end{figure*}

\end{appendix}
\end{document}